\documentclass[12pt]{article}
\usepackage{natbib,epsf,color}
\usepackage{graphicx}
\usepackage{subfigure}
\usepackage{multirow}
\usepackage{booktabs}
\usepackage{amssymb}
\usepackage{amsthm}
\usepackage{amsmath}
\usepackage{multirow,setspace}
\usepackage[figuresright]{rotating}
\begin{document}
\newtheorem{theorem}{Theorem}
\newtheorem{lemma}{Lemma}
\newtheorem{proposition}{Proposition}
\newtheorem{corollary}{Corollary}
\theoremstyle{remark}
\newtheorem{remark}{Remark}
\title{Bayesian Quantile Regression for Partially Linear Additive Models }

\author{Yuao Hu, Kaifeng Zhao and Heng Lian\\
 \small Division of Mathematical Sciences, School of Physical and Mathematical Sciences,\\
\small  Nanyang Technological University, \\
\small  Singapore,  637371\\
  }
\date{}

\maketitle
\begin{abstract}
In this article, we develop a semiparametric Bayesian estimation and model selection approach for partially linear additive models in conditional quantile regression. The asymmetric Laplace distribution provides a mechanism for Bayesian inferences of quantile regression models based on the check loss.  The advantage of this new method is that nonlinear, linear and zero function components can be separated automatically and  simultaneously during model fitting without the need of pre-specification or parameter tuning. This is achieved by spike-and-slab priors using two sets of indicator variables. For posterior inferences, we design an effective partially collapsed Gibbs sampler. Simulation studies are used to illustrate our algorithm. The proposed approach is further illustrated by applications to two real data sets.
\end{abstract}

\emph{\textbf{Keywords}}:  Additive models; Markov chain Monte Carlo; Quantile regression; Variable selection. 

\section{Introduction}
Partially linear additive models (PLAMs) generalize multiple linear regression models. They can also be regarded as a special case of generalized additive regression models (Hastie and Tibshirani 1990). PLAMs, containing both linear and nonlinear additive components, are more flexible than stringent linear models, and are more parsimonious than general nonparametric regression models and they circumvent the difficulty brought by the problem known as ``curse of dimensionality". PLAMs have been widely applied in practice because of these advantages. For example, Liang et al (2008) applied PLAMs to study the relationship between environmental chemical exposures and semen quality, and \cite{anagiotelis2008bayesian} applied the models for intra-day electricity load analysis.

In this article, we propose a Bayesian quantile regression approach for partially linear additive models. At a given quantile level $\tau\in (0,1)$, a partially linear additive model has the following form,
\begin{equation}
\label{Bmodel}
y_i=\mu_{\tau}+\mathop{\sum}\limits_{j=1}\limits^{p}f_{\tau,j}(x_{ij})+\epsilon_{\tau,i}, i=1,\ldots,n,
\end{equation}
where $(y_i, \boldsymbol{x}_i)$ are independent and identically distributed pairs. Here $y_i$ is the response, $\boldsymbol{x_i}=(x_{i1},\ldots, x_{ip})^T$ is the $p$-dimensional predictor, $\mu_{\tau}$ is the intercept, $\epsilon_i, i=1, \ldots, n,$ are random errors with their $\tau$th quantile equal to 0 and $f_{\tau,j}$ is a univariate component function which might be nonlinear, linear or zero. Quantile regression (\cite{koenker1978regression}) has been demonstrated to be valuable by a rapidly expanding literature in economics, social sciences, and biomedical studies (\cite{buchinsky1994changes}, \cite{abrevaya2001effects}, \cite{cade2003gentle}). It provides more robust analyses and more complete descriptions of data structure than traditional mean regression. Quantile regression for additive models has previously been considered in the literature. In the frequentist context, \cite{de2003additive}, \cite{horowitz2005nonparametric}, \cite{yu2004local} all developed methodologies of nonparametric estimation for additive quantile regression. In the Bayesian context, \cite{yue2011bayesian} proposes a Bayesian quantile regression approach for additive mixed models.

There are some works focusing on the selection of significant components using penalization approaches from the frequentist perspective, such as \cite{ravikumar2009sparse}, \cite{meier2009high} and \cite{huang2010variable}. Besides, a number of existing papers are concerned with the selection of function components using Bayesian inferences recently (\cite{anagiotelis2008bayesian},  \cite{shively1999variable}, and \cite{yau2003bayesian}). They express each function as a linear combination of basis terms and assign priors on the coefficients of the basis functions. They all introduce indicator variables to enable variable selection.  However none of the works mentioned above considers linear component identification. These works considered only least squares regression. Usually, pre-specification of linear components is required. Recently, several works consider performing variable selection, parametric component identification, and parameter estimation all at the same time (\cite{zhang2011linear}, \cite{lian2011identification}) from a frequentist perspective, for mean regression. 

The quantile regression approach we propose in this article has the ability of separating function components into those with nonlinear effects, those with linear effects, and those irrelevant to responses, in the context of quantile regression. In a Bayesian context, this separation of components is a soft decision based on the posterior probabilities of the components being selected as nonlinear, linear, or zero. We establish a hierarchical Bayesian model by adopting the asymmetric Laplace distribution for errors (\cite{yu2001bayesian}). Then, extending the Bayesian variable selection approach, we introduce two sets of indicator variables in the spike-and-slab priors which make the separation of components possible. \cite{scheipl12} also conducted a similar study which can separate the components into nonlinear, linear and zero ones, for generalized additive models.

 The remainder of the paper proceeds as follows. In Section 2, we describe our hierarchical Bayesian model for quantile regression based on the additive model structure. We also discuss our prior choices and introduce the posterior sampling algorithm focusing on an efficient partially collapsed sampler. The details of the algorithm are explained in the Appendix. In Section 3, we present numerical illustrations including simulation studies and two real data examples.  In Section 4, we conclude the paper with a discussion.

\section{Hierarchical Bayesian Modeling}
Quantile regression is typically achieved by solving a minimization problem based on the check loss function. With model (\ref{Bmodel}), the specific problem is estimating $\mu_{\tau}$ and the sequence of functions $f_{\tau,j}, j=1,\ldots, p,$ by minimizing the following objective function,
\begin{equation}
\label{risk}
L(\boldsymbol{y}, \boldsymbol{x})=\mathop{\sum}\limits_{i=1}\limits^{n}\rho_{\tau}(y_i-\mu_\tau-\mathop{\sum}\limits_{j=1}\limits^{p}f_{\tau,j}(x_{ij})),
\end{equation}
where $\boldsymbol{y}=(y_1,\ldots, y_{n})^{T }$, $\boldsymbol{x}=(\boldsymbol{x}_1,\ldots, \boldsymbol{x}_n)^{T }$, and $\rho_{\tau}(u)=u(\tau-I(u\leq 0))$ is the so called $check\ function$. In a Bayesian setup, we assume $\epsilon_{\tau,i}$, $i=1,\ldots, n$, are  i.i.d. random variables from an asymmetric Laplace distribution with density
\begin{equation*}
p(\epsilon_{\tau,i})=\frac{\tau(1-\tau)}{\delta_{0}}\exp\{-\frac{1}{\delta_{0}}\rho_{\tau}(\epsilon_{\tau,i})\},
\end{equation*}
where $\delta_{0}$ is the scale parameter.
Then the conditional distribution of  $\boldsymbol{y}$ is in the form of 
\begin{equation}
\label{likelihood}
p(\boldsymbol{y}|\boldsymbol{x})=\frac{\tau^{n}(1-\tau)^{n}}{\delta_{0}^{n}}\exp\{-\frac{1}{\delta_{0}}\mathop{\sum}\limits_{i=1}\limits^{n}\rho_{\tau}(y_i-\mu_{\tau}-\mathop{\sum}\limits_{j=1}\limits^{p}f_{\tau,j}(x_{ij}))\}.
\end{equation}
Hence, maximizing the likelihood (\ref{likelihood}) is equivalent to minimizing (\ref{risk}), giving (\ref{risk}) a likelihood-based interpretation. By introducing the location-scale mixture representation of the asymmetric Laplace distribution (\cite{kozumi2011gibbs}), (\ref{likelihood}) can be equivalently written as 
\begin{equation}
\label{function}
y_i=\mu_{\tau}+\mathop{\sum}\limits_{j=1}\limits^{p}f_{\tau,j}(x_{ij})+k_1e_i+\sqrt{k_2\delta_{0} e_i}z_i,
\end{equation}
where $e_i\sim\exp(1/\delta_{0})$ follows an exponential distribution with mean $\delta_{0}$, $z_i$ follows the standard normal distribution and is independent of $e_{i}$, $k_1=\frac{1-2\tau}{\tau(1-\tau)}$, and $k_{2}=\frac{2}{\tau(1-\tau)}$. For ease of notation, we will omit $\tau$ in the expressions in the following.

We assume the distribution of $x_j,~ j=1,\ldots, p$, is supported on $[0,1]$ and also impose the condition $Ef_j(x_j)=0$ for identifiability. To model each unknown function $f_j$ flexibly, we use the truncated power splines to approximate the functions in this article. Let $t_0=0<t_1<\cdots<t_k<1=t_{k+1}$ partition $[0,1]$ into subintervals $[t_i,t_{i+1})$, $i=0,\ldots, k$ with $k$ internal knots. Here we focus on equally spaced knots although more complicated data-driven choice can be considered.  For a given degree $q$, $B_{0}(x),\ldots, B_{K}(x)$ are used to denote $K+1=q+k$ truncated power spline basis $x, x^{2}, \ldots,  x^{q},(x-t_1)^qI(x>t_1),\ldots, (x-t_{k})^qI(x>t_{k})$. Because of the centering constraint $Ef_j(x_j)=0$, we use the centered basis $\{B_{jk}(x)=B_k(x)-\sum_{i=1}^{n}B_k(x_{ij})/n$, $k=0,\ldots,K\} $ with $ K=q+k-1$. We separate linear basis (a single linear function) and nonlinear basis to enable the identification of linear and nonlinear components. Thus, with splines approximation, we have
\begin{equation*}
y_i=\mu+\mathop{\sum}\limits_{j=1}\limits^{p}\alpha_{j} B_{j0}(x_{ij})+\mathop{\sum}\limits_{j=1}\limits^{p}\mathop{\sum}\limits_{k=1}\limits^{K}\beta_{jk}B_{jk}(x_{ij})+k_1e_i+\sqrt{k_2\delta_0e_i}z_i,~i=1,\ldots,n.
\end{equation*}
We view $e_i,~ i=1,\ldots,n$ as latent variables and denote  $\boldsymbol{e}=(e_1,\ldots,e_n)^{T}$. By defining $\boldsymbol{E}=k_2\delta_{0}\rm{diag}$$(e_1,\ldots,e_n)$,
$\boldsymbol{B}_{0}=(\boldsymbol{B}_{10}, \ldots,\boldsymbol{B}_{p0})$ with
$\boldsymbol{B}_{j0}=(B_{j0}(x_{1j}),\ldots,B_{j0}(x_{nj}))^{T}$, 
$\boldsymbol{\alpha}=(\alpha_1,
\ldots,\alpha_p)^{T}$, $\boldsymbol{\beta}_j=(\beta_{j1},\ldots,\beta_{jK})^{T}$ and  
\begin{displaymath} 
\boldsymbol{B}_j = 
\left( \begin{array}{cccc} 
B_{j1} (x_{1j}) & B_{j2} (x_{1j}) & \ldots &B_{jK} (x_{1j})\\ 
B_{j1} (x_{2j}) & B_{j2} (x_{2j}) & \ldots &B_{jK} (x_{2j})\\ 
\vdots & \vdots & \vdots& \vdots\\
B_{j1} (x_{nj}) & B_{j2} (x_{nj}) & \ldots &B_{jK} (x_{nj})\\ 
\end{array} \right),
\end{displaymath}
the full conditional distribution of $\boldsymbol{y}$ can be expressed as,
\begin{equation}
\label{AdditiveModel}
\begin{split}
&p(\boldsymbol{y}|\boldsymbol{\alpha}, \{\boldsymbol{\beta}_j\}, \boldsymbol{e},\delta_{0},\boldsymbol{x}, \mu)\\
&\propto \exp \{-\frac{1}{2}(\boldsymbol{y}-\boldsymbol{f}-k_1\boldsymbol{e})^{T}\boldsymbol{E}^{-1}(\boldsymbol{y}-\boldsymbol{f}-k_1\boldsymbol{e})\}(\det [\boldsymbol{E}])^{-1/2}.
\end{split}
\end{equation}
where 
\begin{equation*}
\boldsymbol{f}=\mu\boldsymbol{1}_n+\boldsymbol{B}_{0}\boldsymbol{\alpha}+\mathop{\sum}\limits_{j=1}\limits^{p}\boldsymbol{B}_{j} \boldsymbol{\beta}_{j},
\end{equation*}
and $\boldsymbol{1}_n=(1,\ldots, 1)^{T}$ is a vector of dimension $n$ with all components 1.

We choose spike-and-slab priors for
$\alpha_j$ and $ \boldsymbol{\beta}_{j}$ following \cite{george1993variable}, \cite{anagiotelis2008bayesian}, and many others to enable variable selection and linear component selection. We introduce indicator variables
$\boldsymbol{\gamma}^{(\nu)}=(\gamma^{(\nu)}_1,\ldots,\gamma^{(\nu)}_p)^{T}$,
such that $\nu_{j}=0$ if and only if $\gamma^{(\nu)}_j=0$, with $\nu=\alpha,
\boldsymbol{\beta}$.  In other words,  $f_j$ is regarded as a nonlinear function, a linear function or a zero function under the situation that ($\gamma^{(\boldsymbol{\beta})}_j= 1$),  ($\gamma^{(\alpha)}_j=1$, $\gamma^{(\boldsymbol{\beta})}_j= 0$), or ($\gamma^{(\alpha)}_j=0$, $\gamma^{(\boldsymbol{\beta})}_j= 0$), respectively.

For $\boldsymbol{\alpha}$, we choose an independent Gaussian prior on
each component, which is also called $ridge$ $prior$ (\cite{goldstein1974ridge}), 
\begin{equation*}
\label{alpha_prior}
\begin{split}
p(\boldsymbol{\alpha}|\boldsymbol{\gamma}^{(\alpha)},
\boldsymbol{\sigma}) &\sim N(\boldsymbol{0}_p, \boldsymbol{\Sigma}_{\alpha}),\\
\boldsymbol{\Sigma}_{\alpha}&=diag(\gamma^{(\alpha)}_{1} \sigma_1^2,
\gamma^{(\alpha)}_{2} \sigma_2^2,\ldots, \gamma^{(\alpha)}_{p} \sigma_p^2
),
\end{split}
\end{equation*} 
where $\boldsymbol{0}_p$ is a zero vector of dimension $p$ and $\boldsymbol{\sigma}=(\sigma_1,\ldots,\sigma_p)$.

For  $\boldsymbol{\beta}_j$, we take the conjugate prior usually
undertaken in a Bayesian context $p(\boldsymbol{\beta}_j)\propto\exp\{-\frac{1}{2\tau_j^2}\boldsymbol{\beta}_j^{T} \boldsymbol{\Omega}_j
\boldsymbol{\beta}_j\}$ when $\gamma^{(\boldsymbol{\beta})}_j=1$, such as \cite{smith1996nonparametric} and \cite{chib2006inference}. Here the $(k,k^{'})$ entry of $\boldsymbol{\Omega}_j$  is $\int_0^1 B^{''}_{jk}(x)B^{''}_{jk^{'}}(x) dx$ ($B^{''}_{jk}$ is the second derivative of $B_{jk}$). Since $\boldsymbol{\beta}_j =\boldsymbol{0}_K$ when $\gamma^{(\boldsymbol{\beta})}_j=0$, the prior of $\boldsymbol{\beta}_j$ can be described as,  
\begin{equation*}
p(\boldsymbol{\beta}_j|\gamma^{(\boldsymbol{\beta})}_{j}, \tau_j) \sim N(\boldsymbol{0}_K,\gamma^{(\boldsymbol{\beta})}_j\tau_j^{2}\boldsymbol{\Omega}_{j}^{-1}).
\end{equation*}

We use the same prior as in \cite{cripps2005variable} on $\boldsymbol{\gamma}^{(\nu)}$, $\nu=\alpha,\boldsymbol{\beta}$, 
\begin{equation*}
\label{gamma_prior}
p(\boldsymbol{\gamma}^{(\nu)})=\frac{1}{p+1} \left(\begin{array}{c} p\\
 q_{\gamma^{(\nu)}}\end{array}\right)^{-1},
\end{equation*} 
where  $q_{\gamma^{(\nu)}}$ is the number of non-zero $\nu_j$. Under
this prior, equal weights are placed on
$\boldsymbol{\gamma}^{(\nu)}$ with different numbers of non-zero
$\nu_j$.  Alternatively, we can use $p(\boldsymbol{\gamma}^{(\nu)})=\pi^{q_{\gamma^{(\nu)}}}(1-\pi)^{p-q_{\gamma^{(\nu)}}}$ with $\pi$ a constant.  When $\pi=1/2$, we have the uniform prior on $\boldsymbol{\gamma}^{(\nu)}$. We find the results are not sensitive to this choice of prior. 

To summarize, the Bayesian hierarchical formulation is given by,
\begin{equation*}
\begin{split}
\boldsymbol{y}|\boldsymbol{\alpha}, \{\boldsymbol{\beta}_j\},\boldsymbol{e},\delta_{0},\boldsymbol{x}, \mu&\sim N (\boldsymbol{f}+k_1\boldsymbol{e},\boldsymbol{E}),\\
\boldsymbol{\alpha}|\boldsymbol{\gamma}^{(\alpha)},\boldsymbol{\sigma} &\sim N(\boldsymbol{0}_p, \boldsymbol{\Sigma}_{\alpha}), \ \boldsymbol{\gamma}^{(\alpha)}\sim p(\boldsymbol{\gamma}^{(\alpha)}),\\
\boldsymbol{\beta}_j|\gamma^{(\boldsymbol{\beta})}_{j}, \tau_j&\sim N(\boldsymbol{0}_K,\gamma^{(\boldsymbol{\beta})}_j\tau_j^{2}\boldsymbol{\Omega}_{j}^{-1}), \ \boldsymbol{\gamma}^{(\boldsymbol{\beta})}\sim p(\boldsymbol{\gamma}^{(\boldsymbol{\beta})}),\\
e_i&\stackrel{i.i.d.}{\sim}  \exp(1/\delta_{0}),\ \delta_{0}\sim p(\delta_{0}),\\
\sigma_{j}^{2}&\sim p(\sigma_j^{2}),\ \tau_{j}^{2}\sim p(\tau_j^{2}),\ \mu\sim p(\mu),
\end{split}
\end{equation*}
where $p(\delta_{0})$, $p(\sigma_j^{2})$ and $p(\tau_j^{2})$ represent the  hyperpriors of $\delta_{0}$, $\sigma^2_j$ and $\tau^2_j$. They are set to be $IG(a_1,a_2)$, where $IG$ denotes the inverse Gamma distribution, and $a_1$ and $a_2$ are set to be 0.5 in all our numerical experiments as an uninformative choice. Sensitivity analysis reveals that our results are 
not sensitive to these choices. Finally, we use an uninformative prior on $\mu$, $p(\mu)\propto 1$.

We use the Metropolis-within-Gibbs algorithm to sample from the posterior distributions. To improve mixing of the Markov chains, we integrate out $\nu_j$, $\nu=\alpha, \boldsymbol{\beta}$ in some of the sampling steps, resulting in a partially collapsed sampler (\cite{van2008partially}). The readers are referred to the Appendix for details.\\

\section{Numerical Illustrations}
We demonstrate the performance of our proposed quantile regression approach (denoted by BQPLAM) in terms of its estimation accuracy and model selection accuracy. The MCMC algorithm is implemented in {\sf R}, and available upon request. For comparison, we consider the following 4 additional methods. 

\subsubsection*{Method 1:}
The hierarchical Bayesian model in section 2 can be easily revised to deal with partially linear additive mean regression model. Consider the following model, 
\begin{equation*}
y_i=\mu+\mathop{\sum}\limits_{j=1}\limits^{p}f_j(x_{ij})+\epsilon_i,~i=1,\ldots,n,
\end{equation*}   
where $\epsilon_i, i =1,\ldots, n$ are i.i.d. normally distributed with mean zero and variance $\delta_0^2$. The Bayesian hierarchical formulation will then be as follows,
\begin{equation*}
\begin{split}
\boldsymbol{y}|\boldsymbol{\alpha}, \{\boldsymbol{\beta}_j\},\delta_{0}^{2},\boldsymbol{x}, \mu&\sim N (\boldsymbol{f},\delta_{0}^{2}\boldsymbol{I}_{n\times n} ),\\
\boldsymbol{\alpha}|\boldsymbol{\gamma}^{(\alpha)},
\boldsymbol{\sigma} &\sim N(\boldsymbol{0}_{p}, \boldsymbol{\Sigma}_{\alpha}),\  \boldsymbol{\gamma}^{(\alpha)}\sim p(\boldsymbol{\gamma}^{(\alpha)}),\ \\
\boldsymbol{\beta}_j,\tau_{j} &\sim N(\boldsymbol{0}_{K},\gamma^{(\boldsymbol{\beta})}_j\tau_j^{2}\boldsymbol{\Omega}_{j}^{-1}),\ \boldsymbol{\gamma}^{(\boldsymbol{\beta})}\sim p(\boldsymbol{\gamma}^{(\boldsymbol{\beta})}),\ \\
\delta_{0}^{2}&\sim p(\delta_{0}^{2}),\ \tau_{j}^{2}\sim p(\tau_j^{2}),\ \mu\sim p(\mu),
\end{split}
\end{equation*}
where $\boldsymbol{I}_{n\times n}$ represents the $n\times n$ identity matrix. We denote this mean regression method to be BPLAM. 

The other three are quantile regression approaches. 

\subsubsection*{Method 2:}

We consider a Bayesian additive model that combines nonlinear and linear terms together. This only performs variable selection but not linear component identification. To be specific, the Bayesian additive regression is based on model (\ref{AdditiveModel}), with
\begin{equation*}
\boldsymbol{f}=\mu\boldsymbol{1}_n+\mathop{\sum}\limits_{j=1}\limits^{p}\boldsymbol{B}_{j} \boldsymbol{\beta}_{j},
\end{equation*}
where  $\boldsymbol{\beta}_j=(\beta_{j0},\beta_{j1}, \ldots,\beta_{jK})^{T}$ and \begin{displaymath} 
\boldsymbol{B}_j = 
\left( \begin{array}{cccc} 
B_{j0} (x_{1j}) & B_{j1} (x_{1j}) & \ldots &B_{jK} (x_{1j})\\ 
B_{j0} (x_{2j}) & B_{j1} (x_{2j}) & \ldots &B_{jK} (x_{2j})\\ 
\vdots & \vdots & \vdots& \vdots\\
B_{j0} (x_{nj}) & B_{j1} (x_{nj}) & \ldots &B_{jK} (x_{nj})\\ 
\end{array} \right).
\end{displaymath}
With similar prior distributions and hyper-parameters as those described in Section 2, we obtain the following Bayesian hierarchical formulation,
\begin{equation*}
\begin{split}
\boldsymbol{y}|\boldsymbol{\alpha}, \{\boldsymbol{\beta}_j\},\boldsymbol{e},\delta_{0},\boldsymbol{x},\mu&\sim N (\boldsymbol{f}+k_1\boldsymbol{e},\boldsymbol{E}), \\\
\boldsymbol{\beta}_j|\gamma^{(\boldsymbol{\beta})}_j,\tau_{j} &\sim N(\boldsymbol{0}_{K+1},\gamma^{(\boldsymbol{\beta})}_j\tau_j^{2}\boldsymbol{\Omega}_{j}^{-1}),\ \boldsymbol{\gamma}^{(\boldsymbol{\beta})}\sim p(\boldsymbol{\gamma}^{(\boldsymbol{\beta})}),\\
e_i&\stackrel{i.i.d.}{\sim}  \exp(1/\delta_{0}),\  \delta_{0}\sim p(\delta_{0}),\\
\tau_{j}^{2}&\sim p(\tau_j^{2}), ~\mu\sim p(\mu).
\end{split}
\end{equation*}
This method is denoted by $\rm{BQAM_{V}}$. 

\subsubsection*{Method 3:}
We consider Bayesian quantile linear regression based on model (\ref{AdditiveModel}), with
\begin{equation*}
\boldsymbol{f}=\mu\boldsymbol{1}_n+\boldsymbol{B}_{0}\boldsymbol{\alpha},
\end{equation*}
in which all components are assumed to be linear. The Bayesian hierarchical structure is as follows,

\begin{equation*}
\begin{split}
\boldsymbol{y}|\boldsymbol{\alpha}, \{\boldsymbol{\beta}_j\}, \boldsymbol{e},\delta_{0},\boldsymbol{x}, \mu&\sim N (\boldsymbol{f}+k_1\boldsymbol{e},\boldsymbol{E}), \\
\boldsymbol{\alpha}|\boldsymbol{\gamma}^{(\alpha)}, \boldsymbol{\sigma} &\sim N(\boldsymbol{0}_{p}, \boldsymbol{\Sigma}_{\alpha}),\ \boldsymbol{\gamma}^{(\alpha)}\sim p(\boldsymbol{\gamma}^{(\alpha)}), \\
e_i&\stackrel{i.i.d.}{\sim} \exp(1/\delta_{0}), \ \delta_{0}\sim p(\delta_{0}), \\
\sigma_{j}^{2}&\sim p(\sigma_j^{2}),\ \mu\sim p(\mu).
\end{split}
\end{equation*}
The indicator variables $\boldsymbol{\gamma}^{(\alpha)}$ enable component selection. This method is denoted by $\rm{BQLM_{V}}$.

\subsubsection*{Method 4:}

The third quantile regression method for comparison is still based on the model (\ref{AdditiveModel}), except that we fix all indicator variables $\gamma^{(\alpha)}_j,~ j=1, \ldots, p,$ and $\gamma^{(\boldsymbol{\beta})}_j, ~j=1, \ldots, p,$ to be 1. This method can only estimate component functions but cannot select variables. We denote it as $\rm{BQAM_{NV}}$. 

\subsection{Simulation examples}
We generate $n=100$ observations $(\boldsymbol{x}_i, y_i), ~i=1,\ldots, 100, $ from the following heteroscedastic additive model
\begin{equation}
\label{model}
y_i=\mathop{\sum}\limits_{j=1}\limits^{p}f_j(x_{ij})+(0.5+x_{i2})\epsilon_i,
\end{equation}
with $ f_1(x)=\rm{sin}$$(2\pi x)/(2-$$\rm{sin}$$(2 \pi x))$, $f_2(x)=5x(1-x)$, $f_3(x)=2x$, $f_4(x)=x$, $f_5(x)=-x$ and $p=10$. Thus, the first 2 components are nonlinear components, followed by 3 linear components and 5 zero components. The covariates $x_{ij}$ are generated from the standard normal distribution with correlations given by $Cov(x_{ij_1},x_{ij_{2}})=(1/2)^{|j_1-j_2|}$, and then transformed to be marginally uniform on $[0,1]$ by applying the cdf of the standard normal distribution. We consider two distributions of $\epsilon_i$, a normal distribution with mean 0 and standard deviation 0.5, and a Student's t distribution with scale parameter $1/3$ and degrees of freedom 2. We fit the mean regression model and the four quantile regression models at five different quantile levels $\{0.1,0.3,0.5,0.7,0.9\}$. For each scenario, 100 data sets are generated and fitted. For each replicate, the MCMC algorithm is run for 20000 iterations with a burn-in of 10000.

The performance was evaluated by the integrated squared error $(ISE)$, for each component function,  which is approximated over an equally spaced 1000 points $(t_1,\ldots, t_T)$, $T=1000$ on $[0,1]$ by,
\begin{equation}
\widehat{ISE}=\frac{1}{T}\mathop{\sum}\limits_{i=1}\limits^{T}(\hat{f_j}(t_i)-f_j(t_i))^{2},
\end{equation}
where $f_j(t_i)$ is the true value of function $f_j$ at $t_i$, and $\hat{f_j}(t_i)$ is the posterior mean of $f_j$ at $t_i$ based on the 10000 iterations after burn-in.  Tables \ref{ISE_qr_normal} and \ref{ISE_qr_t} summarize the average and standard deviation of $\sqrt{\widehat{ISE}}$ over 100 replicates for the first six components and for the regression function $f=\sum_{j=1}^{10}f_j$. From the results, we can see that BQPLAM is obviously more efficient than $\rm{BQAM_{V}}$ and $\rm{BQAM_{NV}}$ for the parametric components due to the separated linear basis and its associated indicator variables. $\rm{BQLM_{V}}$ performs poorly as expected, since it cannot capture nonlinear effects of the components. BQPLAM and $\rm{BQAM_{V}}$ outperform $\rm{BQAM_{NV}}$ for zero components (note that we only present results of $f_6$ among the five zero components). The results show that  the two sets of indicators can help reduce errors besides selecting components. The results of mean regression are similar to those of median regression when the errors follow a normal distribution. However, the advantage of median regression is more obvious when the errors follow a Student's t distribution.

To measure the prediction accuracy of our method, we generate $n'= 100,000$ independent test samples from the same generating model (\ref{model}) as the training samples and present the test errors in Tables \ref{test_error_median} and \ref{test_error_qr1}. For mean and median regression, we consider two error measures including the root mean squared errors (RMSE) and absolute deviation errors (AD).  For quantile regressions at quantiles $\{0.1,0.3,0.7,0.9\}$, the prediction error refers to the average check loss (ACL). More sepcifically,
\begin{equation*}
\begin{split}
RMSE=&\sqrt{\frac{1}{n'}\mathop{\sum}\limits_{i=1}\limits^{n'}(\hat{y}_i-y_i)^{2}},\\
AD=&\frac{1}{n'}\mathop{\sum}\limits_{i=1}\limits^{n'}|\hat{y}_i-y_i|\\
\end{split}
\end{equation*}
and 
\begin{equation*}
ACL=\frac{1}{n'}\mathop{\sum}\limits_{i=1}\limits^{n'}\rho_{\tau}(\hat{y}_i-y_i),
\end{equation*}
where $y_i, ~i=1,\ldots,n'$, are the responses of the testing data and $\hat{y}_{i}$, $i=1,\ldots,n'$, are the predicted values estimated with 
\begin{equation*}
\hat{y}_i=\hat{\mu}+\mathop{\sum}\limits_{j=1}\limits^{p}\hat{f}_{j}(x_{ij}),
\end{equation*}
where $\hat{\mu}$ and $\hat{f}_{j}(x_{ij})$ are posterior mean based on the 10000 sampled values after burn-in. Our method BQPLAM results in smaller test errors and outperforms the other three quantile regression methods at all quantile levels. The median regression performs similarly as mean regression when the errors are Gaussian, while it outperforms mean regression when errors are Student's t as expected.

\begin{table}
\tiny
\caption{Summary of $\sqrt{\widehat{ISE}}$ for each univariate function and $f=\sum_{j=1}^{10} f_j $ over 100 replicates, when the distribution of error is normal and $p=10$. Standard errors based on simulations are shown as subscripts.}
\label{ISE_qr_normal}
\vspace{0.1in}
\centering 
\begin{tabular}{llccccccc}
\toprule 
\multicolumn{2}{c}{}&$f_1$&$f_2$&$f_3$&$f_4$&$f_5$&$f_6$&$f$\\
\midrule
& $\rm{BPLAM}      $&$0.165_{0.049}$&$0.109_{0.043}$&$0.069_{0.051}$&$0.067_{0.057}$&$0.061_{0.061}$&$0.016_{0.032}$&$0.219_{0.047}$\\ 
\midrule
\multirow{2}{*} {$\tau=0.5$} 
  
& $\rm{BQPLAM}     $&$0.162_{0.048}$&$0.112_{0.049}$&$0.068_{0.050}$&$0.073_{0.076}$&$0.068_{0.066}$&$0.010_{0.029}$&$0.222_{0.048}$\\    
& $\rm{BQAM_{v}}   $&$0.156_{0.046}$&$0.161_{0.055}$&$0.156_{0.054}$&$0.190_{0.076}$&$0.176_{0.080}$&$0.006_{0.011}$&$0.224_{0.072}$\\ 
& $\rm{BQLM_{v}}   $&$0.320_{0.026}$&$0.364_{0.015}$&$0.075_{0.064}$&$0.112_{0.091}$&$0.108_{0.092}$&$0.015_{0.034}$&$0.515_{0.041}$\\      
& $\rm{BQAM_{nv}}  $&$0.157_{0.053}$&$0.127_{0.051}$&$0.101_{0.058}$&$0.104_{0.059}$&$0.123_{0.071}$&$0.114_{0.073}$&$0.320_{0.114}$\\          \cmidrule(r){1-9}
\multirow{2}{*} {$\tau=0.1$} 
& $\rm{BQPLAM}     $&$0.195_{0.066}$&$0.184_{0.074}$&$0.105_{0.068}$&$0.106_{0.091}$&$0.110_{0.092}$&$0.023_{0.059}$&$0.287_{0.084}$\\   
& $\rm{BQAM_{v}}   $&$0.199_{0.074}$&$0.222_{0.073}$&$0.189_{0.063}$&$0.233_{0.064}$&$0.236_{0.060}$&$0.007_{0.016}$&$0.310_{0.096}$\\
& $\rm{BQLM_{v}}   $&$0.359_{0.063}$&$0.377_{0.036}$&$0.107_{0.093}$&$0.133_{0.114}$&$0.147_{0.099}$&$0.046_{0.083}$&$0.526_{0.053}$\\          
& $\rm{BQAM_{nv}}   $&$0.148_{0.045}$&$0.165_{0.060}$&$0.158_{0.062}$&$0.150_{0.052}$&$0.146_{0.054}$&$0.151_{0.059}$&$0.396_{0.139}$\\                   \cmidrule(r){1-9}\multirow{2}{*} {$\tau=0.3$} 
& $\rm{BQPLAM}     $&$0.170_{0.050}$&$0.136_{0.054}$&$0.070_{0.048}$&$0.072_{0.066}$&$0.072_{0.065}$&$0.014_{0.039}$&$0.233_{0.055}$\\    
& $\rm{BQAM_{v}}   $&$0.164_{0.050}$&$0.167_{0.060}$&$0.157_{0.050}$&$0.233_{0.055}$&$0.221_{0.061}$&$0.008_{0.013}$&$0.262_{0.089}$\\ 
& $\rm{BQLM_{v}}   $&$0.335_{0.044}$&$0.362_{0.048}$&$0.092_{0.072}$&$0.134_{0.098}$&$0.131_{0.101}$&$0.020_{0.046}$&$0.522_{0.046}$\\       
& $\rm{BQAM_{nv}} $&$0.142_{0.042}$&$0.158_{0.064}$&$0.160_{0.068}$&$0.145_{0.053}$&$0.146_{0.053}$&$0.147_{0.076}$&$0.383_{0.136}$\\         \cmidrule(r){1-9} \multirow{2}{*} {$\tau=0.7$} 
& $\rm{BQPLAM}     $&$0.166_{0.045}$&$0.132_{0.055}$&$0.070_{0.052}$&$0.084_{0.076}$&$0.073_{0.71}$&$0.010_{0.030}$&$0.239_{0.064}$\\    
& $\rm{BQAM_{v}}   $&$0.160_{0.049}$&$0.179_{0.065}$&$0.156_{0.058}$&$0.240_{0.055}$&$0.223_{0.062}$&$0.007_{0.010}$&$0.278_{0.096}$\\
& $\rm{BQLM_{v}}   $&$0.359_{0.063}$&$0.375_{0.033}$&$0.077_{0.068}$&$0.114_{0.086}$&$0.114_{0.090}$&$0.014{0.029}$&$0.517_{0.043}$\\         
& $\rm{BQAM_{nv}}  $&$0.150_{0.056}$&$0.166_{0.045}$&$0.153_{0.056}$&$0.145_{0.056}$&$0.144_{0.049}$&$0.155_{0.055}$&$0.379_{0.126}$\\          \cmidrule(r){1-9} \multirow{2}{*} {$\tau=0.9$} 
& $\rm{BQPLAM}     $&$0.196_{0.053}$&$0.204_{0.078}$&$0.101_{0.066}$&$0.115_{0.101}$&$0.117_{0.093}$&$0.026_{0.058}$&$0.311_{0.092}$\\    
& $\rm{BQAM_{v}}   $&$0.181_{0.055}$&$0.226_{0.081}$&$0.179_{0.069}$&$0.246_{0.056}$&$0.234_{0.057}$&$0.003_{0.004}$&$0.357_{0.108}$\\
& $\rm{BQLM_{v}}   $&$0.316_{0.025}$&$0.405_{0.075}$&$0.101_{0.081}$&$0.119_{0.092}$&$0.121_{0.095}$&$0.028_{0.064}$&$0.534_{0.063}$\\        
& $\rm{BQAM_{nv}}  $&$0.159_{0.052}$&$0.181_{0.060}$&$0.161_{0.057}$&$0.155_{0.059}$&$0.160_{0.051}$&$0.164_{0.059}$&$0.424_{0.141}$\\          \bottomrule
\end{tabular}
\end{table}

\begin{table}
\tiny
\caption{Summary of $\sqrt{\widehat{ISE}}$ for each univariate function  and  $f=\sum_{j=1}^{10} f_j $  over 100 replicates, when the distribution of error is Student's t and $p=10$. }
\vspace{0.1in}
\centering 
\begin{tabular}{llccccccc}
\toprule 
\multicolumn{2}{c}{}&$f_1$&$f_2$&$f_3$&$f_4$&$f_5$&$f_6$&$f$\\
 \midrule
 & $\rm{BPLAM}      $&$0.278_{0.081}$&$0.232_{0.104}$&$0.129_{0.106}$&$0.134_{0.095}$&$0.135_{0.084}$&$0.032_{0.043}$&$0.411_{0.161}$\\ 
 \midrule
  \multirow{2}{*} {$\tau=0.5$} 
  
& $\rm{BQPLAM}     $&$0.176_{0.055}$&$0.126_{0.067}$&$0.072_{0.050}$&$0.088_{0.089}$&$0.087_{0.089}$&$0.007_{0.023}$&$0.244_{0.073}$\\    
& $\rm{BQAM_{v}}   $&$0.145_{0.051}$&$0.154_{0.068}$&$0.137_{0.058}$&$0.193_{0.074}$&$0.186_{0.069}$&$0.032_{0.053}$&$0.269_{0.098}$\\ 
& $\rm{BQLM_{v}}   $&$0.319_{0.029}$&$0.361_{0.029}$&$0.075_{0.064}$&$0.136_{0.098}$&$0.121_{0.099}$&$0.015_{0.041}$&$0.517_{0.041}$\\       
& $\rm{BQAM_{nv}}  $&$0.160_{0.057}$&$0.161_{0.060}$&$0.158_{0.065}$&$0.165_{0.062}$&$0.161_{0.058}$&$0.147_{0.059}$&$0.439_{0.160}$\\          \cmidrule(r){1-9}
\multirow{2}{*} {$\tau=0.1$} 
& $\rm{BQPLAM}     $&$0.295_{0.104}$&$0.274_{0.129}$&$0.157_{0.129}$&$0.176_{0.099}$&$0.165_{0.098}$&$0.027_{0.059}$&$0.405_{0.144}$\\    
& $\rm{BQAM_{v}}   $&$0.230_{0.089}$&$0.283_{0.157}$&$0.244_{0.155}$&$0.238_{0.105}$&$0.241_{0.124}$&$0.073_{0.176}$&$0.492_{0.301}$\\
& $\rm{BQLM_{v}}
$&$0.373_{0.063}$&$0.370_{0.027}$&$0.144_{0.123}$&$0.191_{0.105}$&$0.186_{0.096}$&$0.048_{0.088}$&$0.529_{0.055}$\\        
& $\rm{BQAM_{nv}}  $&$0.259_{0.239}$&$0.251_{0.178}$&$0.258_{0.296}$&$0.261_{0.244}$&$0.247_{0.265}$&$0.231_{0.235}$&$0.610_{0.392}$\\               \cmidrule(r){1-9}\multirow{2}{*} {$\tau=0.3$} 
& $\rm{BQPLAM}     $&$0.201_{0.080}$&$0.142_{0.065}$&$0.076_{0.064}$&$0.097_{0.092}$&$0.093_{0.086}$&$0.008_{0.029}$&$0.268_{0.087}$\\    
& $\rm{BQAM_{v}}   $&$0.153_{0.062}$&$0.166_{0.072}$&$0.154_{0.061}$&$0.193_{0.076}$&$0.186_{0.071}$&$0.037_{0.058}$&$0.274_{0.108}$\\
& $\rm{BQLM_{v}}   $&$0.339_{0.049}$&$0.359_{0.022}$&$0.098_{0.077}$&$0.136_{0.100}$&$0.120_{0.103}$&$0.020_{0.050}$&$0.523_{0.041}$\\        
& $\rm{BQAM_{nv}}  $&$0.183_{0.082}$&$0.185_{0.071}$&$0.178_{0.081}$&$0.185_{0.079}$&$0.170_{0.072}$&$0.164_{0.065}$&$0.464_{0.173}$\\          \cmidrule(r){1-9} \multirow{2}{*} {$\tau=0.7$} 
& $\rm{BQPLAM}     $&$0.182_{0.057}$&$0.146_{0.085}$&$0.078_{0.061}$&$0.098_{0.094}$&$0.096_{0.094}$&$0.006_{0.021}$&$0.266_{0.089}$\\    
& $\rm{BQAM_{v}}   $&$0.163_{0.054}$&$0.167_{0.076}$&$0.154_{0.073}$&$0.205_{0.069}$&$0.195_{0.067}$&$0.032_{0.050}$&$0.308_{0.129}$\\
& $\rm{BQLM_{v}}   $&$0.317_{0.028}$&$0.369_{0.027}$&$0.083_{0.077}$&$0.130_{0.093}$&$0.130_{0.099}$&$0.013_{0.037}$&$0.519_{0.048}$\\        
& $\rm{BQAM_{nv}}  $&$0.191_{0.070}$&$0.205_{0.110}$&$0.189_{0.081}$&$0.175_{0.080}$&$0.168_{0.078}$&$0.178_{0.076}$&$0.505_{0.020}$\\           \cmidrule(r){1-9} \multirow{2}{*} {$\tau=0.9$} 
& $\rm{BQPLAM}     $&$0.279_{0.100}$&$0.293_{0.134}$&$0.146_{0.112}$&$0.160_{1.09}$&$0.156_{0.116}$&$0.031_{0.080}$&$0.408_{0.155}$\\    
& $\rm{BQAM_{v}}  $&$0.254_{0.130}$&$0.264_{0.134}$&$0.272_{0.174}$&$0.247_{0.091}$&$0.235_{0.072}$&$0.083_{0.131}$&$0.575_{0.362}$\\
& $\rm{BQLM_{v}}   $&$0.345_{0.054}$&$0.396_{0.084}$&$0.130_{0.107}$&$0.172_{0.100}$&$0.175_{0.105}$&$0.048_{0.085}$&$0.576_{0.111}$\\      
& $\rm{BQAM_{nv}}  $&$0.253_{0.130}$&$0.258_{0.139}$&$0.259_{0.165}$&$0.268_{0.189}$&$0.220_{0.097}$&$0.217_{0.123}$&$0.679_{0.387}$\\                  \bottomrule
\end{tabular}
\label{ISE_qr_t}
\end{table}

\begin{center}
\begin{table*}
\footnotesize
\caption{Summary of testing errors for mean estimators and median estimators over 100 replicates, when $p=10$. ``Normal" and ``Student's t" indicate the distribution of $\epsilon_i$.}
\label{test_error_median}
\vspace{0.1in}
\centering 
\begin{tabular}{llcc}
\toprule 
\multicolumn{2}{c}{}&RMSE&AD\\

\cmidrule(r){1-4} 
\multirow{5}{*} {Normal}
&$\rm{BPLAM}$&                                   $0.586_{0.030} $        &$0.457_{0.026} $ \\   
& $\rm{BQPLAM}$ &                              $0.587_{0.031} $         & $0.459_{0.026} $ \\           
&$\rm{BQAM_{v}}$&                             $0.646_{0.032} $        & $0.505_{0.026} $ \\
&$\rm{BQLM_{v}}$ &                              $0.762_{0.030} $       &$0.604_{0.025} $\\
&$\rm{BQAM_{nv}}$&                               $0.696_{0.251}$        &$0.540_{0.165}$\\     
\cmidrule(r){1-4} \multirow{5}{*} {Student's t} 
&$\rm{BPLAM}$&                                    $1.224_{0.079}$        &$0.648_{0.101}$ \\   
& $\rm{BQPLAM}$ &                                $1.160_{0.032}$      &$0.551_{0.054}$  \\           
&$\rm{BQAM_{v}}$&                              $1.350_{0.022}$       &$0.589_{0.031}$ \\
&$\rm{BQLM_{v}}$ &                                 $1.354_{0.015}$      &$0.679_{0.022}$  \\
 &$\rm{BQAM_{nv}}$ &                               $1.236_{0.043}$       &$0.666_{0.058}$\\ 
 \bottomrule
\end{tabular}
\end{table*}
\end{center}

\begin{center}
\begin{table*}
\footnotesize
\caption{Summary of testing errors for quantile estimators at different quantile levels over 100 replicates, when $p=10$. ``Normal" and ``Student's t" indicate the distribution of $\epsilon_i$.}
\label{test_error_qr1}
\vspace{0.1in}
\centering 
\begin{tabular}{llcccc}
\toprule 
\multicolumn{2}{c}{}&$\tau=0.1$&$\tau=0.3$&$\tau=0.7$&$\tau=0.9$\\
\midrule
\multirow{4}{*} {Normal} &$\rm{BQPLAM}$ & $0.634_{0.100} $&    $0.311_{0.041}$    &$0.322_{0.046}$ &$0.660_{0.128}$\\             
&$\rm{BQAM_{v}}$ &$0.652_{0.083}        $&$0.338_{0.045}$&$0.351_{0.048}$&$0.669_{0.096}$\\             
&$\rm{BQLM_{v}}$ &$0.868_{0.119}        $&$0.415_{0.058}$&$0.424_{0.064}$&$0.846_{0.113}$\\        
&$\rm{BQAM_{nv}}$ &$0.684_{0.084}        $&$0.356_{0.065}$ &$0.355_{0.048}$&$0.664_{0.089}$\\ 
\midrule
\multirow{4}{*} {Student's t} &$\rm{BQPLAM}$&$0.819_{0.219}$ &    $0.360_{0.061}$   &$0.365_{0.063}$ &$0.843_{0.182}$\\                 
&$\rm{BQAM_{v}}$&                        $0.831_{0.261}        $&$0.393_{0.063}$&$0.391_{0.058}$&$0.853_{0.228}$\\             
&$\rm{BQLM_{v}}$&                        $0.988_{0.206}        $&$0.464_{0.060}$&$0.464_{0.566}$&$0.967_{0.146}$\\                      
&$\rm{BQAM_{nv}}$&                       $0.835_{0.514}        $&$0.445_{0.077}$&$0.456_{0.075}$&$0.850_{0.309}$\\ 
\bottomrule
\end{tabular}
\end{table*}
\end{center}

\begin{figure}[htp]
\centerline{
\subfigure[Mean]{\includegraphics[width=1.8in,trim=0 30 0 0]{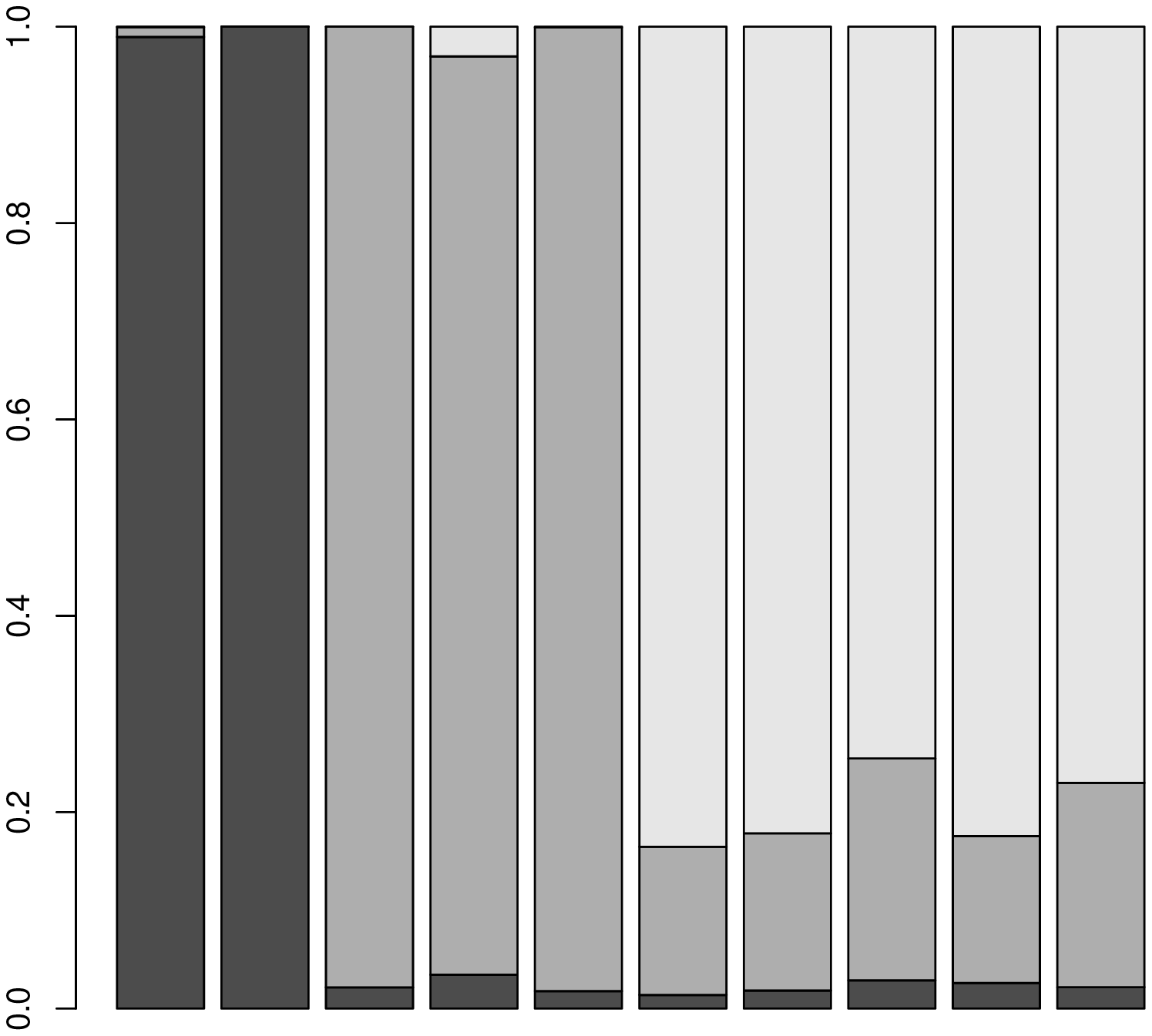}}
\hfil
\subfigure[$\tau=0.1$]{\includegraphics[width=1.8in,trim=0 30 0 0]{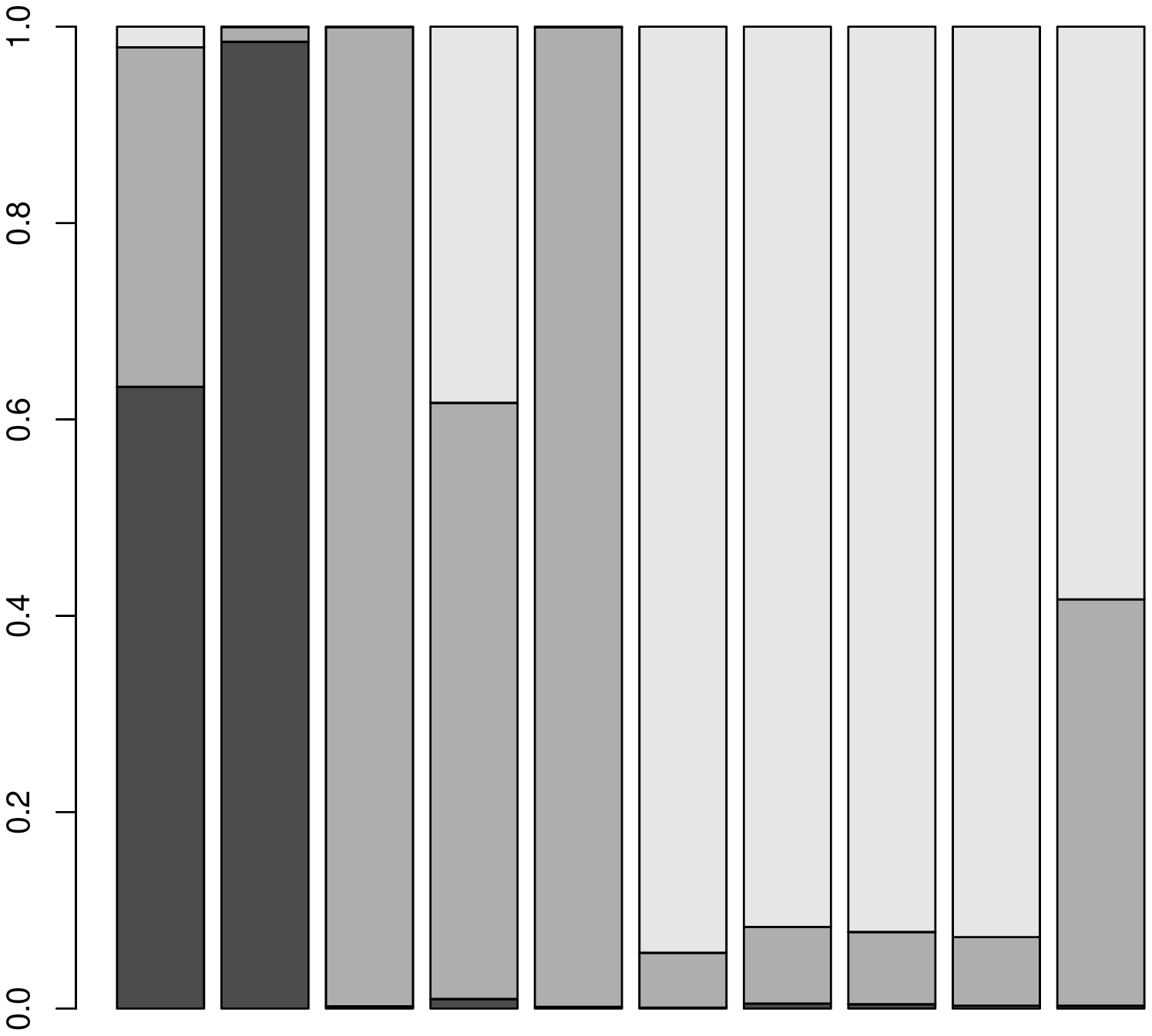}}
\hfil
\subfigure[$\tau=0.3$]{\includegraphics[width=1.8in,trim=0 30 0 0]{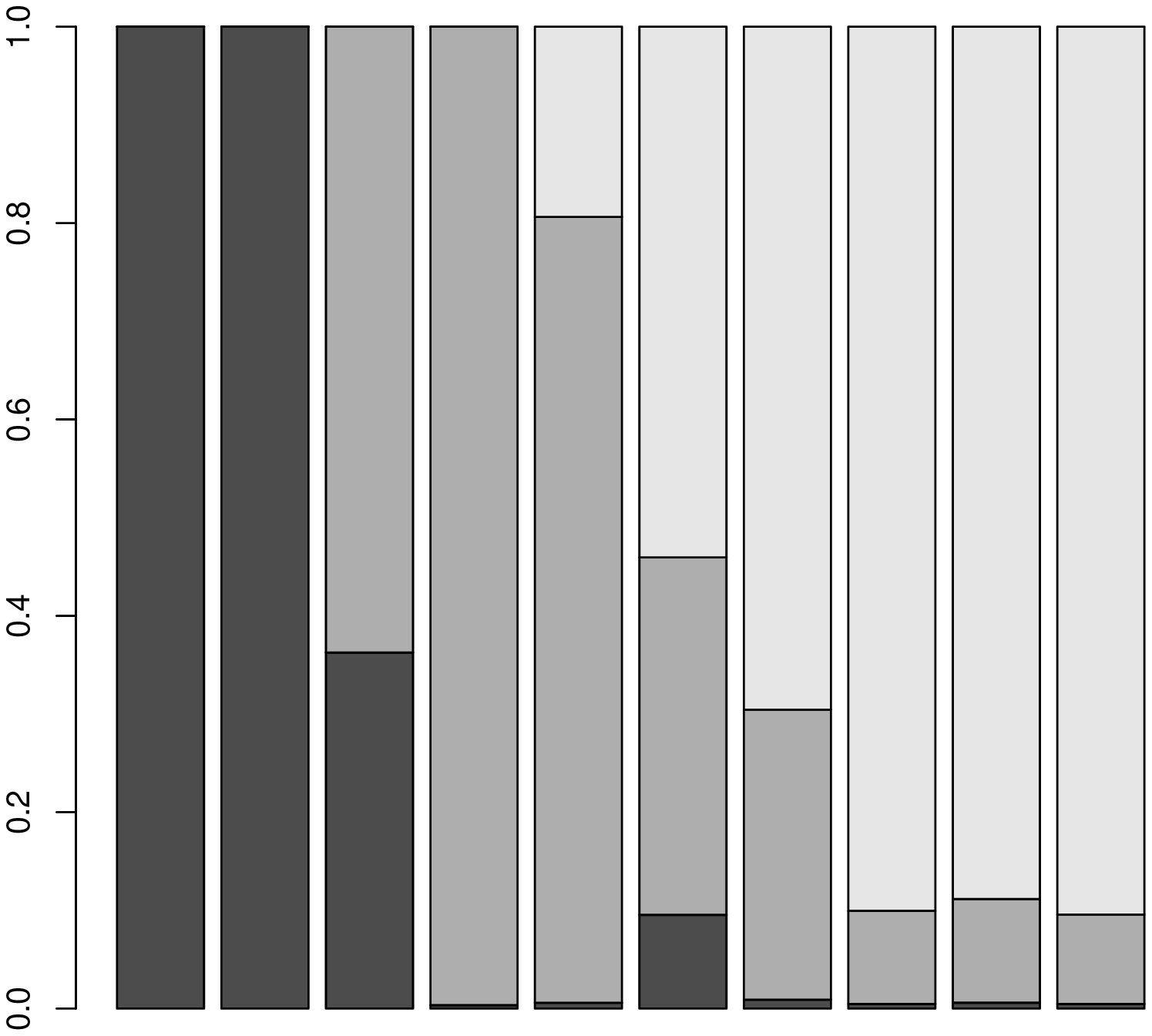}}
}

\centerline{
\subfigure[$\tau=0.5$]{\includegraphics[width=1.8in,trim=0 30 0 0]{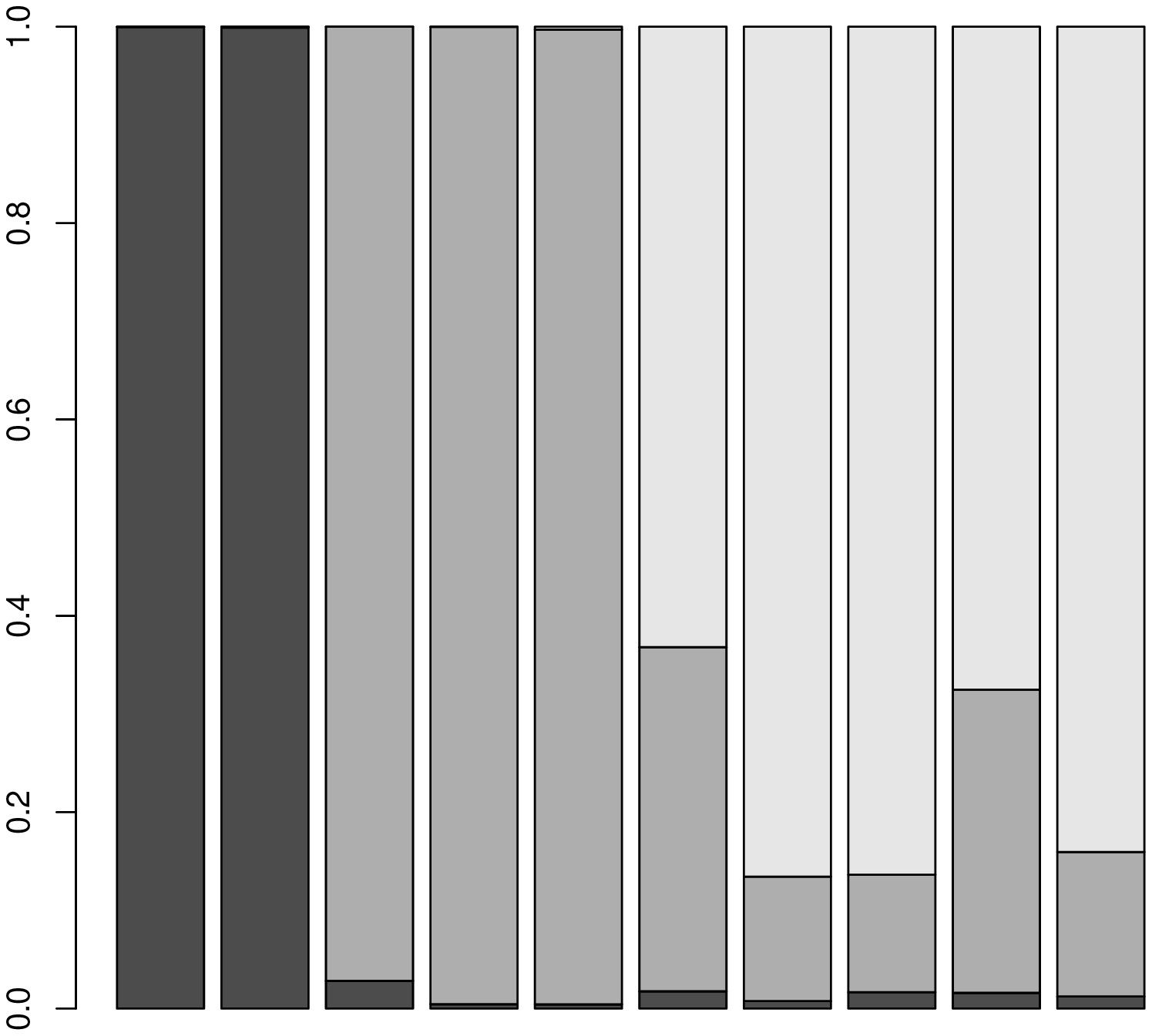}}
\hfil
\subfigure[$\tau=0.7$]{\includegraphics[width=1.8in,trim=0 30 0 0]{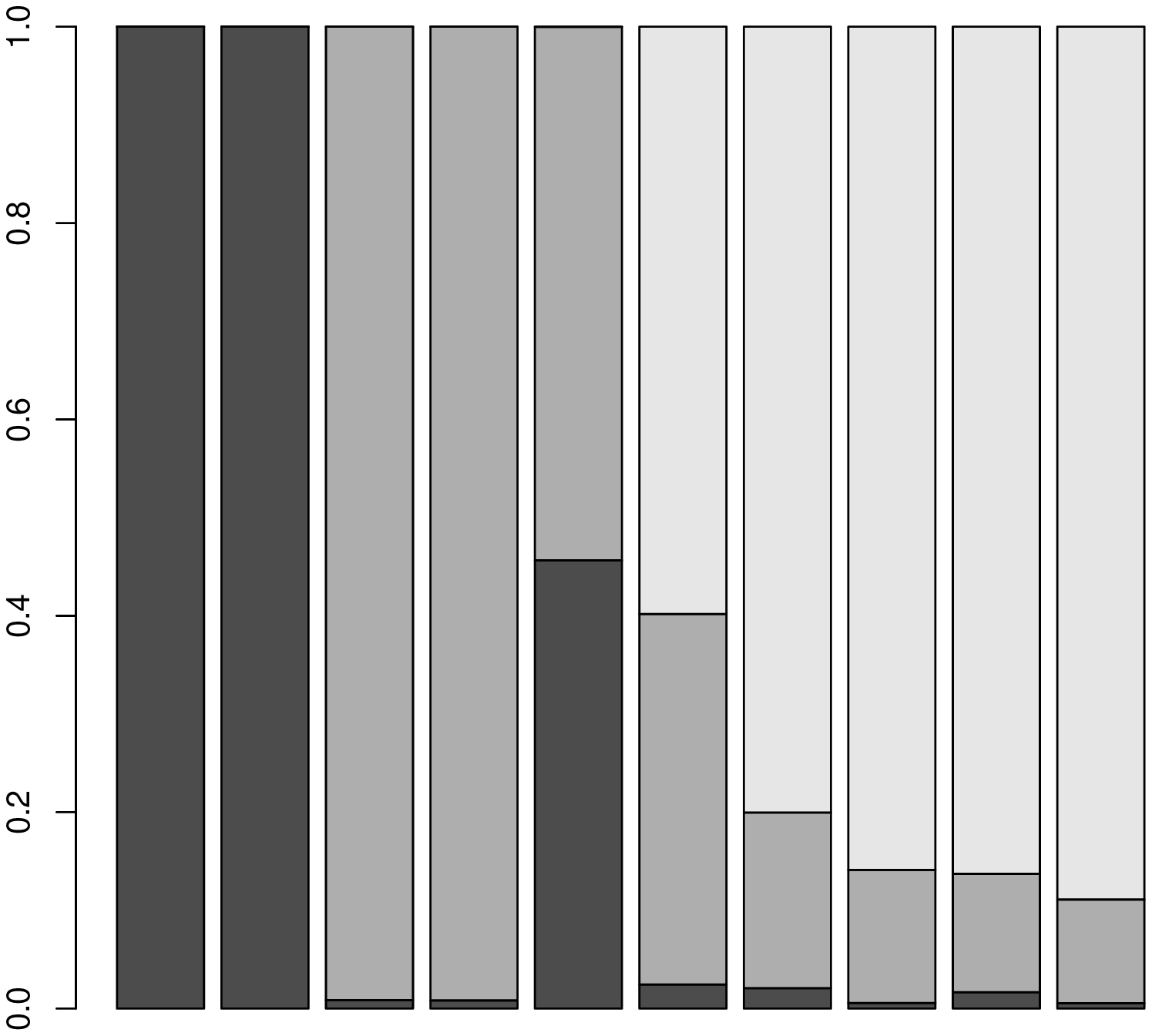}}
\hfil
\subfigure[$\tau=0.9$]{\includegraphics[width=1.8in,trim=0 30 0 0]{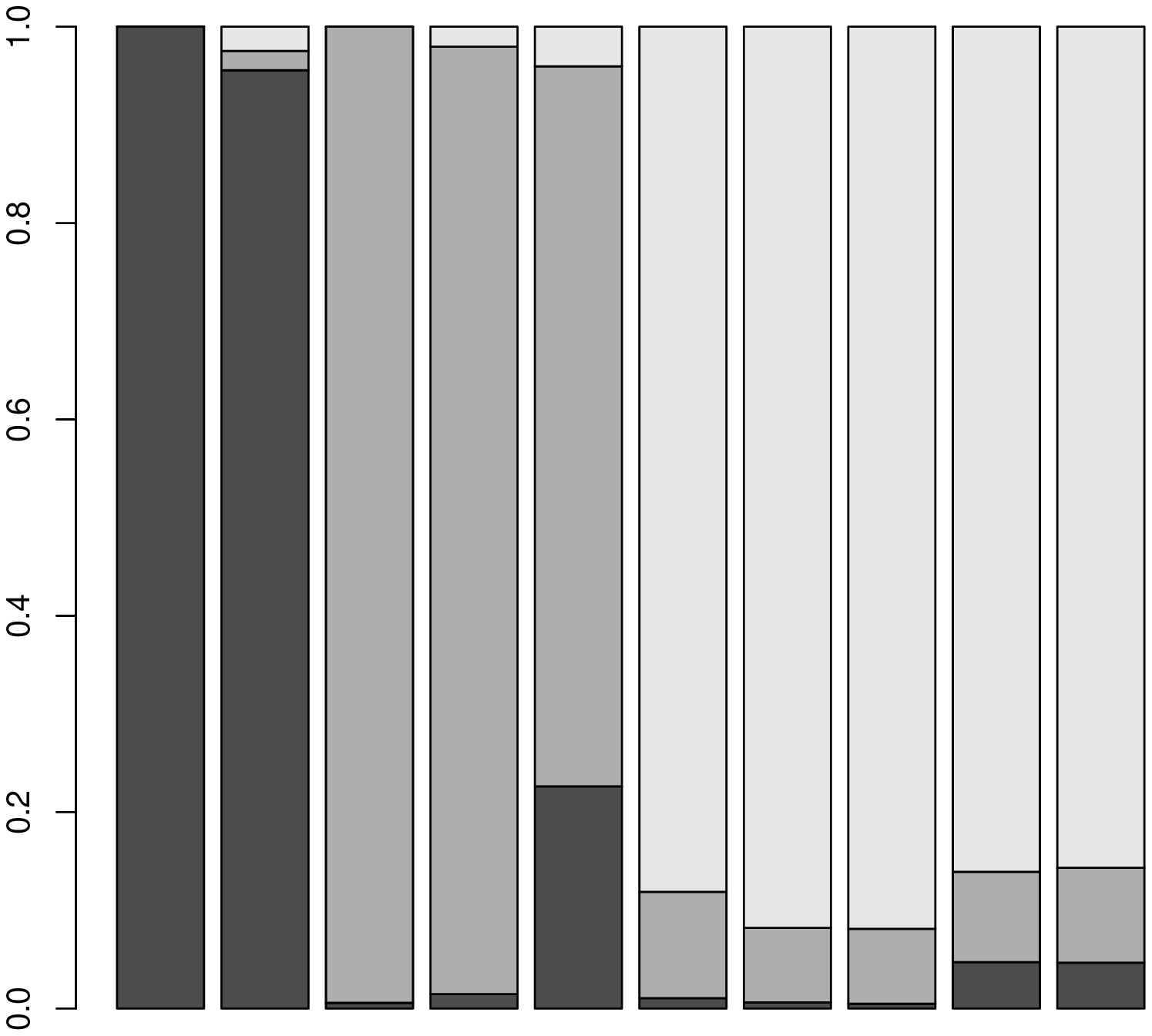}}
}
\caption{Component selection results for one randomly selected replicate, when the distribution of errors is Gaussian and $p=10$. The black, dark grey, and light grey areas represent the posterior probabilities of the component being nonlinear, linear and zero, respectively. }
\label{indicator_posterior_1}
\end{figure}

\begin{figure}[htp]
\centerline{
\subfigure[Mean]{\includegraphics[width=1.8in,trim=0 30 0 0]{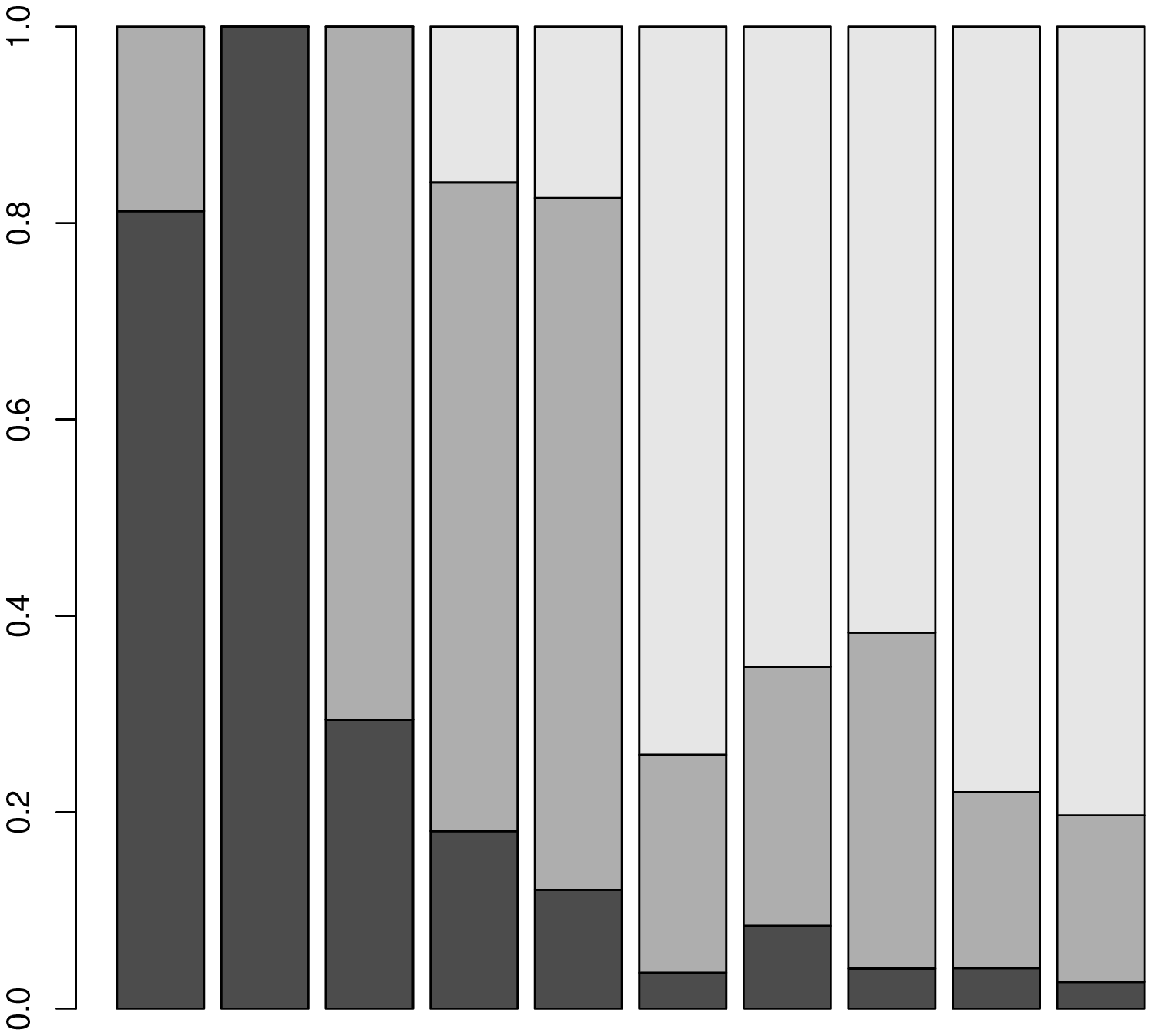}}
\hfil
\subfigure[$\tau=0.1$]{\includegraphics[width=1.8in,trim=0 30 0 0]{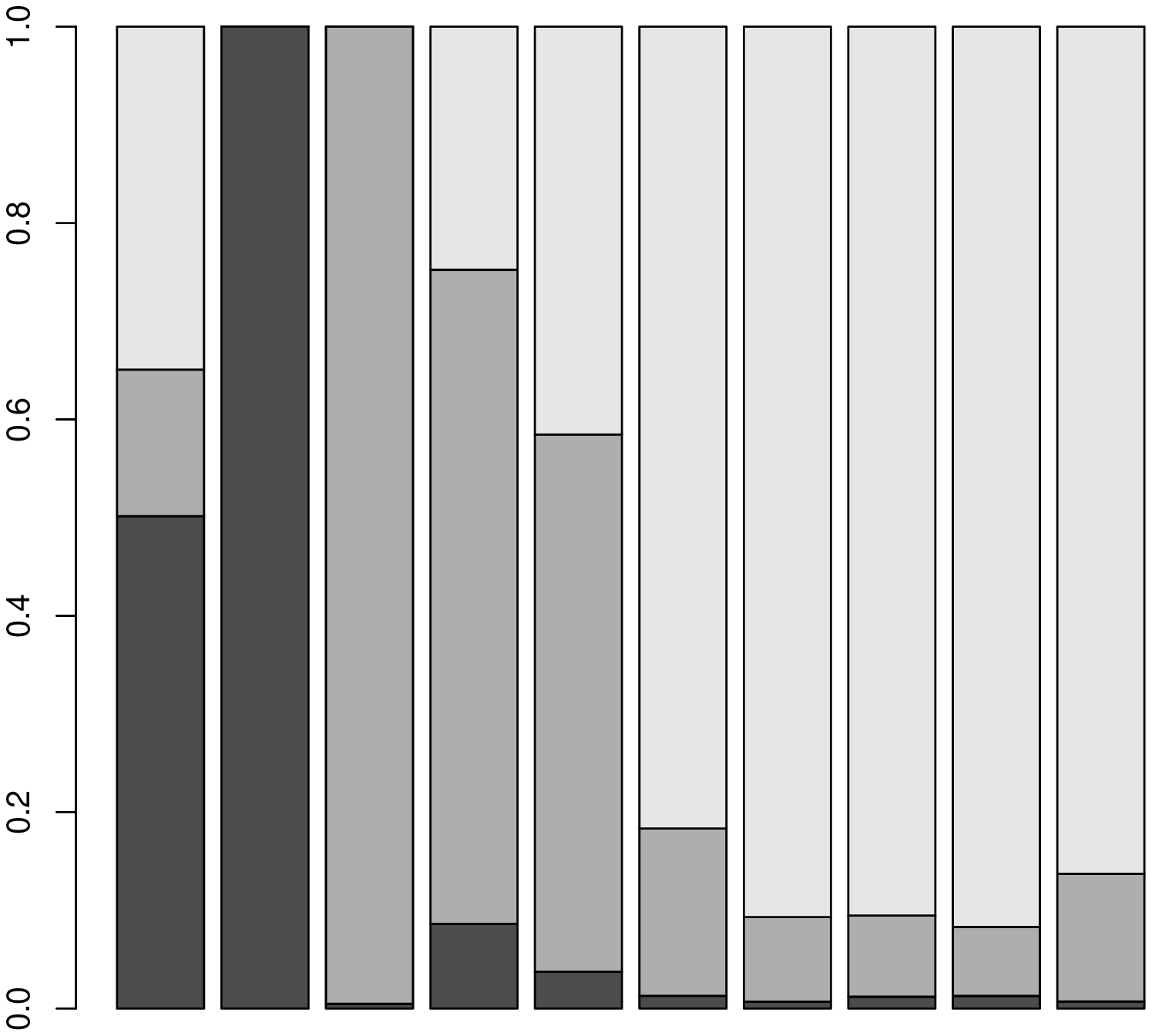}}
\hfil
\subfigure[$\tau=0.3$]{\includegraphics[width=1.8in,trim=0 30 0 0]{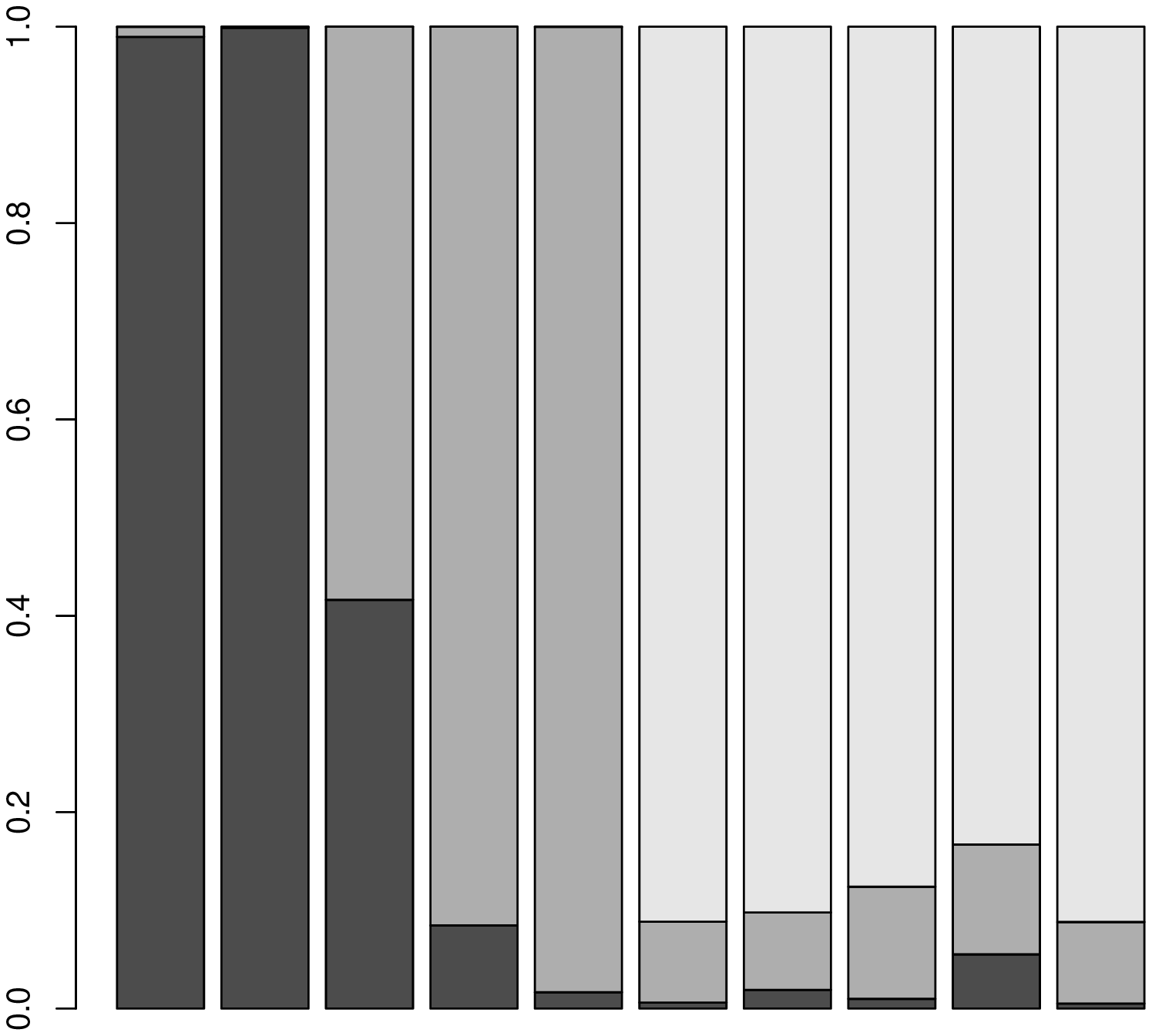}}
}

\centerline{
\subfigure[$\tau=0.5$]{\includegraphics[width=1.8in,trim=0 30 0 0]{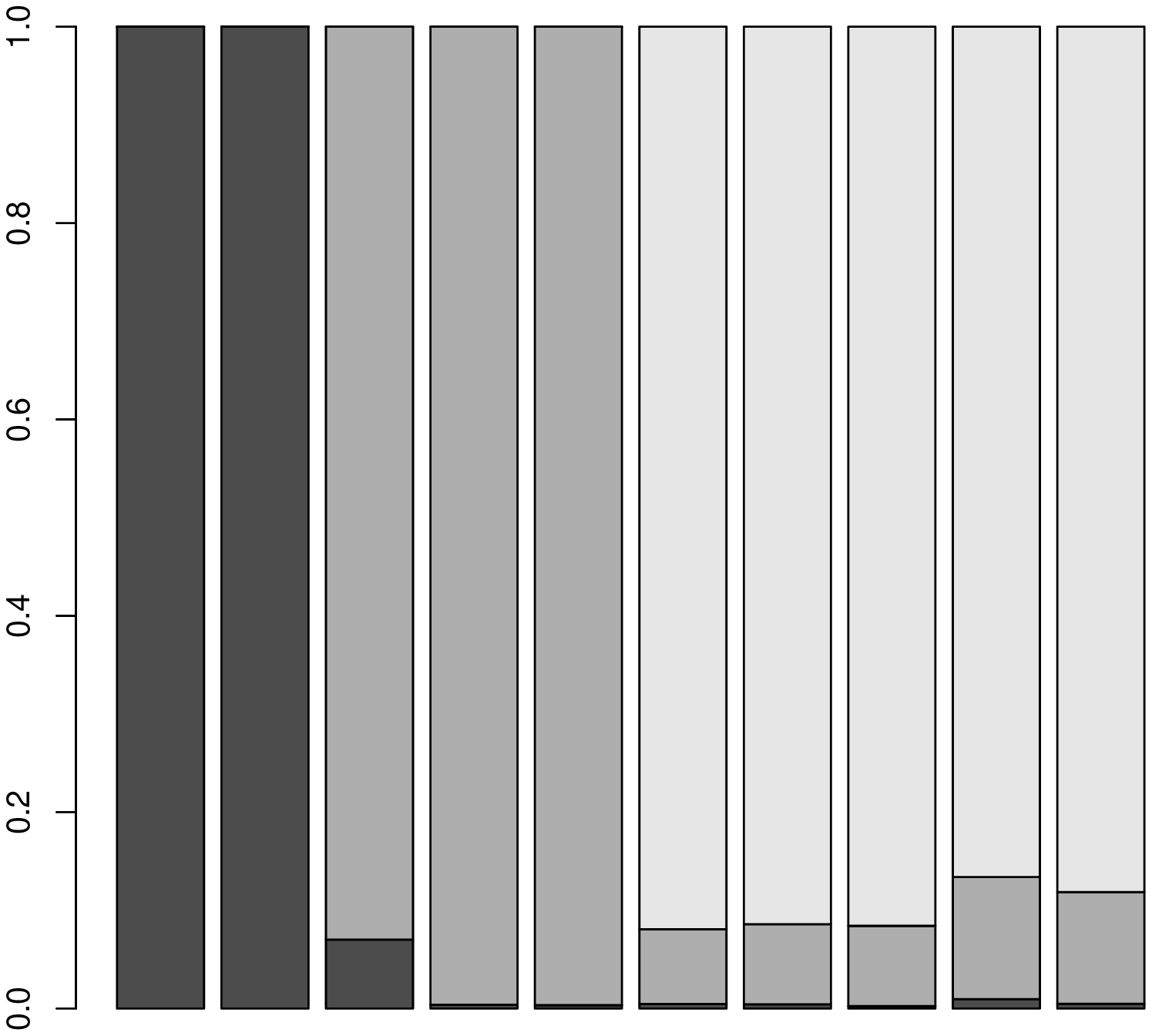}}
\hfil
\subfigure[$\tau=0.7$]{\includegraphics[width=1.8in,trim=0 30 0 0]{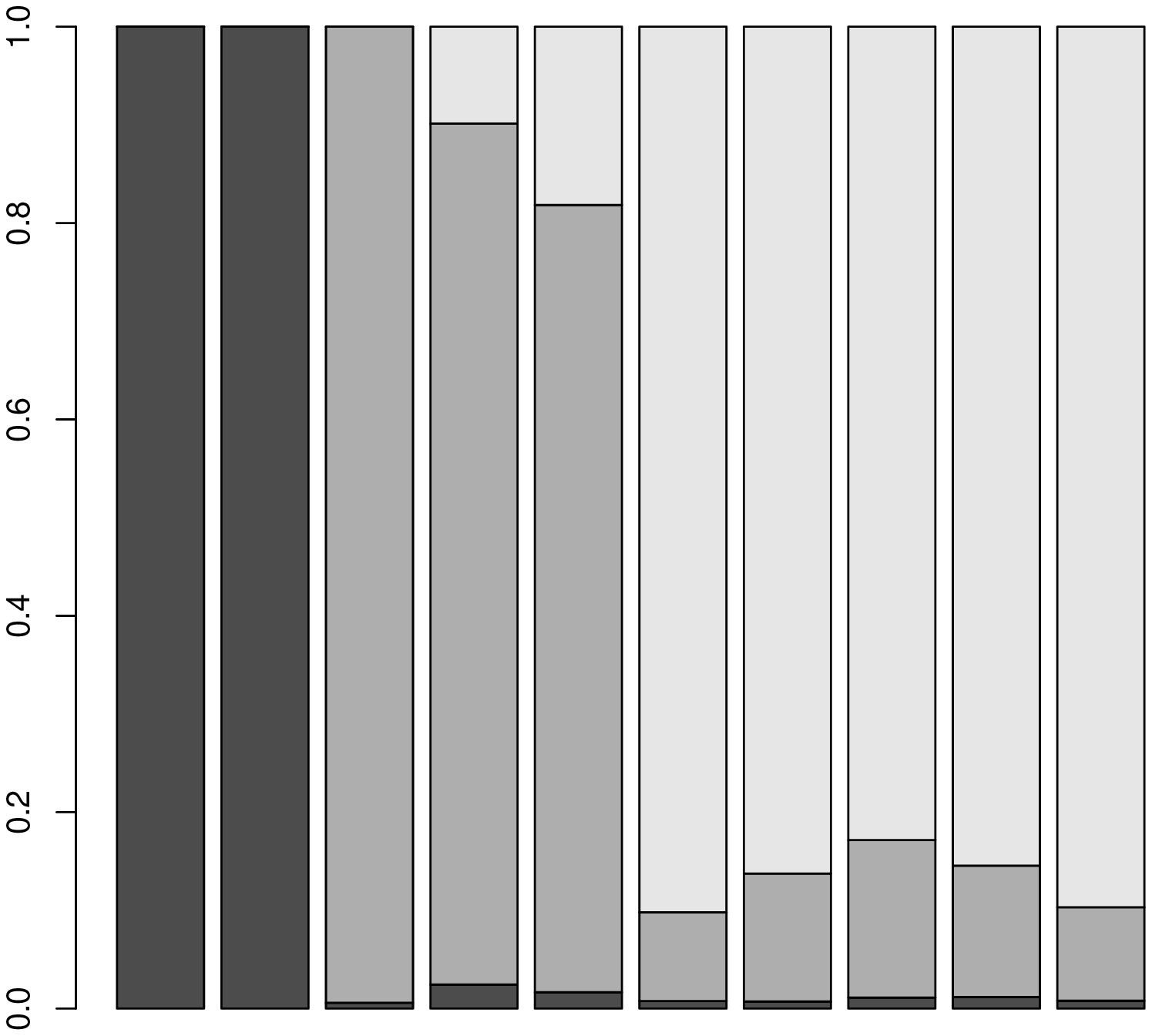}}
\hfil
\subfigure[$\tau=0.9$]{\includegraphics[width=1.8in,trim=0 30 0 0]{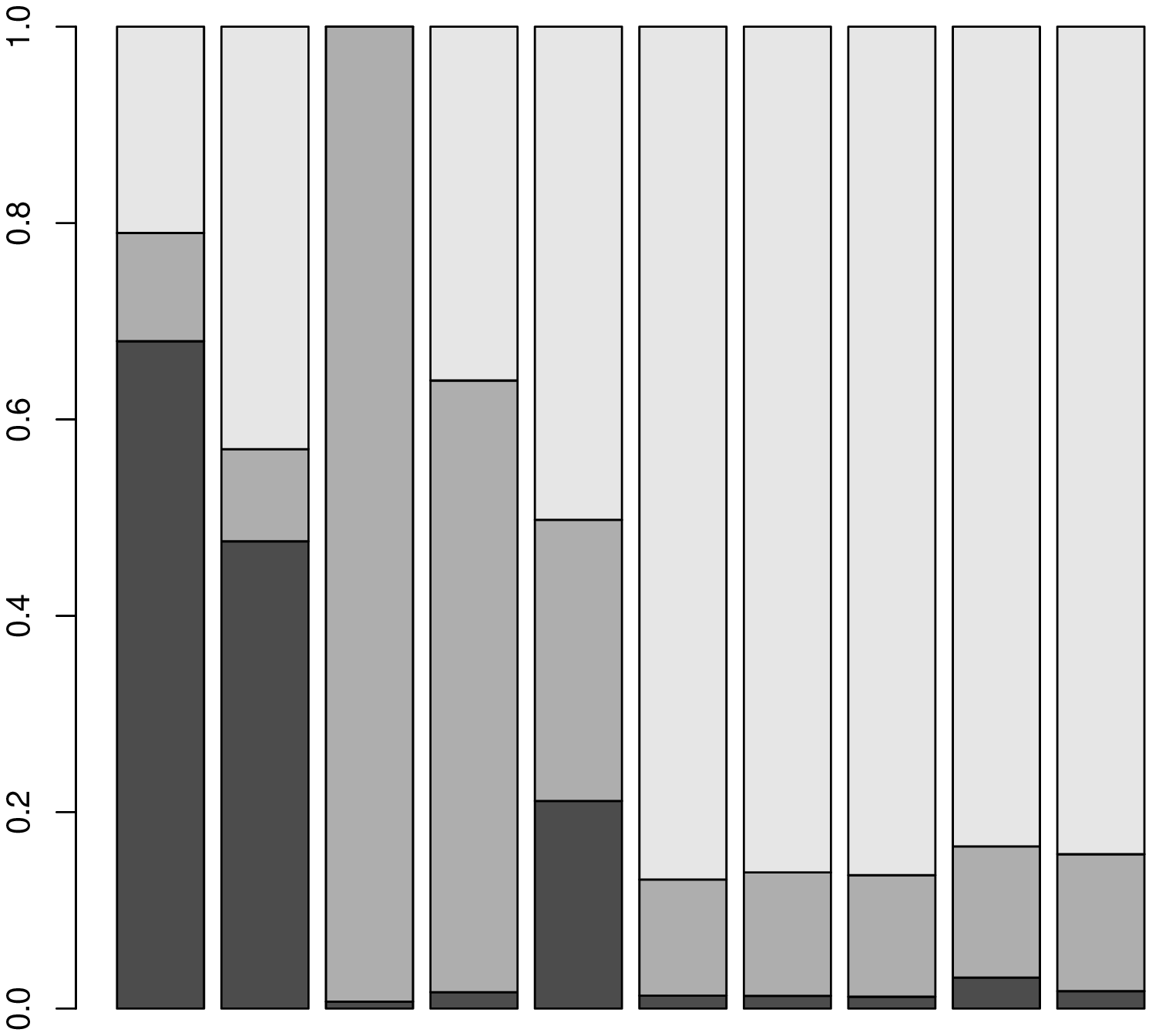}}
}
\caption{Component selection results for one randomly selected replicate, when the distribution of errors is Student's t and $p=10$.  }
\label{indicator_posterior_2}
\end{figure}

The results of zero and linear component selection are based on the estimation of posterior distribution of $\boldsymbol{\gamma}^{(\alpha)}$ and $\boldsymbol{\gamma}^{(\boldsymbol{\beta})}$. To be specific, we compare the estimated values of the posterior probabilities $p(\gamma_j^{(\boldsymbol{\beta})}=1|\boldsymbol{y},\boldsymbol{x})$, $p(\gamma_j^{(\alpha)}=1, \gamma_j^{(\boldsymbol{\beta})}=0|\boldsymbol{y},\boldsymbol{x})$ and $p(\gamma_j^{(\alpha)}=0,\gamma_j^{(\boldsymbol{\beta})}=0|\boldsymbol{y},\boldsymbol{x})$. The component will be identified as nonlinear, linear or zero according to which of the three probabilities is the largest (this rule is used only for illustration and Bayesians usually do not perform this step of making hard decision). In Tables 5-7, we report the number of nonzero components selected, number of nonzero components selected that are nonzero in the true model, number of linear components selected, and the number of linear components selected that are linear in the true model for BQPLAM and BPLAM. We report the number of variables selected and the number of variables selected that are truly nonzero for $\rm{BQAM_{v}}$ and $\rm{BLAM_{v}}$, since these two methods can only select nonzero components.  Figures \ref{indicator_posterior_1} and \ref{indicator_posterior_2} display estimates of the three posterior probabilities for each component based on one randomly selected replicate among the 100. The black, dark grey, and light grey areas represent the percentage of times the component is selected as nonlinear, linear and zero, respectively. It is seen from the figures that BQPLAM can detect linear and nonlinear components quite accurately.  

\begin{center}
\begin{table*}
\footnotesize
\caption{Summary of the component selection results for mean estimators and median estimators over 100 replicates, when $p=10$. ``Normal" and ``Student's t" indicate the distribution of $\epsilon_i$. }
\label{Variable_Selection_median}
\vspace{0.1in}
\centering 
\begin{tabular}{llcc}
\toprule 
\multicolumn{2}{c}{}&Normal&Student's t\\

\cmidrule(r){1-4} \multirow{2}{*} {$\rm{BPLAM}$}  
&$\rm{\#of\ Nonzero\ Variables }                       $&$5.21_{0.67} $       &$4.69_{1.89}$    \\  
&$\rm{\#of\ Correct\ Nonzero\ Variables }               $&$4.91_{0.32}$         &$4.00_{1.21}$     \\ 
&$\rm{\#of\ Linear\ Variables }                          $&$3.19_{0.82} $       &$3.22_{1.99}$     \\  
&$\rm{\#of\ Correct\ Linear\ Variables }                  $&$2.81_{0.46}$         &$2.16_{1.03}$\\  
\midrule    
\multirow{2}{*} {$\rm{BQPLAM}$} 
&$\rm{\#of\ Nonzero\ Variables }                       $&$5.14_{0.69} $       &$4.69_{1.01}$    \\  
&$\rm{\#of\ Correct\ Nonzero\ Variables }               $&$4.88_{0.39}$         &$4.54_{0.93}$     \\ 
&$\rm{\#of\ Linear\ Variables }                          $&$2.90_{0.94} $       &$2.58_{0.98}$     \\  
&$\rm{\#of\ Correct\ Linear\ Variables }                  $&$2.56_{0.69}$         &$2.35_{0.92}$\\     
\cmidrule(r){1-4} \multirow{2}{*} {$\rm{BQAM_{v}}$}                 
&$\rm{\#of\ Nonzero\ Variables}                          $&$4.51_{0.93}$       &$5.66_{2.23}$\\ 
&$\rm{\#of\ Correct\ Nonzero\ Variables }                 $&$4.43_{0.87}$         &$4.31_{0.85}$\\ 
\cmidrule(r){1-4} \multirow{2}{*} {$\rm{BQLM_{v}}$}                   
&$\rm{\#of\ Nonzero\ Variables}                          $&$4.39_{1.18}$       &$3.82_{1.26}$\\ 
&$\rm{\#of\ Correct\ Nonzero\ Variables }                 $&$3.90_{0.89}$         &$3.49_{0.97}$\\ 
\bottomrule
\end{tabular}
\end{table*}
\end{center}

\begin{center}
\begin{table*}
\scriptsize
\caption{Summary of the component selection results for quantile estimators at different quantile levels over 100 replicates,  when the distribution of errors is Gaussian and $p=10$. }
\label{Variable_Selection_qr1}
\vspace{0.1in}
\centering 
\begin{tabular}{llcccc}
\toprule 
\multicolumn{2}{c}{}&$\tau=0.1$&$\tau=0.3$&$\tau=0.7$&$\tau=0.9$\\
\midrule
\multirow{2}{*} {$\rm{BQPLAM}$}          
&$\rm{\#of\ Nonzero\ Variables }                       $&$5.00_{1.00} $&    $5.01_{0.77}$&$4.92_{0.73}$ & $4.92_{1.05}$   \\  
&$\rm{\#of\ Correct\ Nonzero\ Variables }              $&$4.56_{0.75}$&  $4.77_{0.52}$&$4.73_{0.61}$ &$4.53_{0.79}$     \\ 
&$\rm{\#of\ Linear\ Variables }                          $&$2.40_{1.39} $   & $2.67_{1.05}$   &$2.58_{1.05} $& $2.45_{1.39}$     \\  
&$\rm{\#of\ Correct\ Linear\ Variables }                  $&$1.97_{1.04}$    & $2.40_{0.84}$ &$2.35_{0.88}$ &$2.00_{1.00}$\\     
\cmidrule(r){1-6} \multirow{2}{*} {$\rm{BQAM_{v}}$} 
&$\rm{\#of\ Nonzero\ Variables}                          $&$4.17_{0.91}$       &$4.37_{0.91}$ &$4.28_{0.81}$&$4.21_{0.83}$\\ 
&$\rm{\#of\ Correct\ Nonzero\ Variables }                 $&$4.12_{0.90}$         &$4.24_{0.89}$ &$4.26_{0.81}$&$4.17_{0.83}$\\ 
\cmidrule(r){1-6} \multirow{2}{*} {$\rm{BQLM_{v}}$} 
&$\rm{\#of\ Nonzero\ Variables}                          $&$4.44_{1.92}$       &$3.88_{1.91}$ &$4.19_{1.18}$&$4.88_{1.54}$\\ 
&$\rm{\#of\ Correct\ Nonzero\ Variables }                 $&$3.45_{1.07}$         &$3.49_{1.05}$ &$3.75_{0.90}$&$3.52_{0.96}$\\         
\bottomrule
\end{tabular}
\end{table*}
\end{center}

\begin{center}
\begin{table*}
\scriptsize
\caption{Summary of the component selection results for quantile estimators at different quantile levels over 100 replicates, when the distribution of errors is Student's t  and $p=10$. }
\label{Variable_Selection_qr2}
\vspace{0.1in}
\centering 
\begin{tabular}{llcccc}
\toprule 
\multicolumn{2}{c}{}&$\tau=0.1$&$\tau=0.3$&$\tau=0.7$&$\tau=0.9$\\
\midrule
\multirow{2}{*} {$\rm{BQPLAM}$} 
&$\rm{\#of\ Nonzero\ Variables }                          $&$4.15_{1.53}$&    $4.58_{0.96}$&$4.57_{0.94}$ & $4.31_{1.66}$   \\  
&$\rm{\#of\ Correct\ Nonzero\ Variables }                 $&$3.72_{1.36}$&  $4.51_{0.93}$&$4.49_{0.87}$ &$3.76_{1.07}$     \\ 
&$\rm{\#of\ Linear\ Variables }                          $&$2.31_{1.38}$  & $2.49_{1.11}$  &$2.49_{1.12} $& $2.63_{1.96}$     \\  
&$\rm{\#of\ Correct\ Linear\ Variables }                  $&$1.76_{1.03}$    & $2.28_{0.92}$ &$2.29_{1.10}$ &$1.65_{1.02}$\\     
\cmidrule(r){1-6} \multirow{2}{*} {$\rm{BQAM_{v}}$} 
&$\rm{\#of\ Nonzero\ Variables}                          $&$4.20_{2.07}$       &$5.45_{2.23}$&$5.18_{2.07}$&$4.22_{1.97}$\\ 
&$\rm{\#of\ Correct\ Nonzero\ Variables }                 $&$3.69_{0.93}$       &$4.21_{0.82}$ &$4.13_{0.83}$&$3.66_{0.85}$\\ 

\cmidrule(r){1-6} \multirow{2}{*} {$\rm{BQLM_{v}}$}    
&$\rm{\#of\ Nonzero\ Variables}                          $&$3.68_{1.93}$       &$3.78_{1.56}$&$3.94_{1.39}$&$4.07_{2.08}$\\ 
&$\rm{\#of\ Correct\ Nonzero\ Variables }                 $&$2.81_{1.26}$       &$3.35_{1.08}$ &$3.58_{1.05}$&$3.02_{1.29}$\\          
\bottomrule
\end{tabular}
\end{table*}
\end{center}

\begin{figure}[htp]
\centerline{
\subfigure[Mean]{\includegraphics[width=2.8in,trim=0 30 0 0]{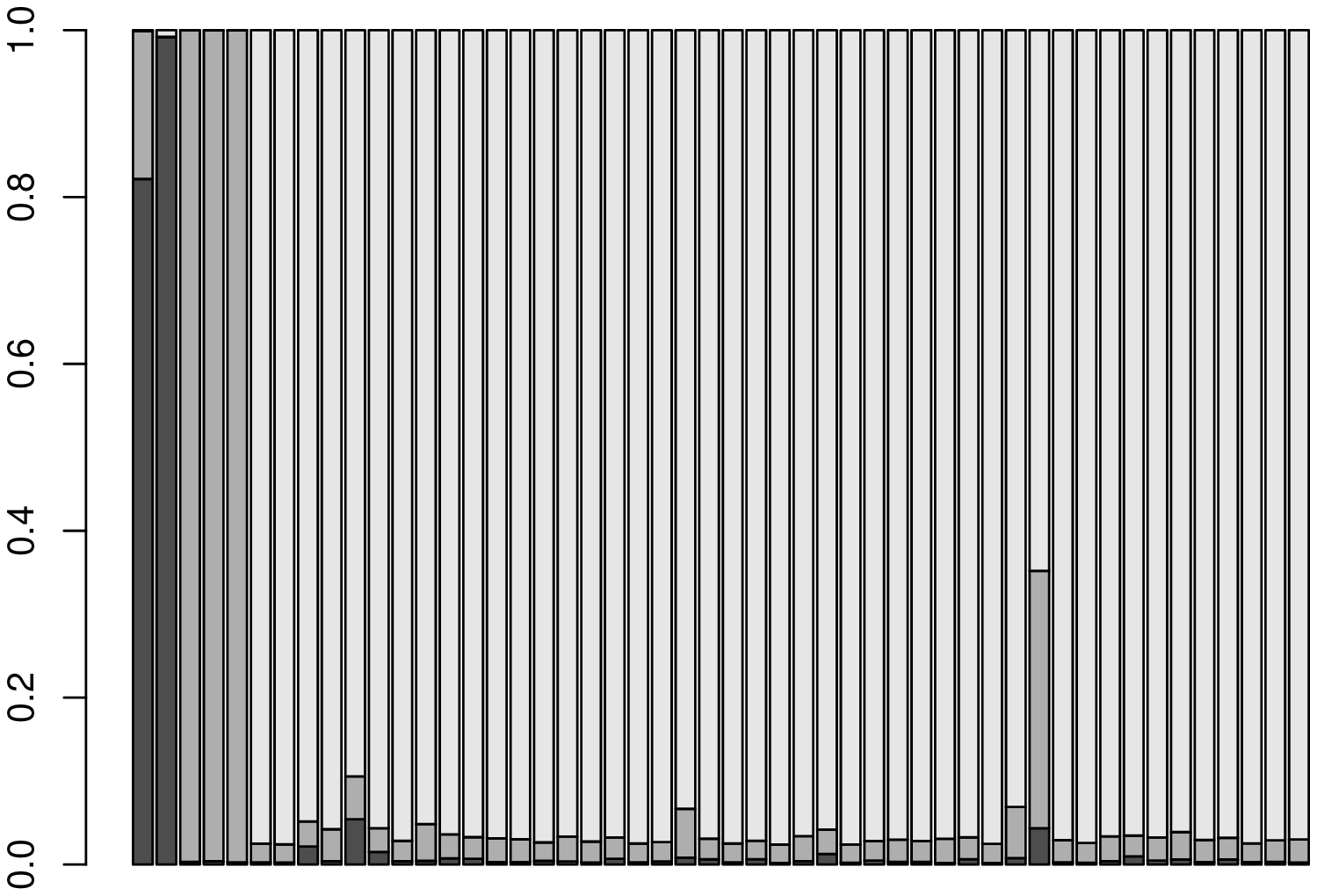}}
\hfil
\subfigure[$\tau=0.5$]{\includegraphics[width=2.8in,trim=0 30 0 0]{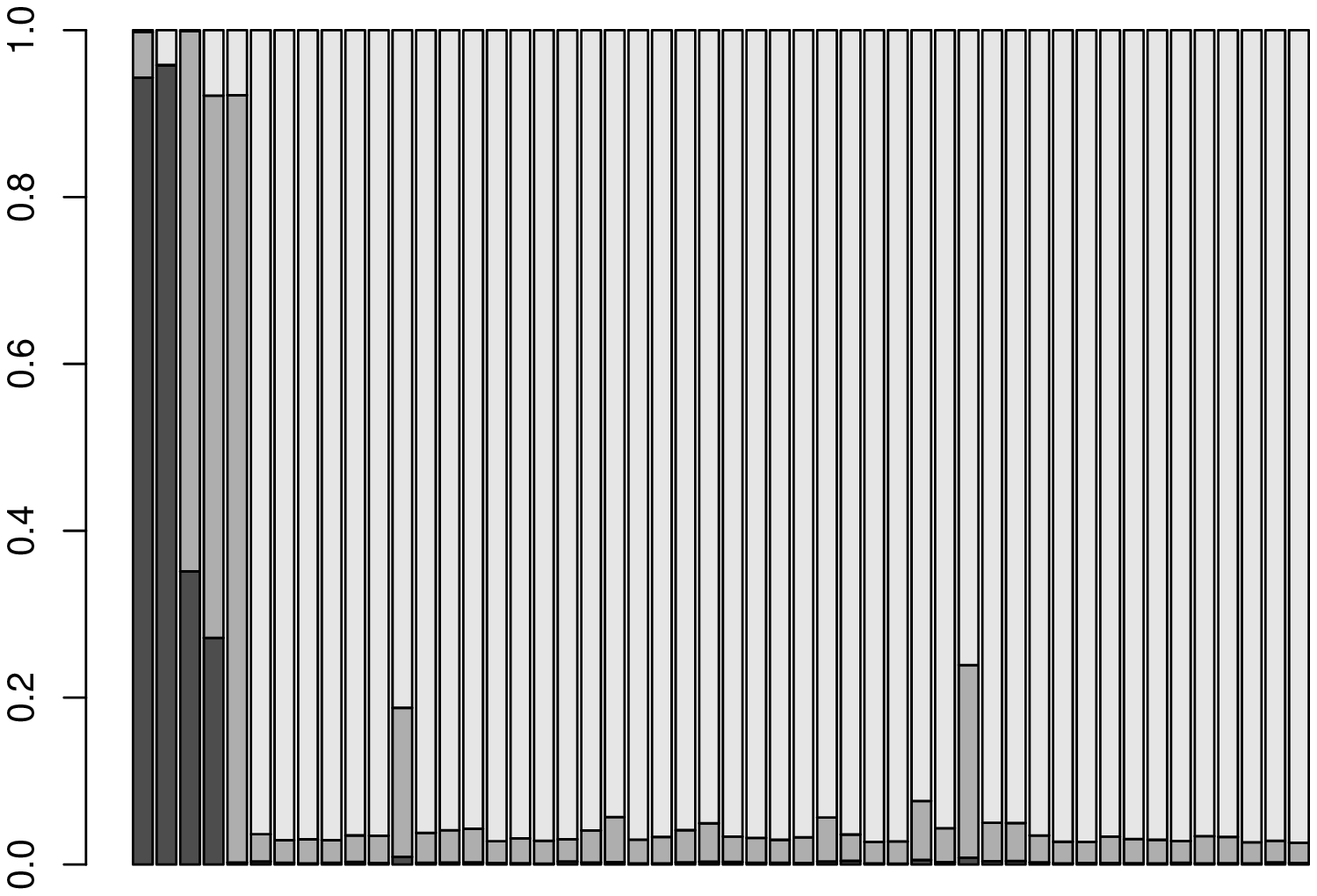}}
}
\centerline{
\subfigure[Mean]{\includegraphics[width=2.8in,trim=0 30 0 0]{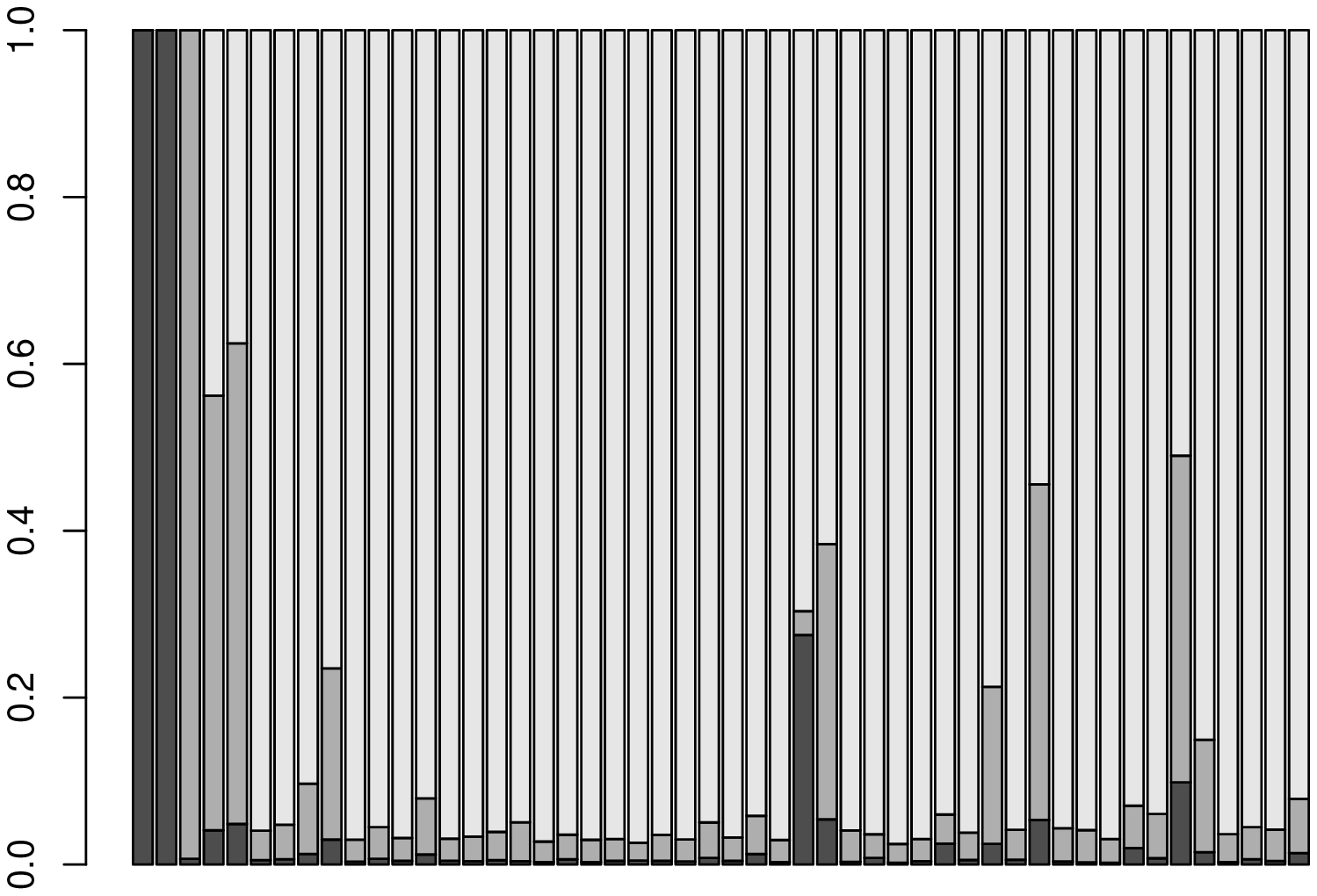}}
\hfil
\subfigure[$\tau=0.5$]{\includegraphics[width=2.8in,trim=0 30 0 0]{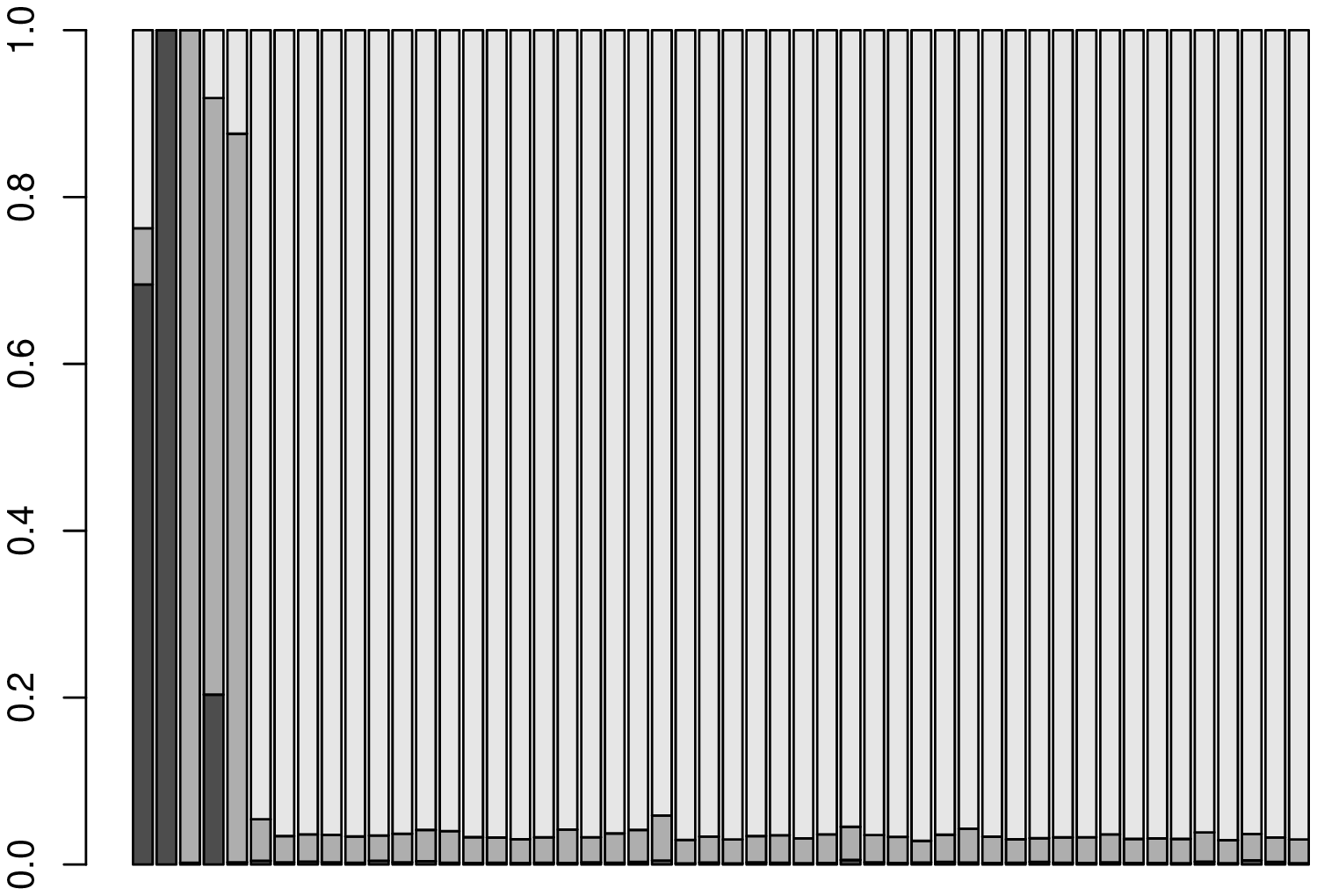}}
}
\caption{Component selection results for one randomly selected replicate with $p=50$. The upper panels are results when error distribution is Gaussian, and the lower panels are results when the error distribution is Student's t.}
\label{indicator_posterior_3}
\end{figure}

\begin{figure}[htp]
\centerline{
\subfigure[Mean]{\includegraphics[width=2.8in,trim=0 30 0 0]{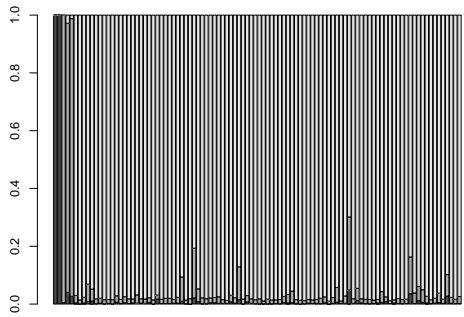}}
\hfil
\subfigure[$\tau=0.5$]{\includegraphics[width=2.8in,trim=0 30 0 0]{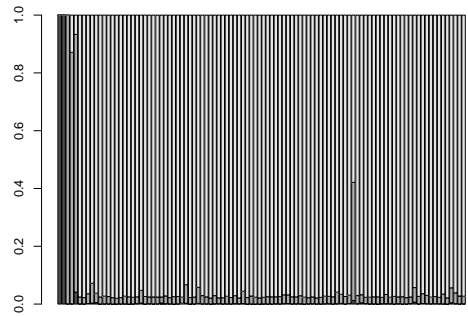}}
}

\centerline{
\subfigure[Mean]{\includegraphics[width=2.8in,trim=0 30 0 0]{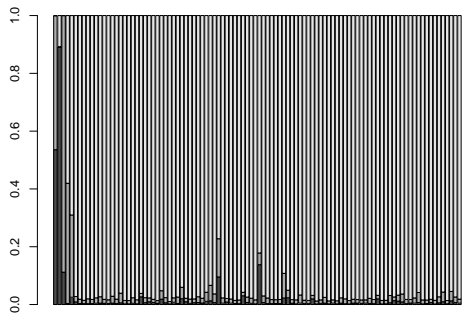}}
\hfil
\subfigure[$\tau=0.5$]{\includegraphics[width=2.8in,trim=0 30 0 0]{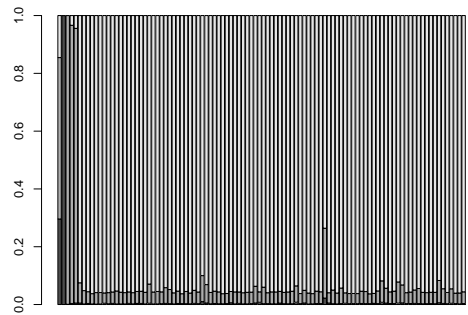}}
}
\caption{Component selection results for one randomly selected replicate with $p=100$. The upper panels are results when error distribution is Gaussian, and the lower panels are results when the error distribution is Student's t.}
\label{indicator_posterior_4}
\end{figure}

Finally, we increase the dimension in the example (by adding zero components) to demonstrate the performance of our proposed method in higher dimensions. We consider the dimension $p=50$ and $100$. To save space and time, we only consider BQPLAM at $\tau=0.5$ and BPLAM. Test errors and variable selection results are presented in Tables \ref{test_error_high} and \ref{Variable_Selection_high}. The estimates of three posterior probabilities for each component (for one data set) are displayed in Figures \ref{indicator_posterior_3} and \ref{indicator_posterior_4}. The results show that our proposed method still performs well.

\begin{center}
\begin{table*}
\footnotesize
\caption{Summary of  testing errors for mean and median estimators over 100 replicates, when $p=50$ and $100$. ``Normal" and ``Student's t" indicate the distribution of $\epsilon_i$. }
\label{test_error_high}
\vspace{0.1in}
\centering 
\begin{tabular}{llcccc}
\toprule 
\multicolumn{2}{c}{}&\multicolumn{2}{c}{$p=50$}&\multicolumn{2}{c}{$p=100$}\\
\cmidrule(r){3-4} \cmidrule(r){5-6} 
\multicolumn{2}{c}{}&RMSE&AD&RMSE&AD\\
\cmidrule(r){1-6} \multirow{2}{*} {Normal}
&$\rm{BPLAM}$&                $0.624_{0.055} $        &$0.482_{0.046} $ &$0.638_{0.060}$&$0.497_{0.050}$\\   
&$\rm{BQPLAM}$&            $0.625_{0.057} $        &$0.483_{0.048} $&$0.643_{0.054}$ &$0.506_{0.045}$\\             
\midrule    
\multirow{2}{*} {Student's t} 
&$\rm{BPLAM}$&                           $1.251_{0.069}$          &$0.685_{0.088}$  &$1.563_{0.657}$ &$0.709_{0.108}$ \\   
&$\rm{BQPLAM}$&                       $1.176_{0.039}$    &$0.569_{0.065}$  &$1.185_{0.052}$  &$0.588_{0.081}$ \\             
\bottomrule
\end{tabular}
\end{table*}
\end{center}

\begin{center}
\begin{table*}
\scriptsize
\caption{Summary of the component selection results for mean and median estimators over 100 replicates, when $p=50$ and $100$. ``Normal" and ``Student's t" indicate the distribution of $\epsilon_i$.}
\label{Variable_Selection_high}
\vspace{0.1in}
\centering 
\begin{tabular}{llcccc}
\toprule 
\multicolumn{2}{c}{}&\multicolumn{2}{c}{$p=50$}&\multicolumn{2}{c}{$p=100$}\\
\cmidrule(r){3-4} \cmidrule(r){5-6} 
\multicolumn{2}{c}{}&Normal&Student's t&Normal&Student's t\\
\cmidrule(r){1-6} \multirow{2}{*} {$\rm{BPLAM}$}
&$\rm{\#of\ Nonzero\ Variable }                       $&$4.32_{1.12} $       &$2.73_{1.45}$   &$4.82_{1.11}$   &$2.42_{1.73}$    \\  
&$\rm{\#of\ Correct\ Nonzero\ Variable }               $&$4.19_{0.99}$         &$2.57_{1.38}$  &$4.10_{1.11}$    &$2.21_{1.49}$       \\ 
&$\rm{\#of\ Linear\ Variable }                          $&$2.38_{1.30} $       &$1.92_{1.21}$ &$2.26_{1.04}$&$1.61_{1.40}$    \\  
&$\rm{\#of\ Correct\ Linear\ Variable }                  $&$2.10_{1.07}$         &$1.45_{0.93}$&$1.96_{0.99}$&$1.31_{1.32}$ \\  
\midrule    
\multirow{2}{*} {$\rm{BQPLAM}$} 
&$\rm{\#of\ Nonzero\ Variable }                       $&$4.41_{0.17} $          &$4.05_{1.17}$ &$4.29_{1.07}$ &$3.72_{1.41}$     \\  
&$\rm{\#of\ Correct\ Nonzero\ Variable }               $&$4.30_{1.05}$                 &$4.02_{1.16}$  &$4.20_{1.04}$    &$3.72_{1.43}$         \\ 
&$\rm{\#of\ Linear\ Variable }                        $&$2.24_{1.26} $          &$2.13_{0.97}$    &$2.16_{1.07}$    &$2.06_{1.10}$      \\  
&$\rm{\#of\ Correct\ Linear\ Variable }                $&$2.06_{1.03}$         &$1.91_{1.05}$ &$1.98_{1.06}$    &$1.81_{1.01}$ \\
\bottomrule
\end{tabular}
\end{table*}
\end{center}

\subsection{Real data examples}
Now we illustrate the methodology with two data sets. In the following examples, we fix $\boldsymbol{\gamma}^{(\boldsymbol{\beta})}$ associated with the discrete dummy variables to be zero all the time, which means that dummy variables can only be modeled as linear or zero  based on the posterior estimate of $\boldsymbol{\gamma}^{(\alpha)}$. All covariates are standardized by a linear transformation to lie in $[0,1]$ before analysis.
\subsubsection{GDP growth data}
We apply our method to a GDP growth data. This data set  includes data for 74 countries over two 20-year time intervals: 1960-1979 and 1980-1999, with sample size $n=147$. This data set is used by \cite{tan2010no} to uncover the interplay between geography, institutions, and fractionalization in economic development by employing a regression tree analysis. Here we use our method to detect the determinants of economic growth. The dependent variable is the difference between the logarithmic values of real per capita GDP for the start and end years of each 20-year time interval. We consider four categories of independent variables, a total of 13. The first covariate we consider is a variable measuring the quality of expropriation risk of government across the years 1984-1997, which is denoted by EXPROP8497. \cite{tan2010no} also considers another variable ICRG8497 to measure the quality of institution. Since the correlation between the two variables is very high, at over 0.8, we drop the second one and only keep EXPROP8497.  The second category of independent variables includes:  proportion of a country's land area that experiences more than 5 frost days per month in winter (FROST5), percentage of a country's land area that is classified as a tropical eco-zone (ZTROPICS), and percentage of a country's land area within 100 km of an ice-free coast (LCR100KM). This category is used as proxy for climate. The third category measures degree of ethnic fractionalization, including ELF60, ETHNIC, LANG and RELIG. ELF60 measures the probability in 1960 that two randomly selected people from a given country will not belong to the same ethno-linguistic subgroup. ETHNIC combines racial and linguistic characteristics, LANG is based on data for shares of languages spoken as `mother tongues', and RELIG describes differences in religion. The last category of independent variables includes several familiar neoclassical determinants:  log net depreciation rate (MNGD), log investment share (MINV), log schooling (MSCH15), a dummy variable for the period 1960-1979 (DUM6079), and log initial per capita income (MGDP0). 

\begin{figure}[htp]
\centerline{
\subfigure[Mean]{\includegraphics[width=1.8in,trim=0 30 0 0]{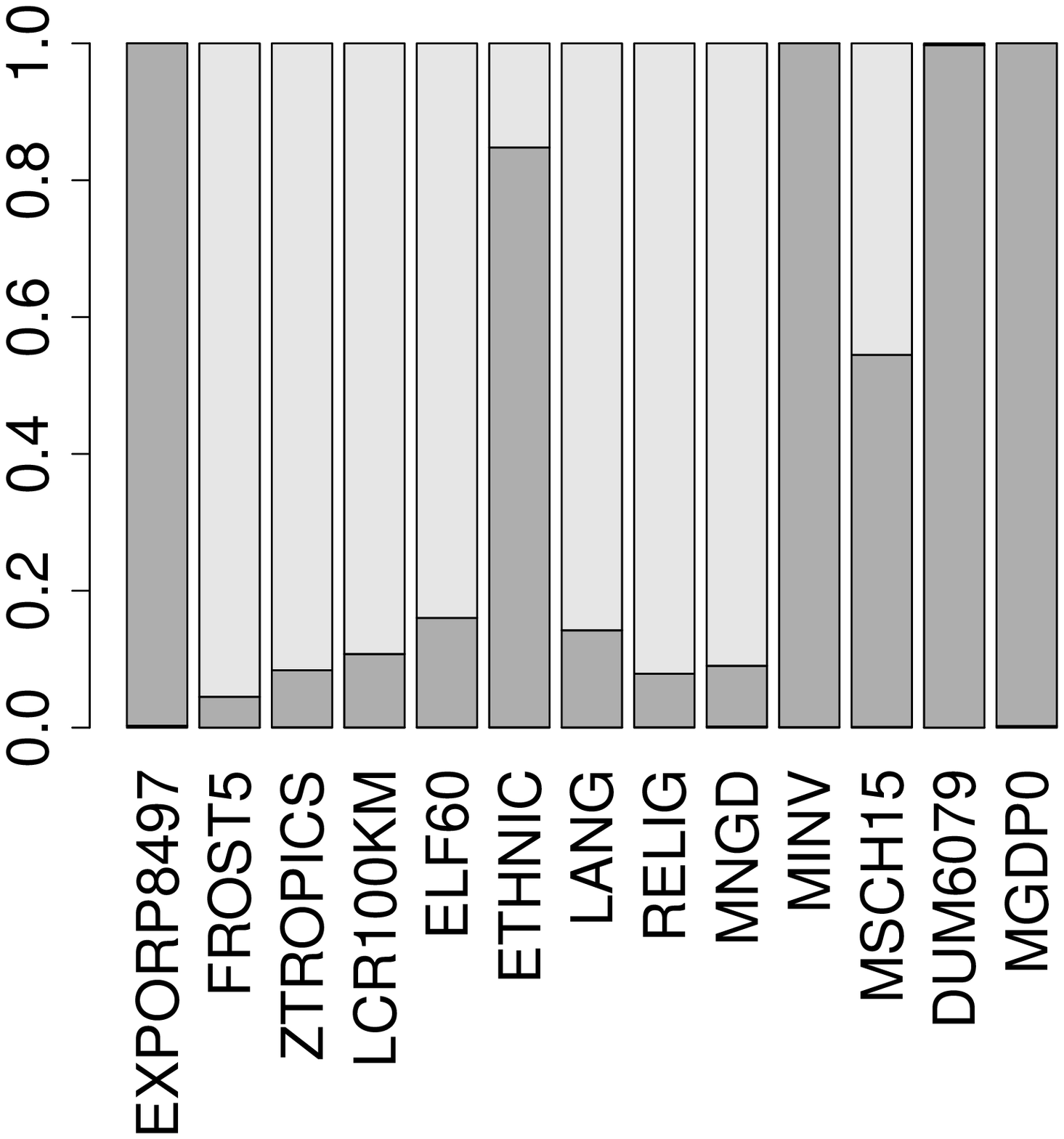}}
\subfigure[$\tau=0.1$]{\includegraphics[width=1.8in,trim=0 30 0 0]{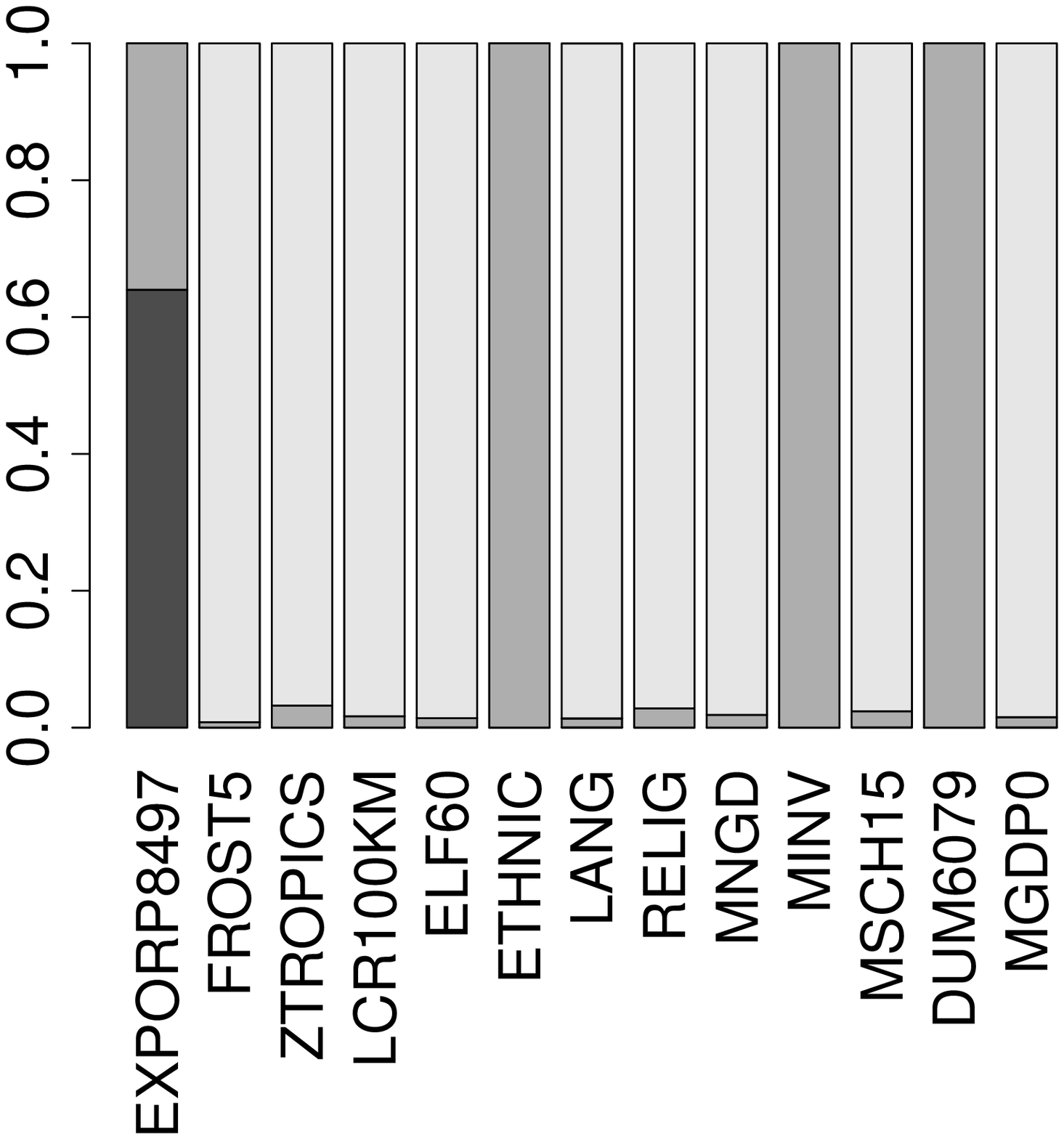}}
\subfigure[$\tau=0.3$]{\includegraphics[width=1.8in,trim=0 30 0 0]{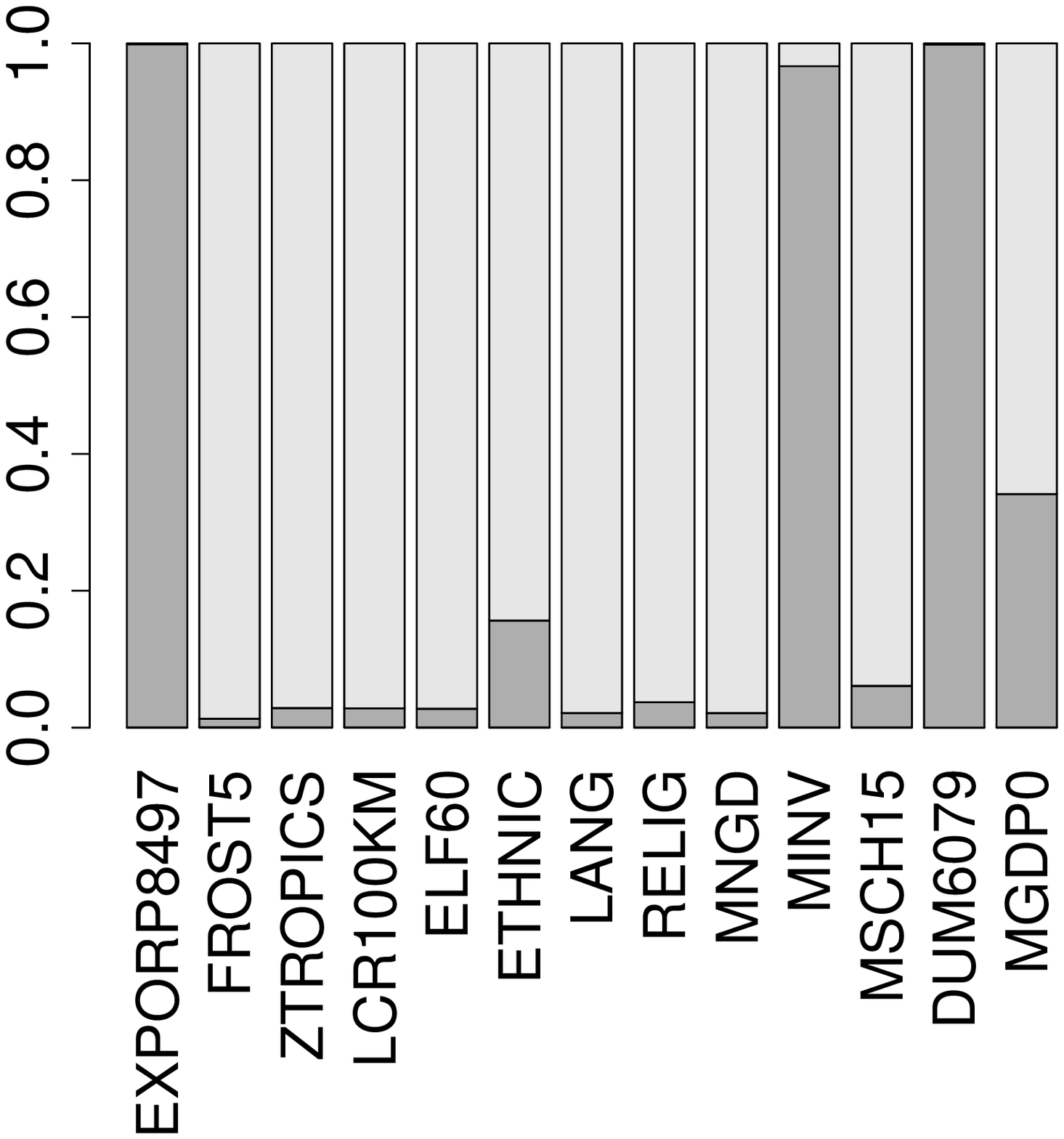}}
}
\centerline{
\subfigure[$\tau=0.5$]{\includegraphics[width=1.8in,trim=0 30 0 0]{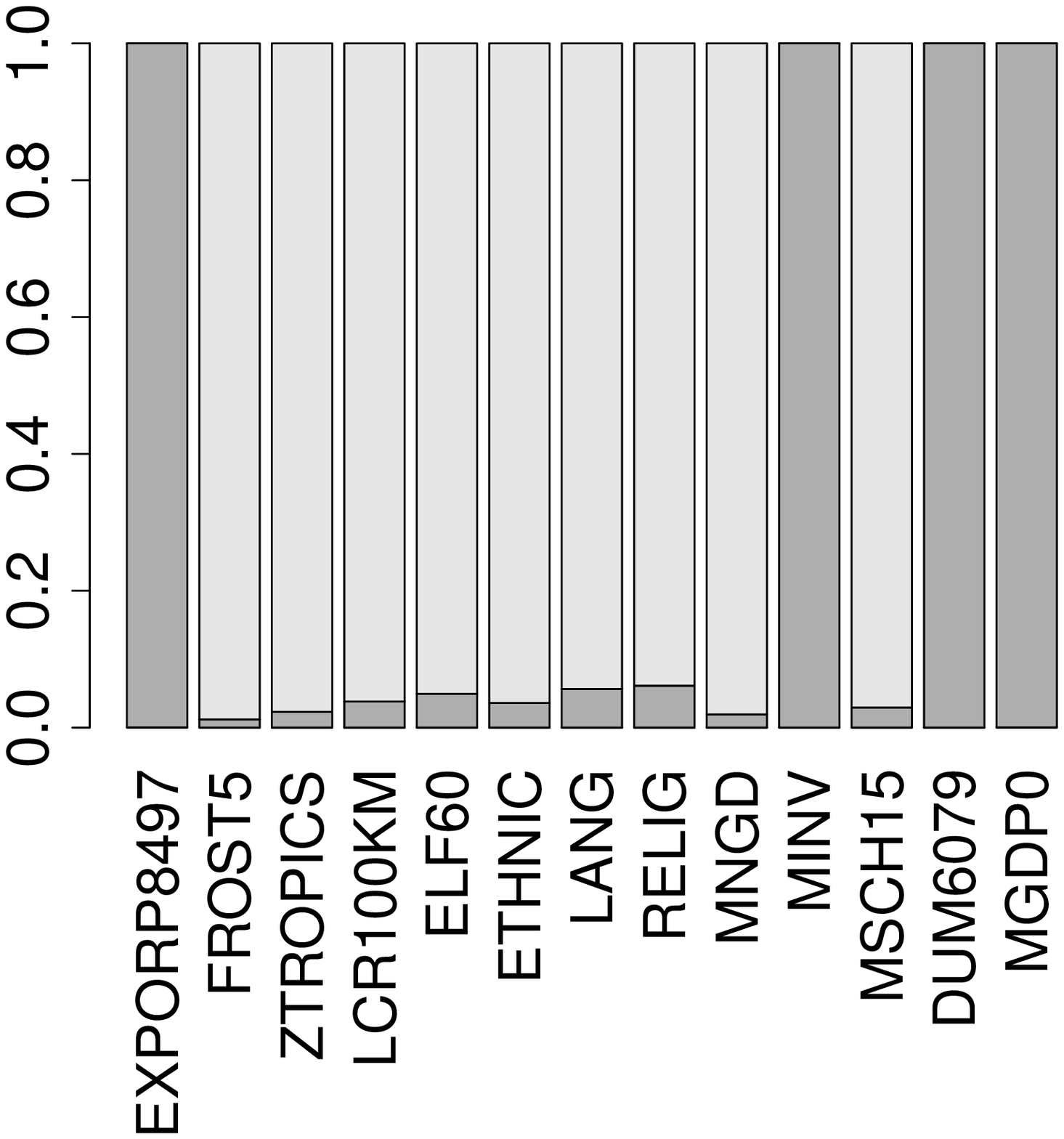}}
\subfigure[$\tau=0.7$]{\includegraphics[width=1.8in,trim=0 30 0 0]{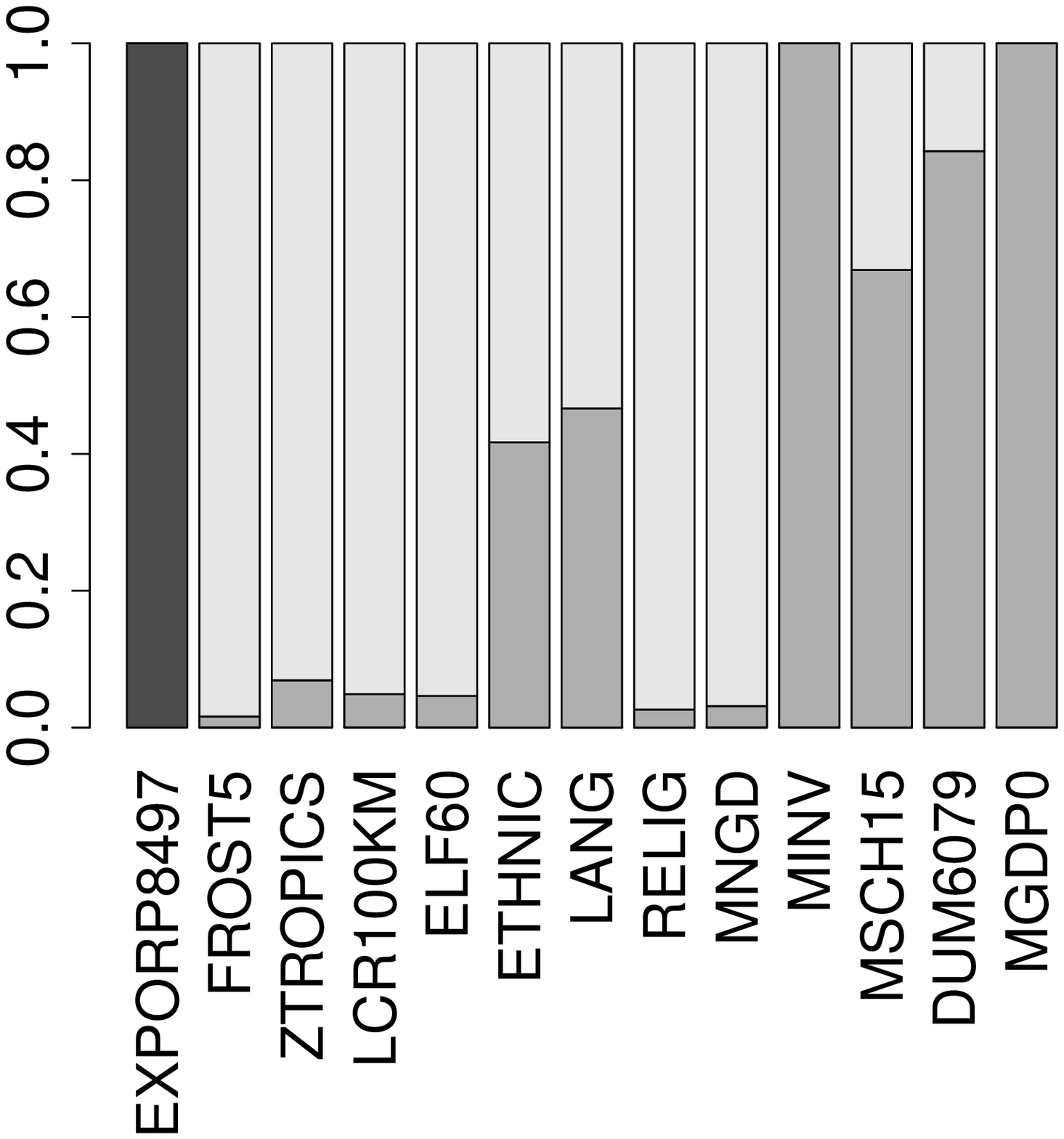}}
\subfigure[$\tau=0.9$]{\includegraphics[width=1.8in,trim=0 30 0 0]{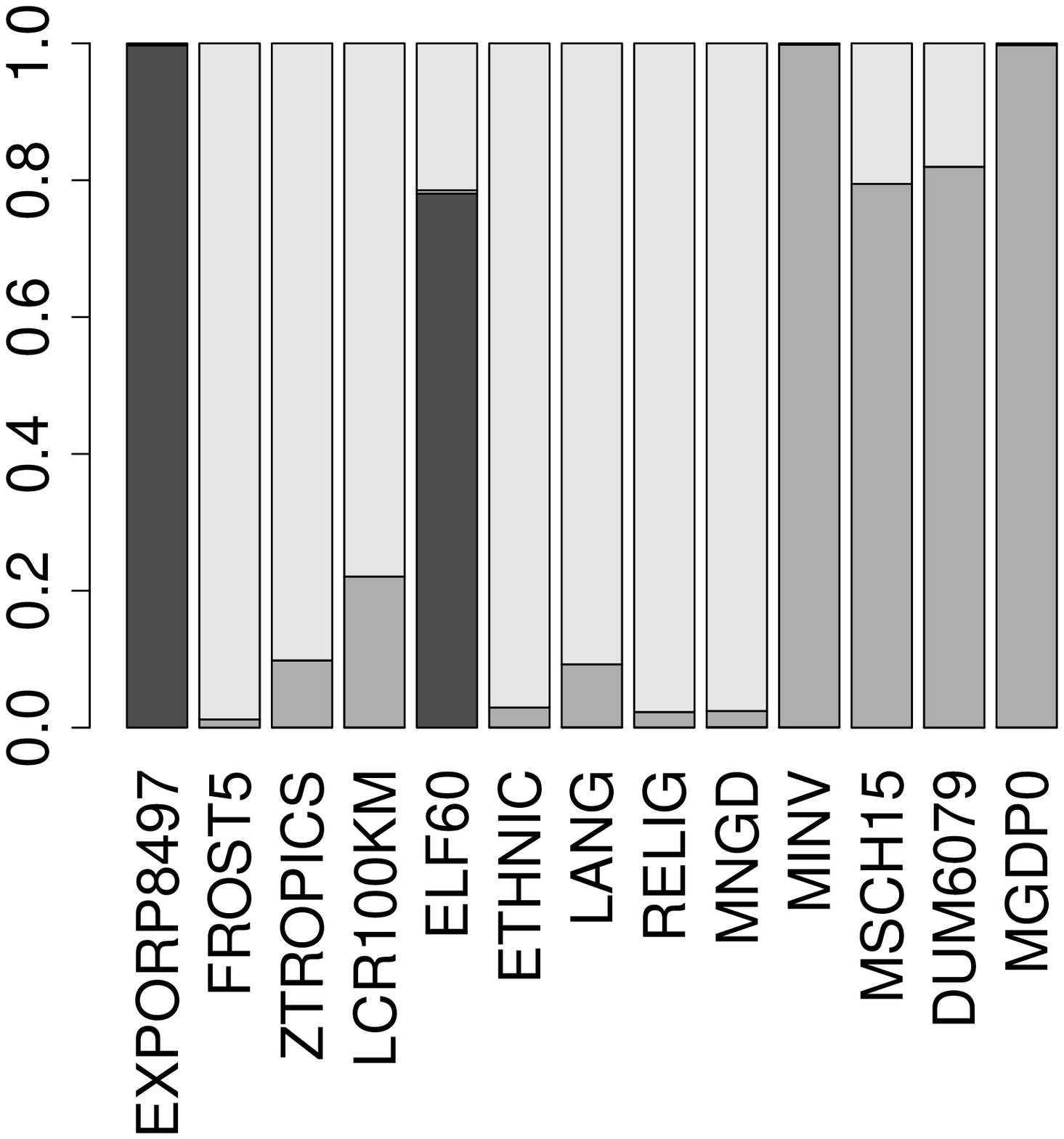}}
}
\caption{Component selection results for the GDP growth data.}
\label{indicator_posterior_GDP}
\end{figure}

\begin{figure}[htp]
\centerline{
\subfigure{\includegraphics[width=1.5in,trim=0 30 0 0]{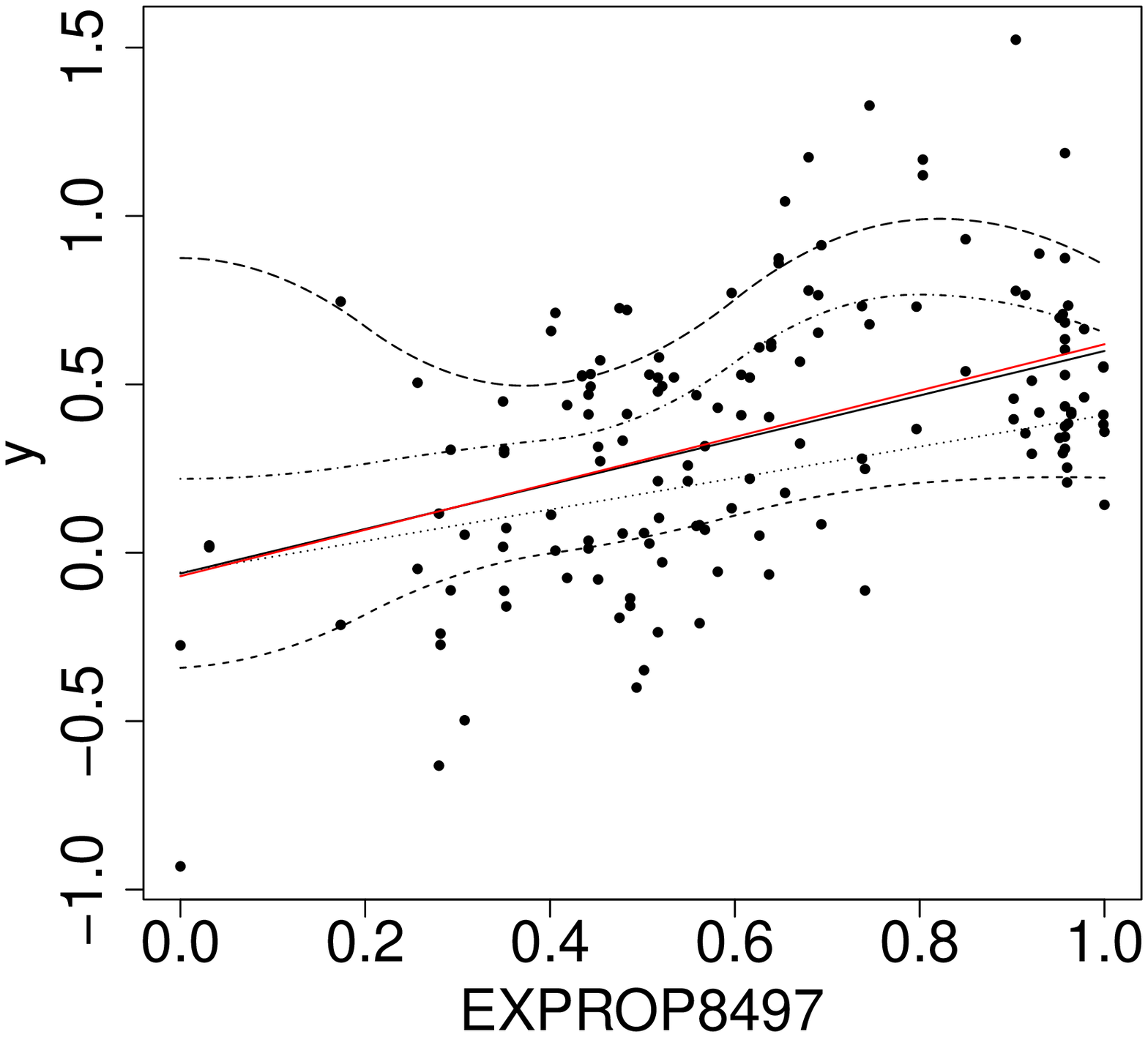}}
\hfil
\subfigure{\includegraphics[width=1.5in,trim=0 30 0 0]{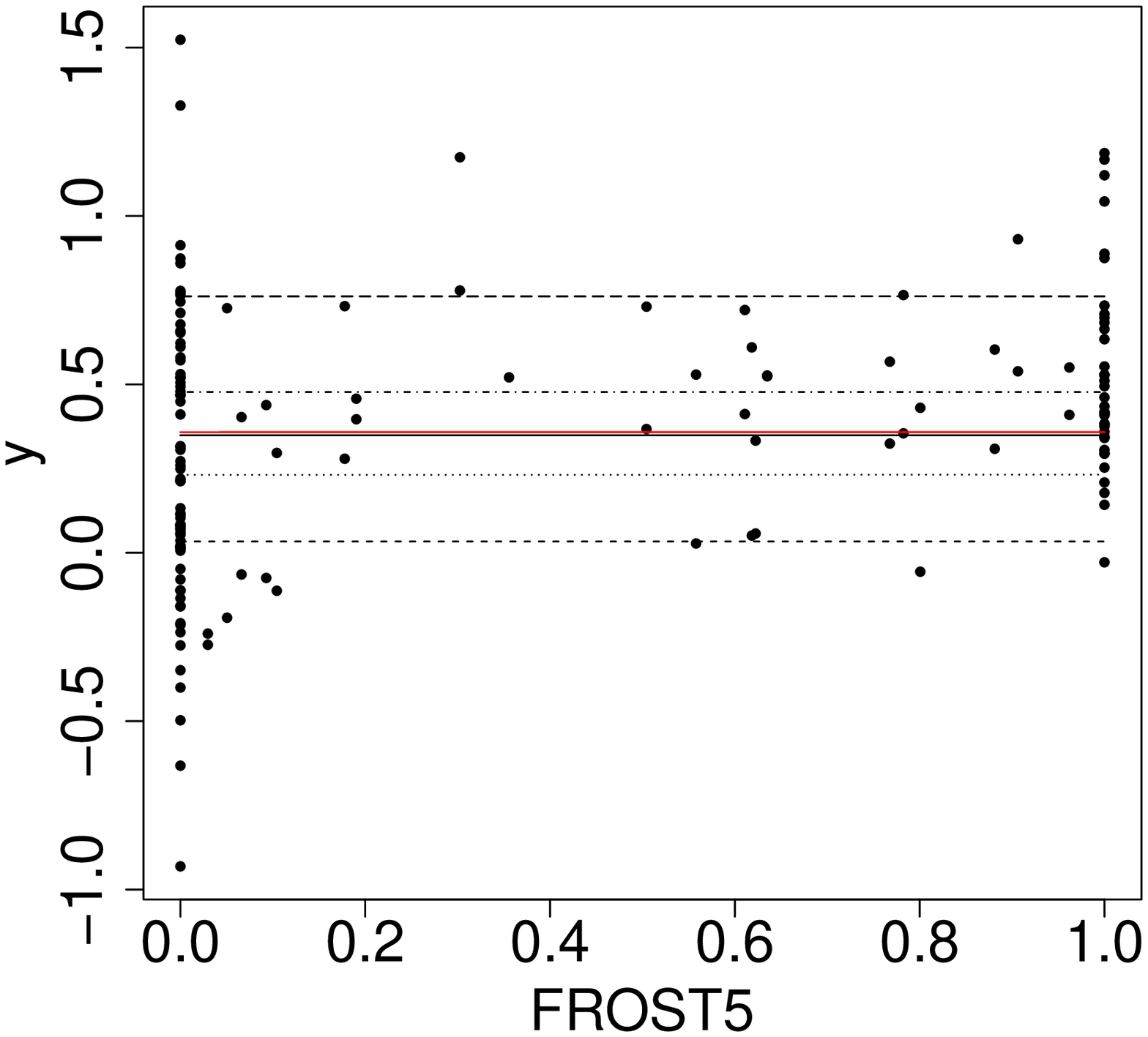}}
\hfil
\subfigure{\includegraphics[width=1.5in,trim=0 30 0 0]{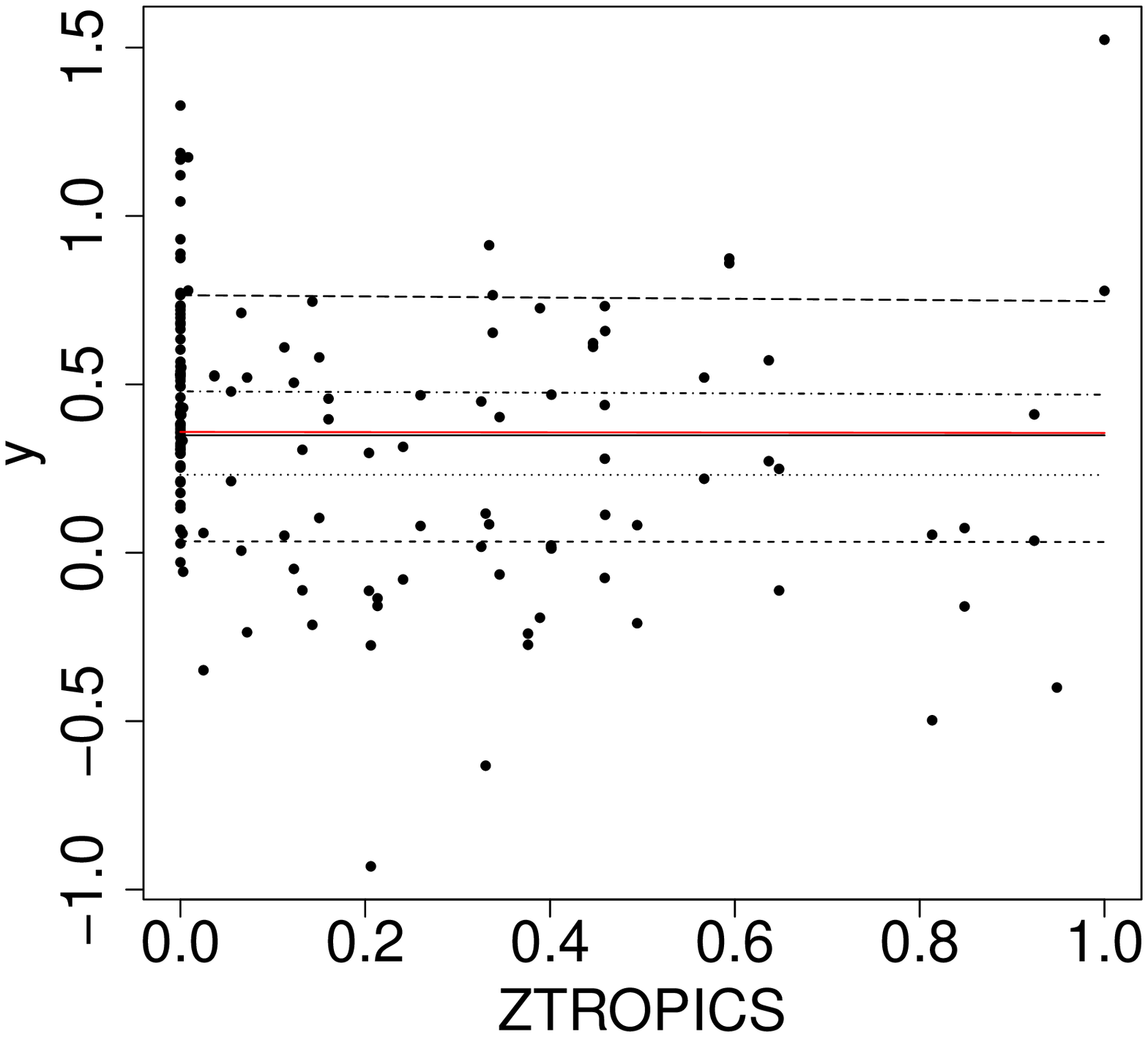}}
}
\vspace{-0.2in}
\centerline{
\subfigure{\includegraphics[width=1.5in,trim=0 30 0 0]{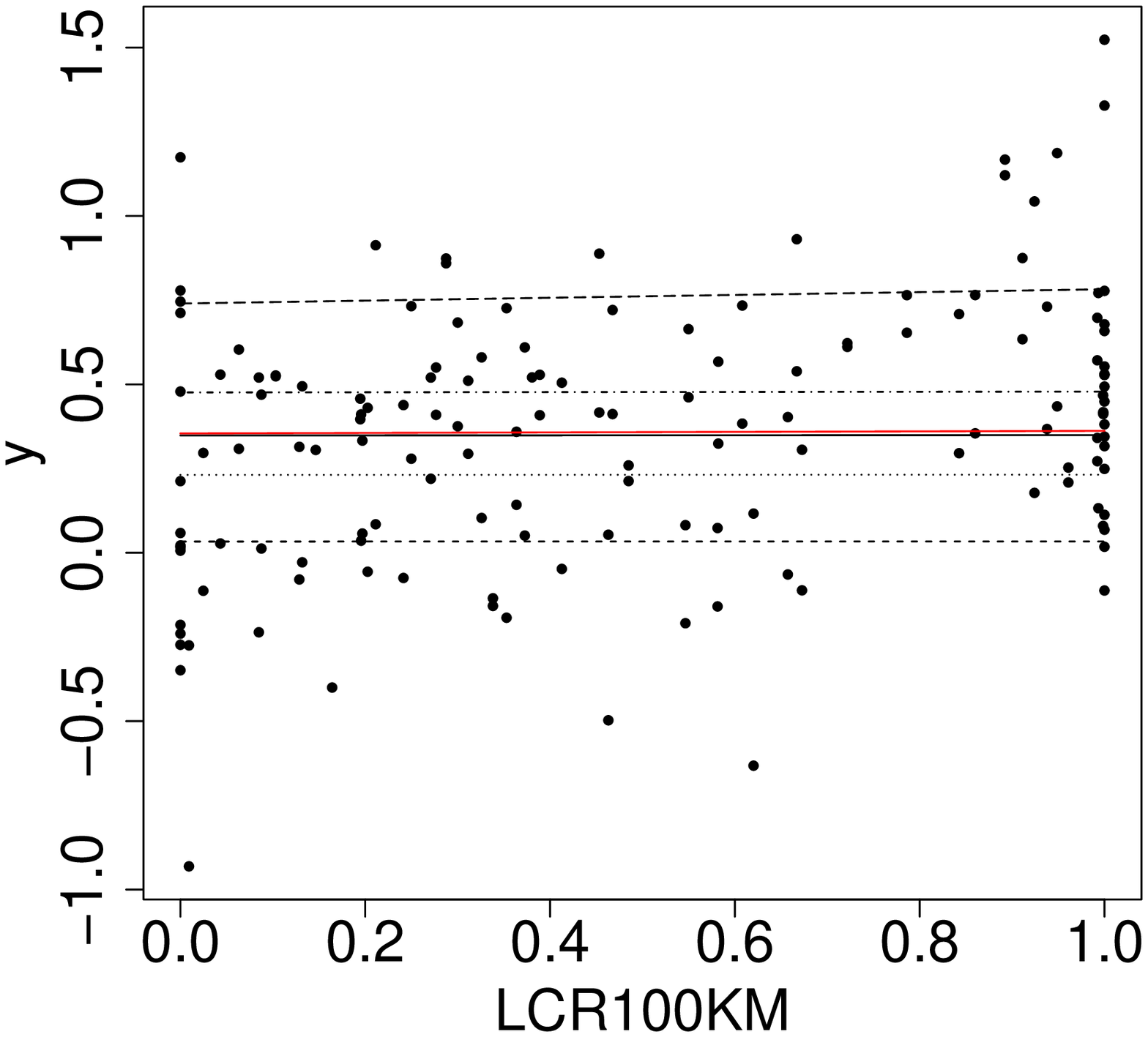}}
\hfil
\subfigure{\includegraphics[width=1.5in,trim=0 30 0 0]{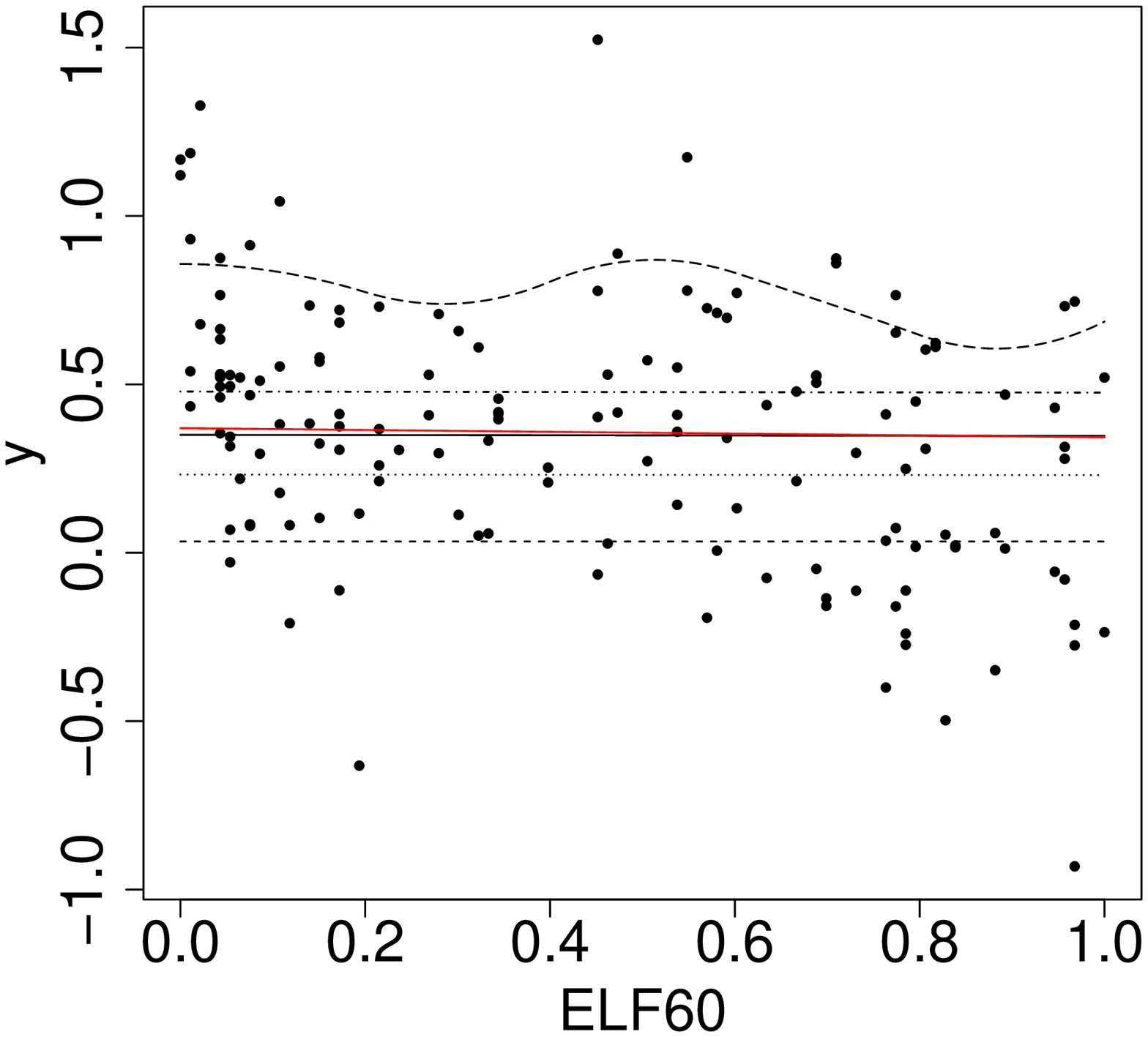}}
\hfil
\subfigure{\includegraphics[width=1.5in,trim=0 30 0 0]{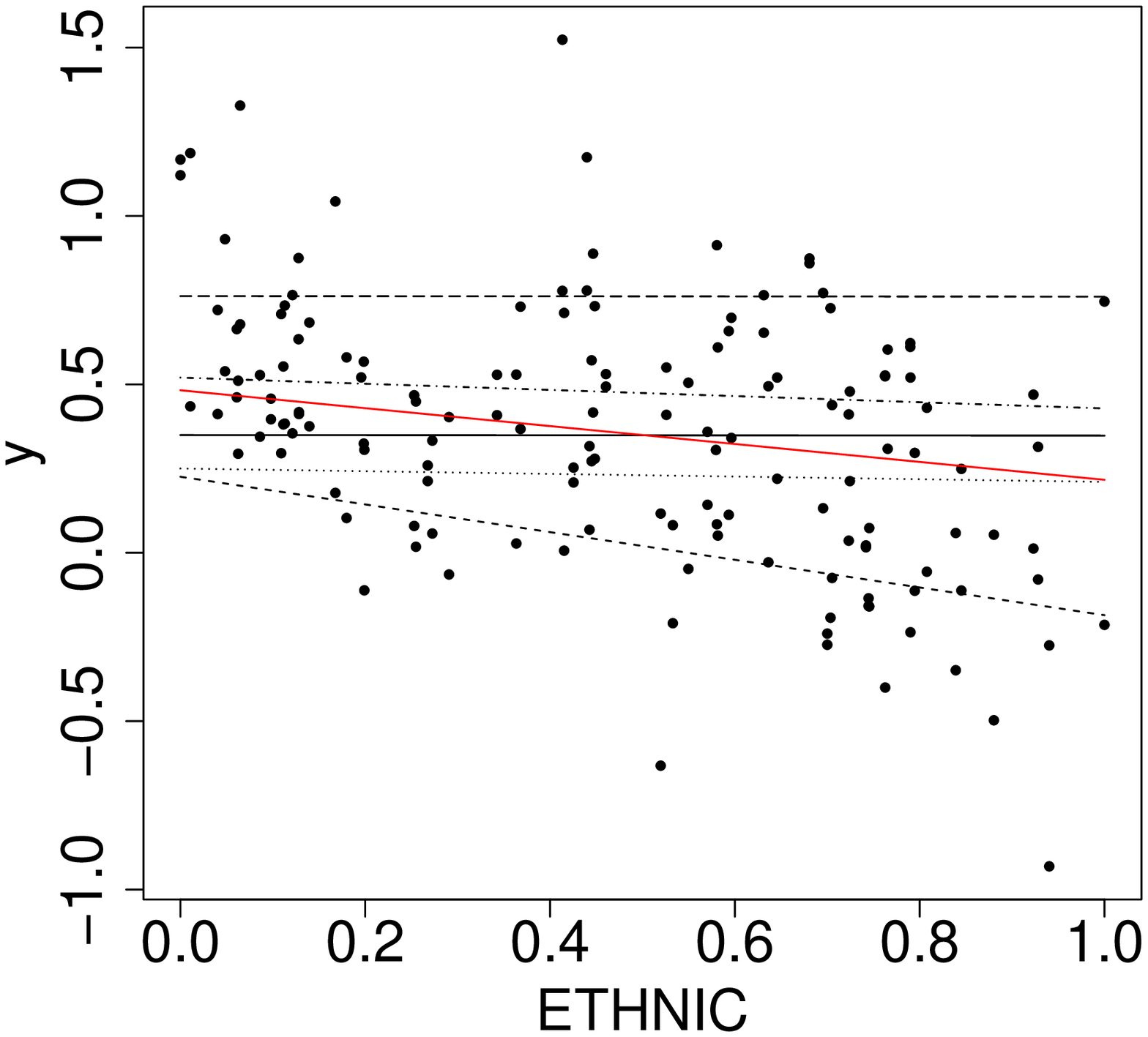}}
}
\vspace{-0.2in}
\centerline{
\subfigure{\includegraphics[width=1.5in,trim=0 30 0 0]{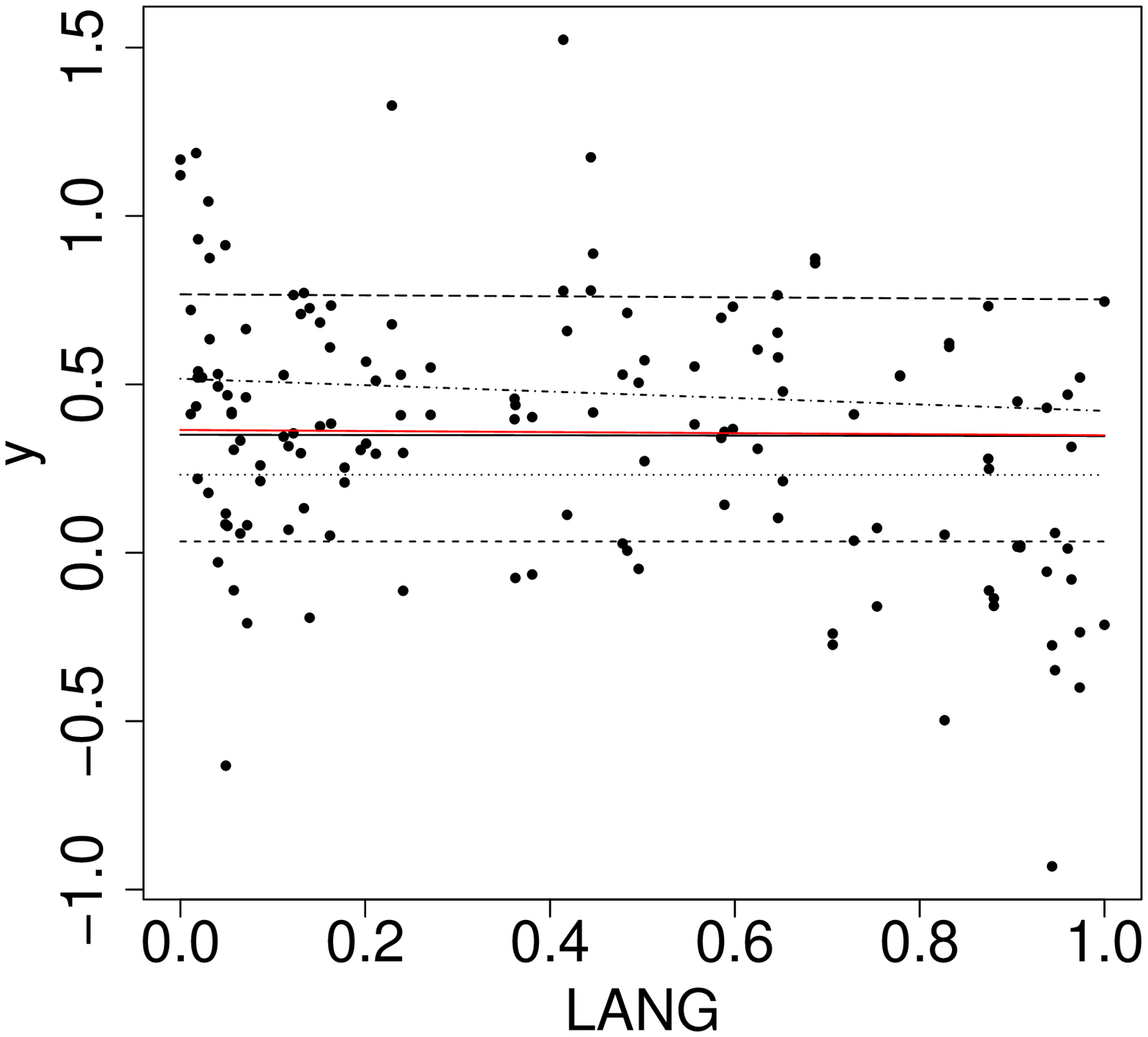}}
\hfil
\subfigure{\includegraphics[width=1.5in,trim=0 30 0 0]{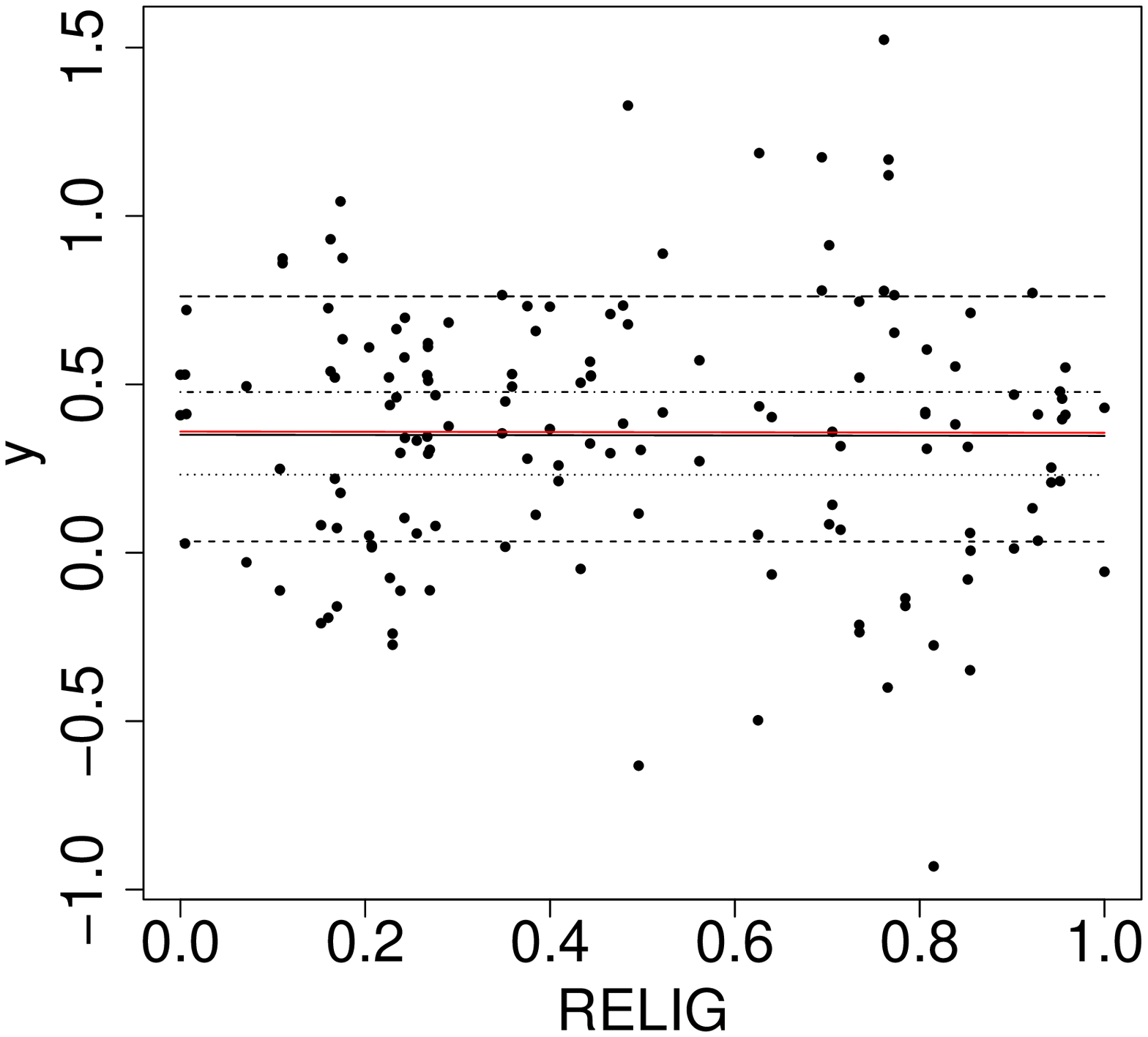}}
\hfil
\subfigure{\includegraphics[width=1.5in,trim=0 30 0 0]{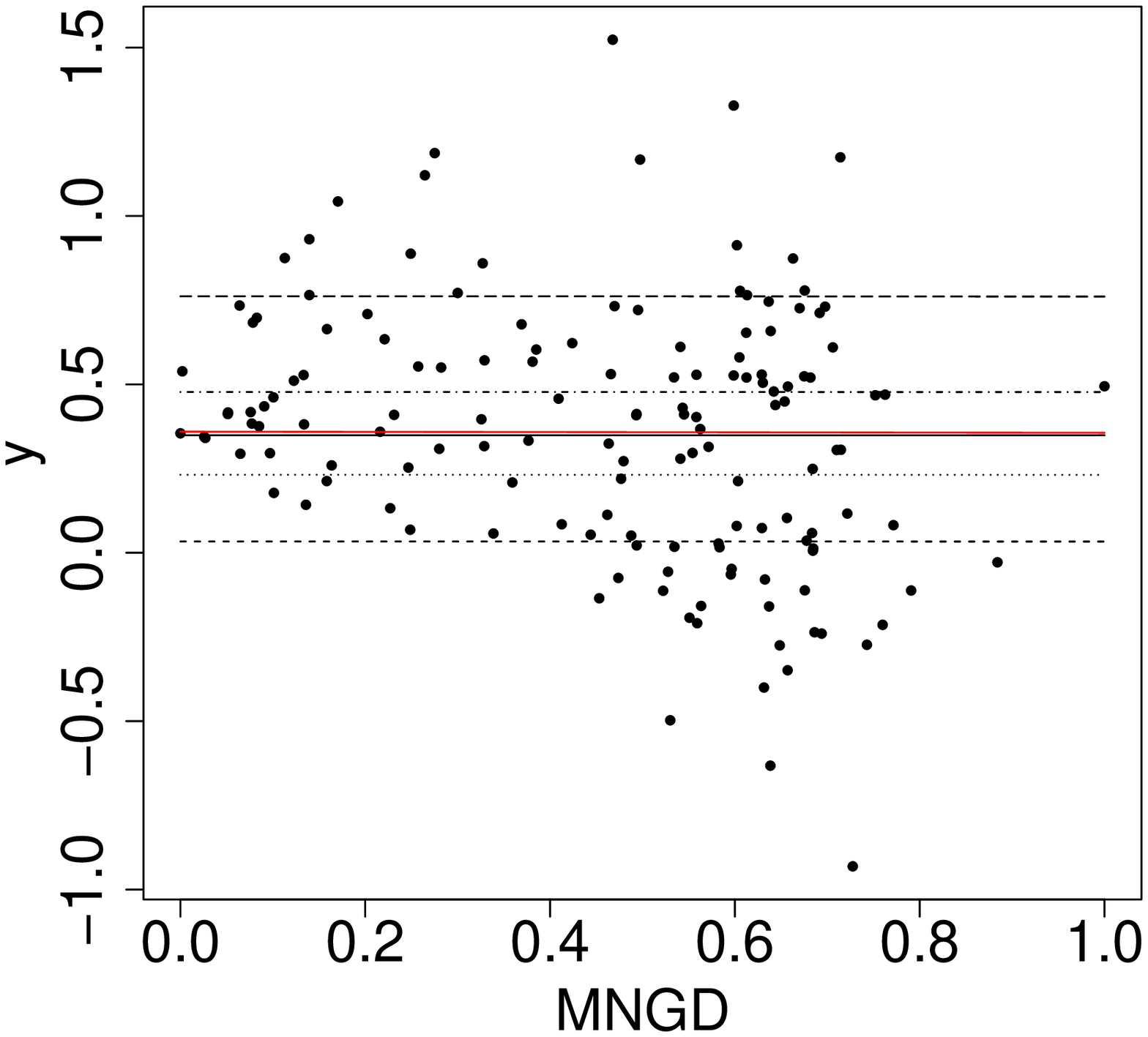}}
}
\vspace{-0.2in}
\centerline{
\subfigure{\includegraphics[width=1.5in,trim=0 30 0 0]{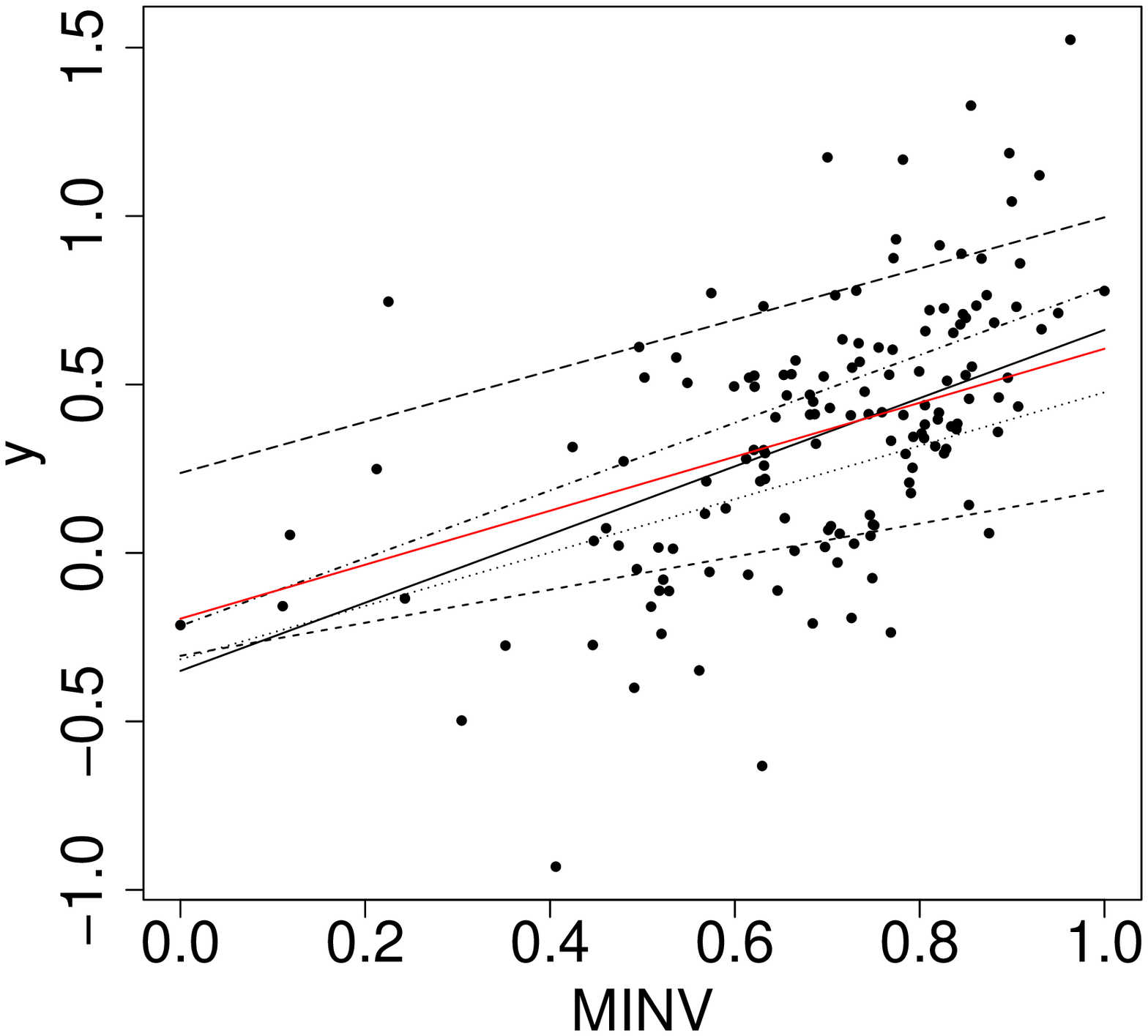}}
\hfil
\subfigure{\includegraphics[width=1.5in,trim=0 30 0 0]{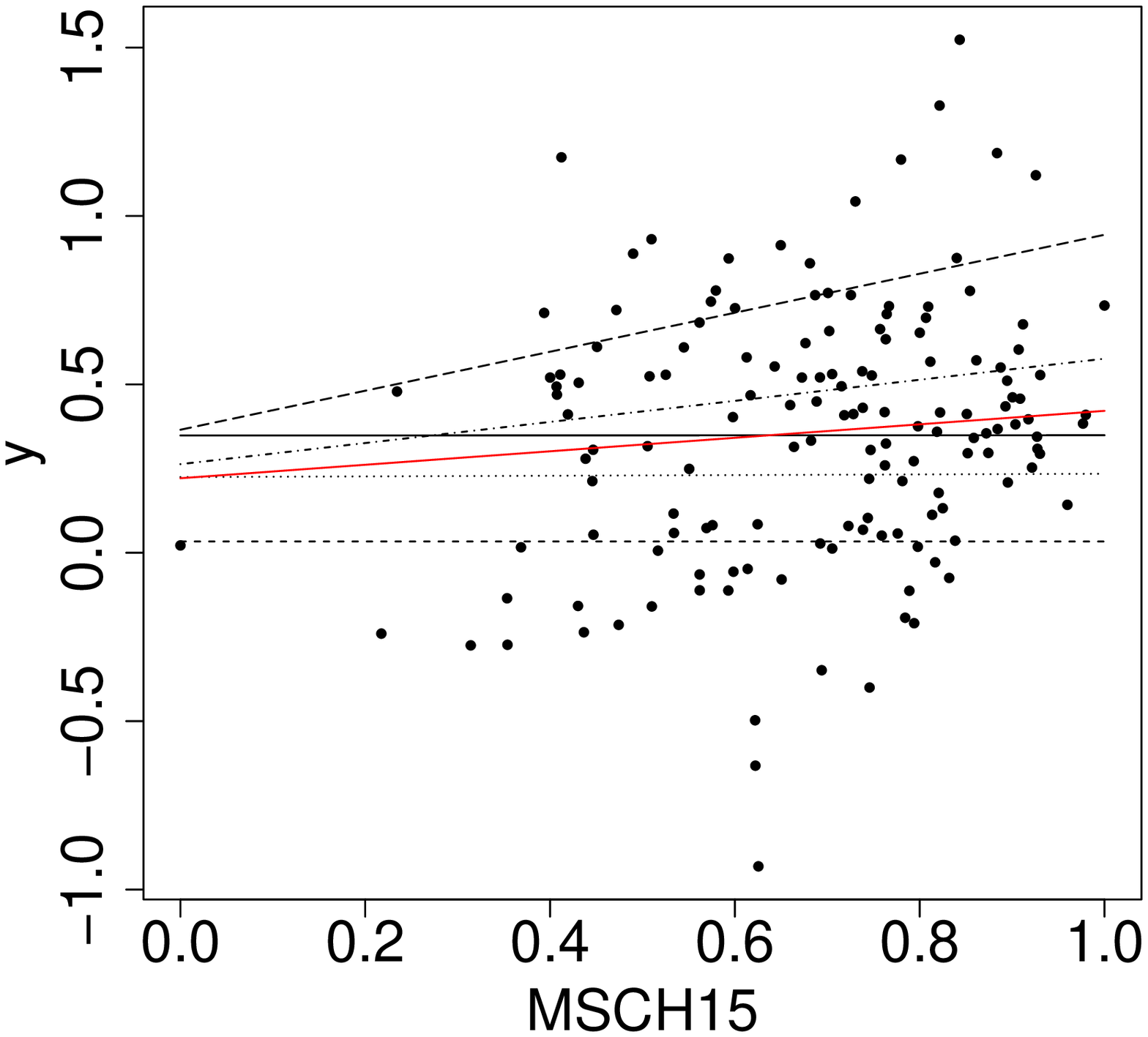}}
}
\vspace{-0.2in}
\centerline{
\subfigure{\includegraphics[width=1.5in,trim=0 30 0 0]{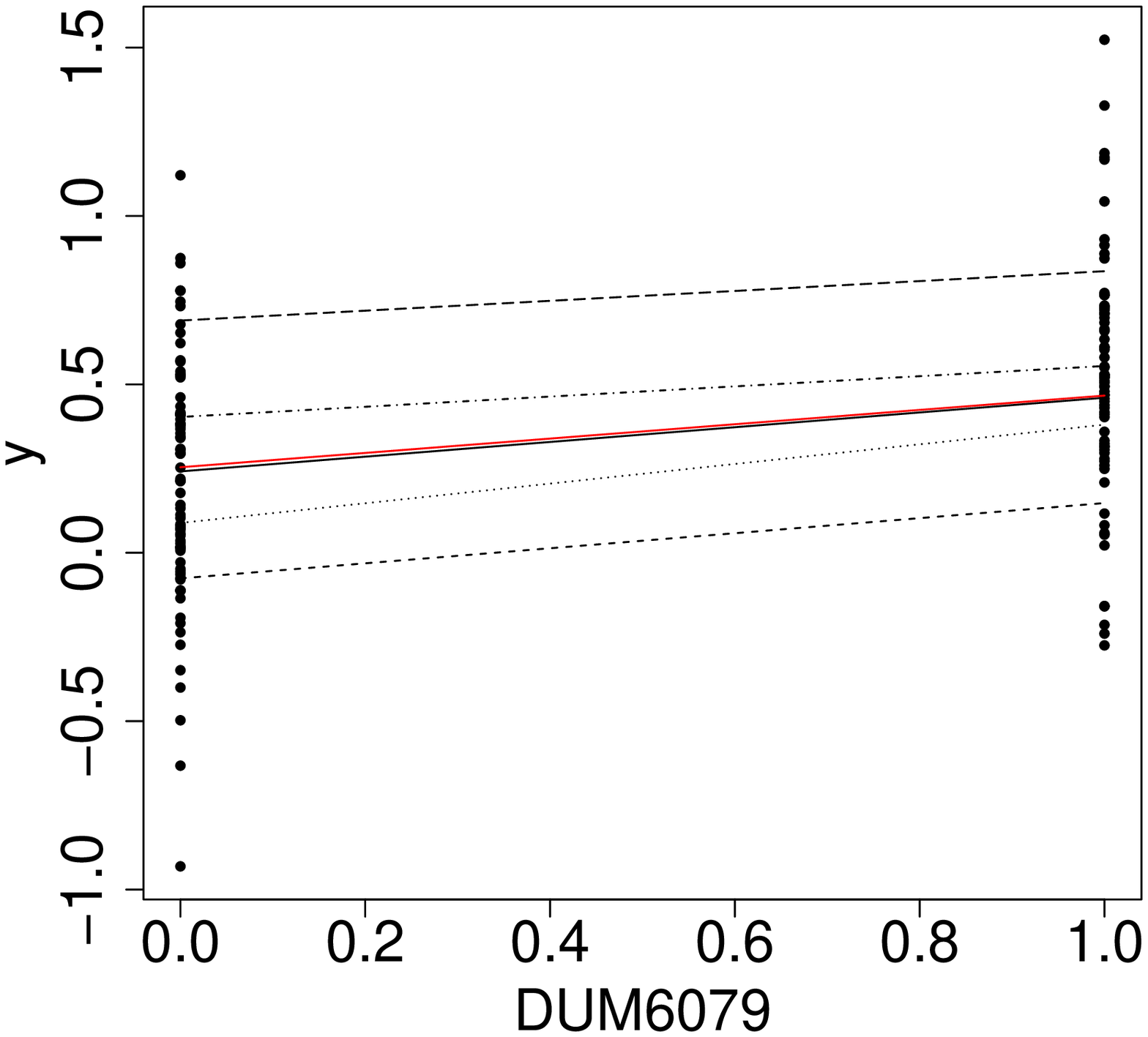}}
\hfil
\subfigure{\includegraphics[width=1.5in,trim=0 30 0 0]{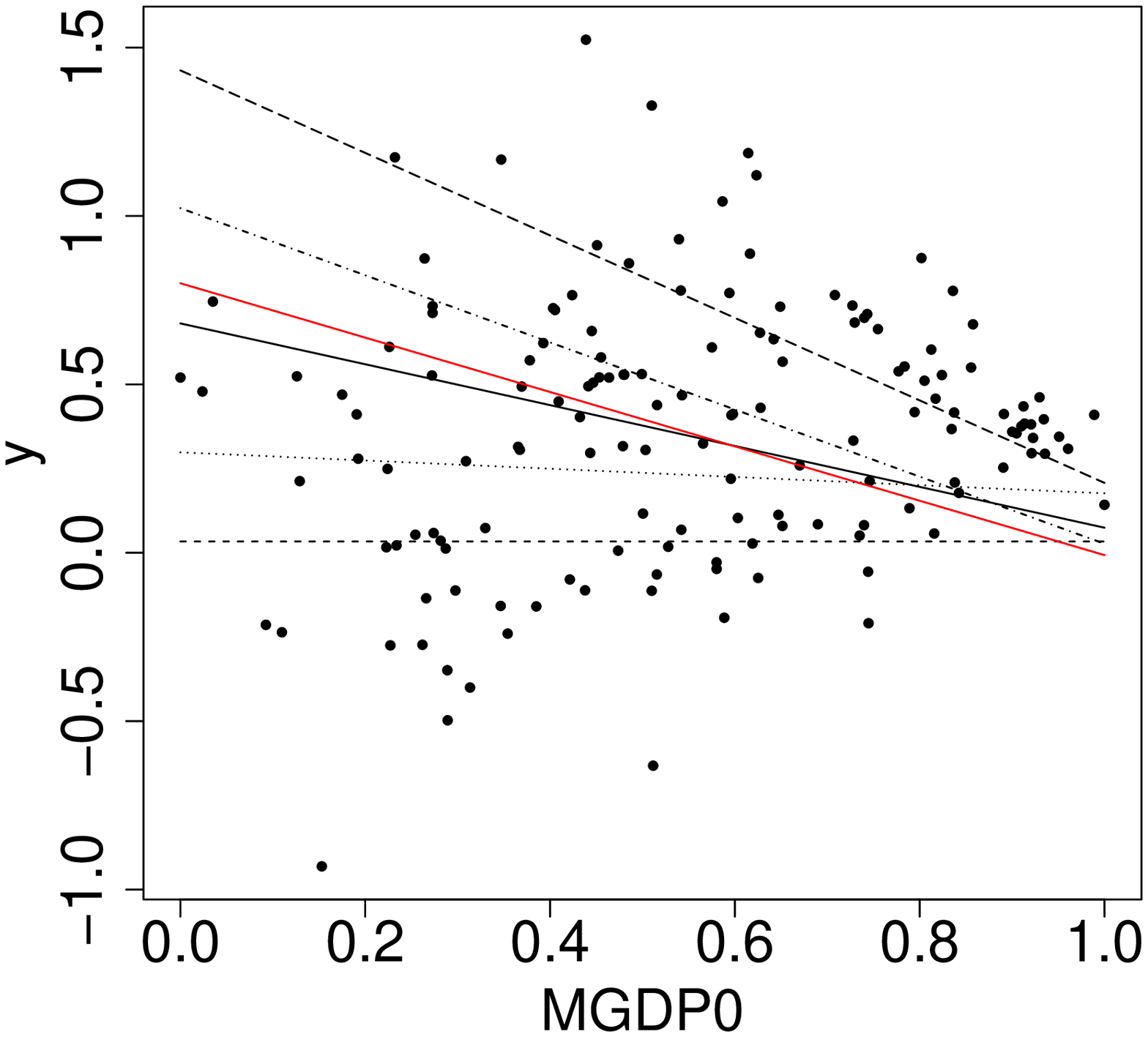}}
}
\caption{The fitted regression functions for the GDP growth data when one covariate varies and others are fixed at 0.5, at quantile levels $\tau=0.1, 0.3, 0.5, 0.7, 0.9$. The red solid lines are the fitted curves of mean regression.}
\label{factor_plot_GDP}
\end{figure}

\begin{figure}[htp]
\centerline{
\subfigure{\includegraphics[width=1.5in,trim=0 30 0 0]{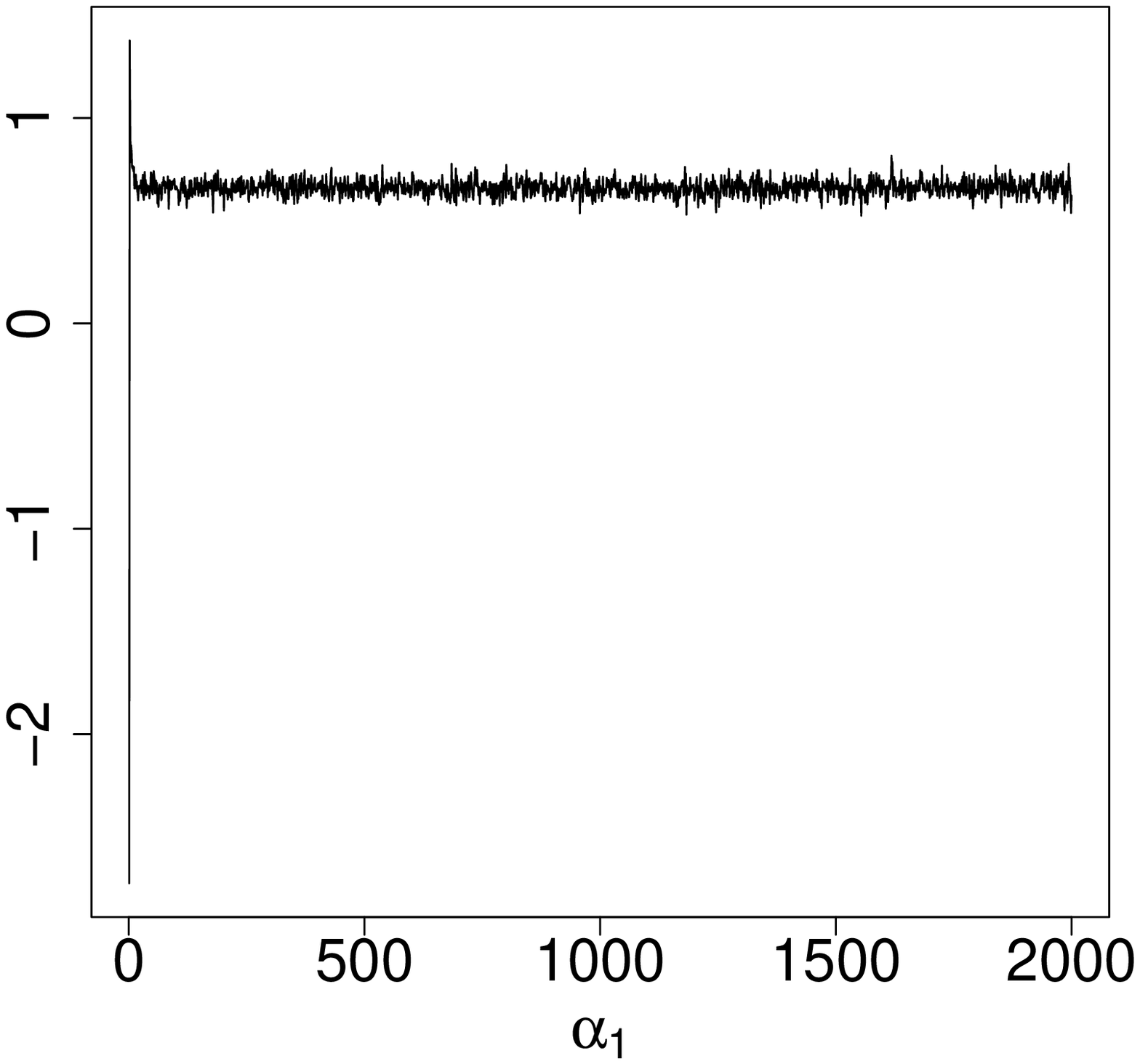}}
\hfil
\subfigure{\includegraphics[width=1.5in,trim=0 30 0 0]{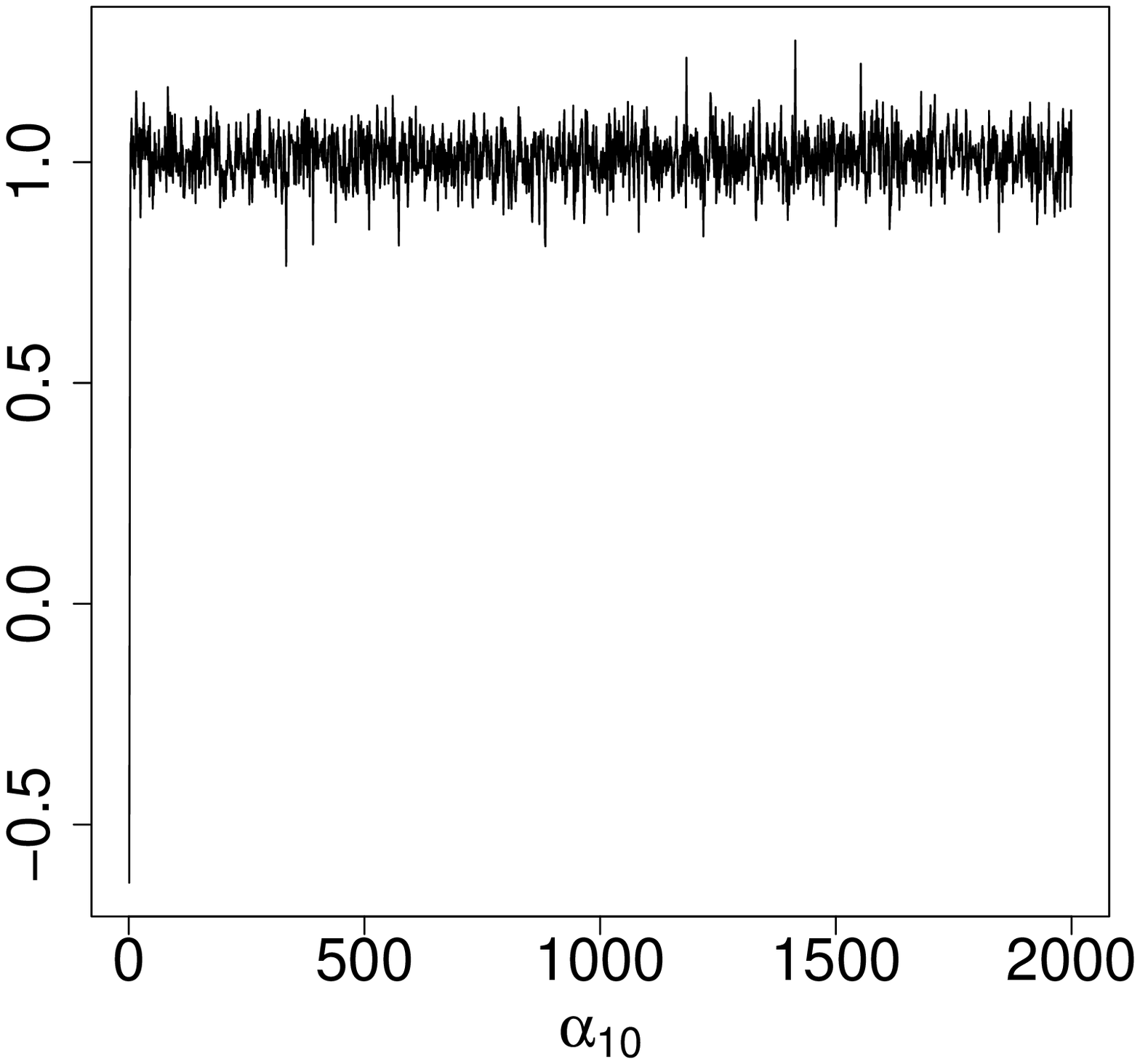}}
\hfil
\subfigure{\includegraphics[width=1.5in,trim=0 30 0 0]{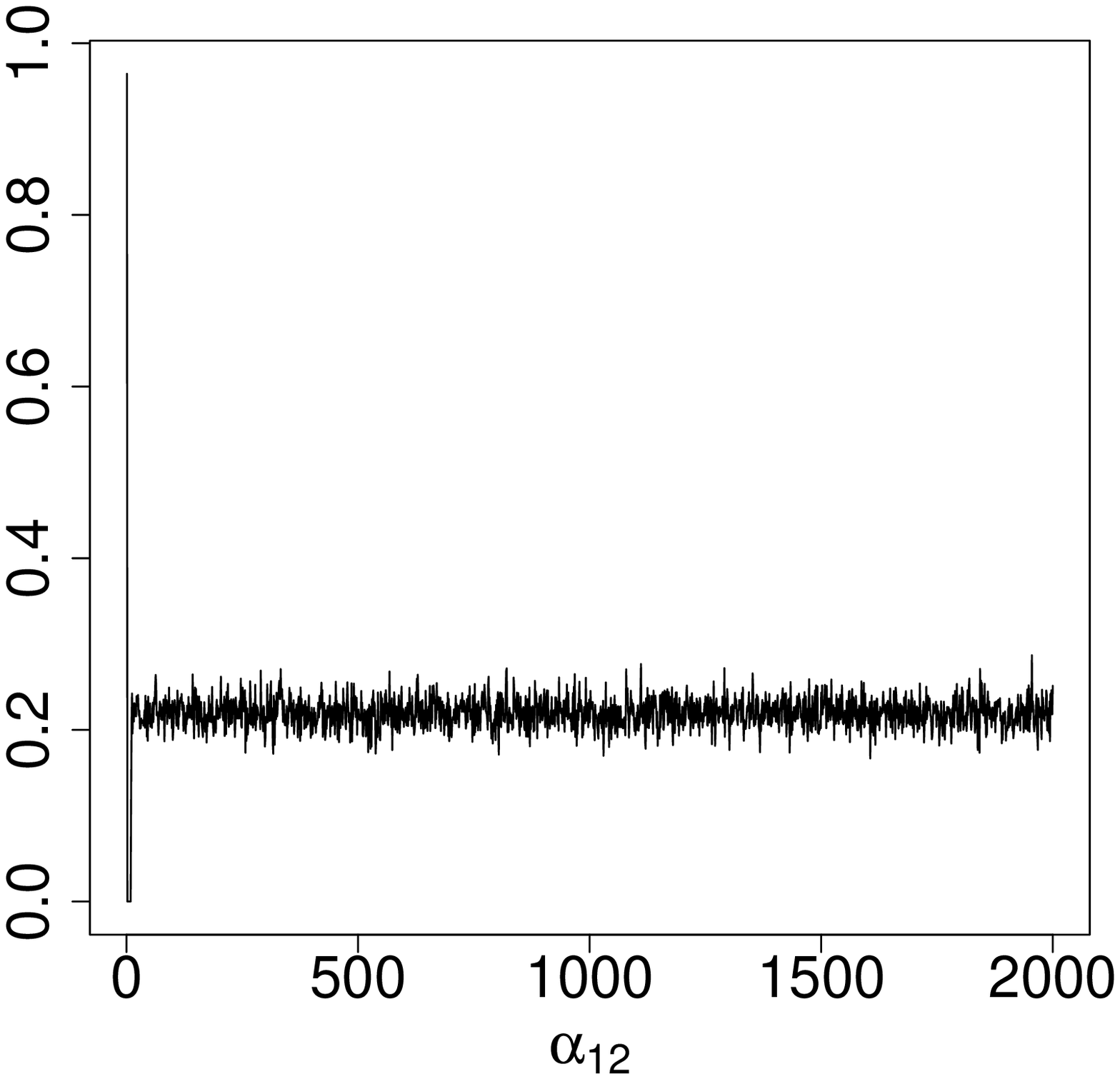}}
}
\centerline{
\subfigure{\includegraphics[width=1.5in,trim=0 30 0 0]{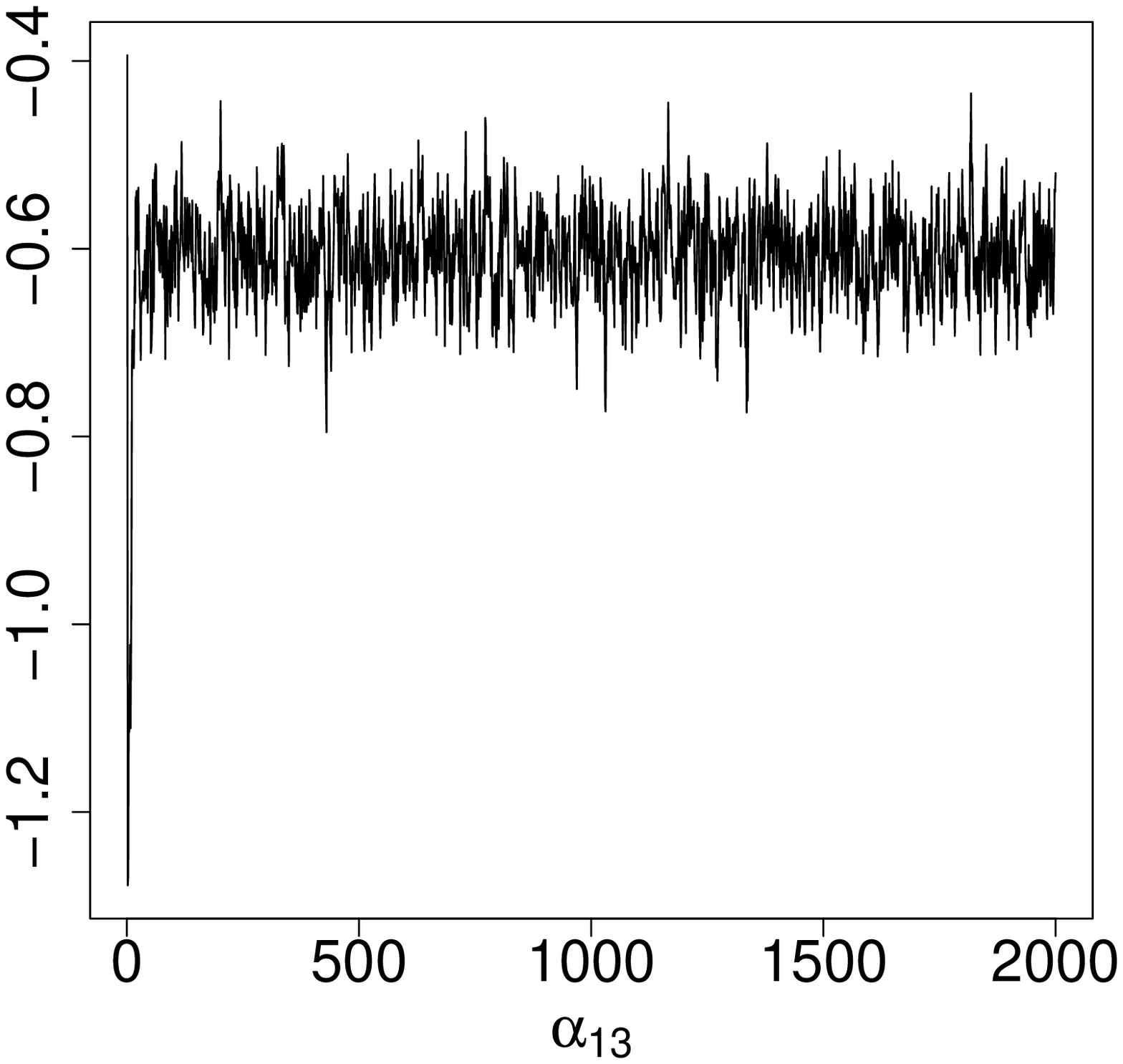}}
\hfil
\subfigure{\includegraphics[width=1.5in,trim=0 30 0 0]{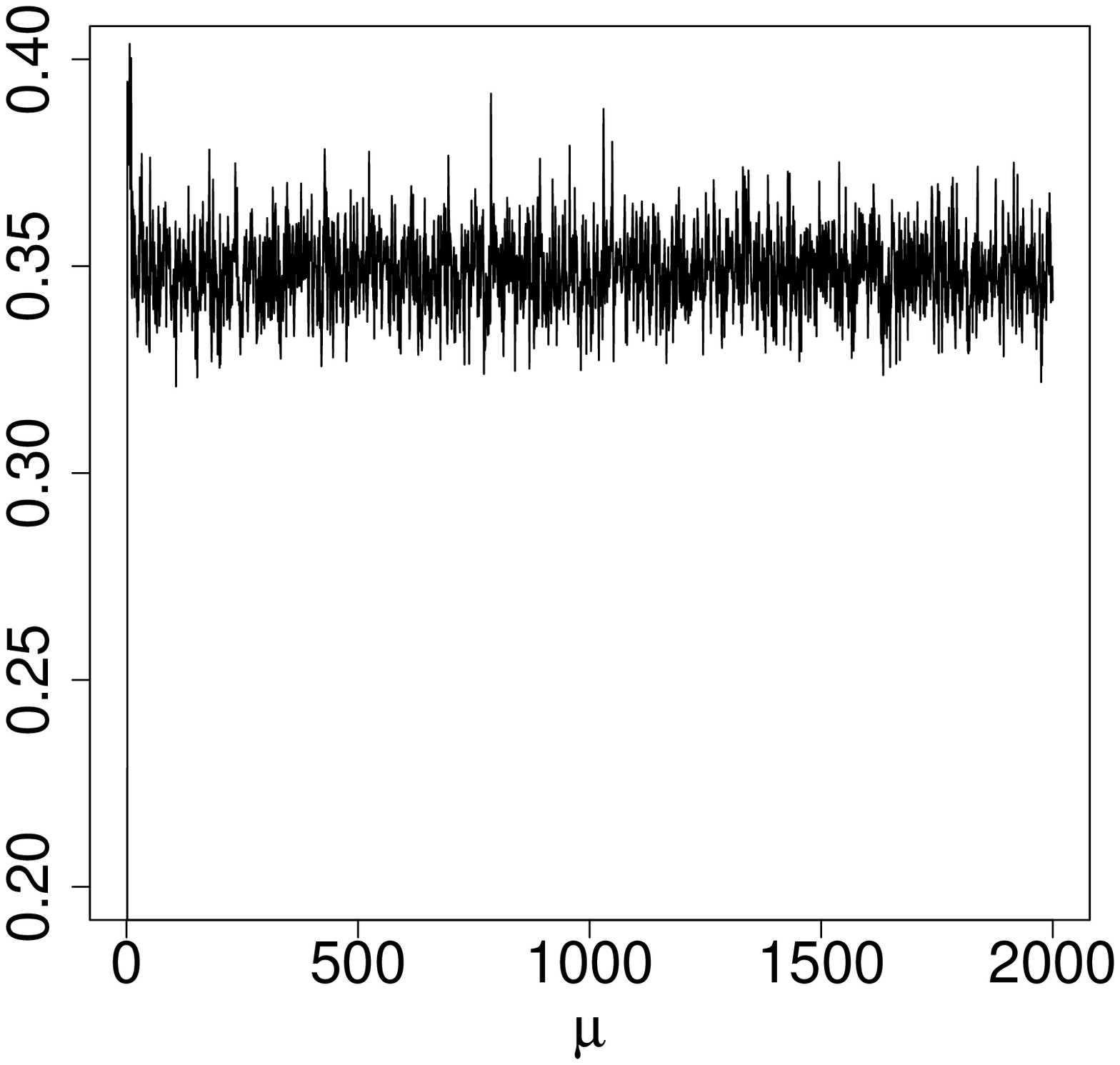}}
\hfil
\subfigure{\includegraphics[width=1.5in,trim=0 30 0 0]{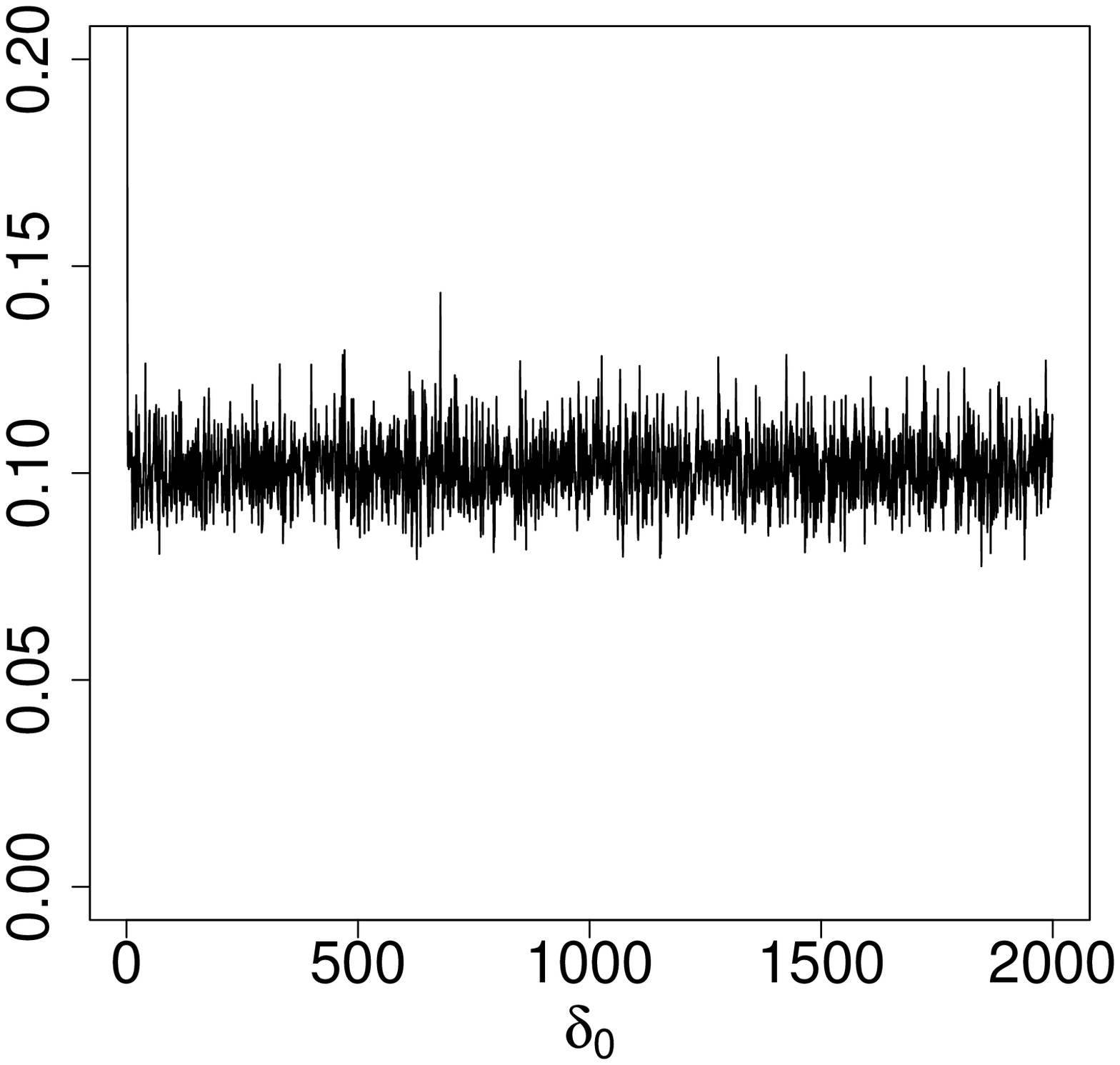}}
}
\caption{Trace plots of selected $\alpha$, $\delta_{0}$, and $\mu$  of median regression for GDP growth data. }
\label{trace_plot}
\end{figure}

In Figure \ref{indicator_posterior_GDP}, we show the variable selection results of BPLAM and  BQPLAM. For mean regression, we identified 6 nonzero components, all of which are linear components. For quantile regression, we identified 4, 3, 4, 5 and 6 nonzero components, of which 1, 0, 0, 1, 2  components are nonlinear at 5 quantile levels respectively. The barplots show that some covariates only have effects at lower or upper quantiles. For example, MSCH15 and ELF60 only have effects at upper quantiles, and ETHNIC only affects lower quantiles. In Figure \ref{factor_plot_GDP}, we show the fitted curves of the estimated regression functions (posterior mean  based on the 10000 iterations after burn-in) as one covariate varies while others are fixed at 0.5. For convergence diagnosis, we present trace plots of $\alpha$ (we only present $\alpha$ of those covariates detected to be nonzero components), $\delta_{0}$, and $\mu$  for the median regression in Figure \ref{trace_plot} (we plot one sampled point  in every ten to reduce the image size).

 To summarize, besides those neoclassicial determinants, which are fundamental determinants used in the production function (Mankiw et al 1992), both institutions and ethnic fractionalization are found to have effects on economical development while geography is found to play no roles. These results are consistent with the results of \cite{tan2010no}, and lead to the same conclusion that more attention should be paid on the impact of quality of institutions and degree of ethnic fractionalization on the development of economics. 
\subsubsection{Housing price data}
Another data set we use is a housing price data. This data set is used to investigate the relationship between house price and several features, including physical attributes of the residential houses,  characteristics of the surrounding rural district and environmental nuisances. This application concerns transaction prices ($n=2053$) for the residential houses sold during $1996$ and $1997$ in Brittany. We consider four physical characteristics of a house and its adjacent lot: age of the house (AGE), state of repair (REPAIR)(dummy), number of rooms (ROOMS), and lot size (LOT).  We use another four indicators on the characteristics of the surrounding rural district: the population of the district (POP),  average taxable family income (AVINC),  proportion of vacant houses (VACANT), and a dummy variable indicating whether or not the surrounding district is located in the department of Ille et Vilaine (COUNTRY). The environmental nuisances are expressed by two indicators, amount of nitrogen emissions from livestock farming per hectare of arable land in the rural district where the residential house is located (NITRO), and proportion of permanent grassland converted into cultivated grassland (TMEAD). This data set was used by \cite{bontemps2008semiparametric} to investigate the effects of agricultural pollution on house prices.  \cite{bontemps2008semiparametric} pre-specifes all the explanatory variables (AGE, REPAIR, ROOMS, LOT, COUNTY, VACANT, POP and AVINC) except the two environment indicators (TMEAD and NITRO) to be linear components in the semiparametric models, and incorporated the two environmental indicators in a nonparametric way.

\begin{figure}[htp]
\centerline{
\subfigure[Mean]{\includegraphics[width=1.8in,trim=0 30 0 0]{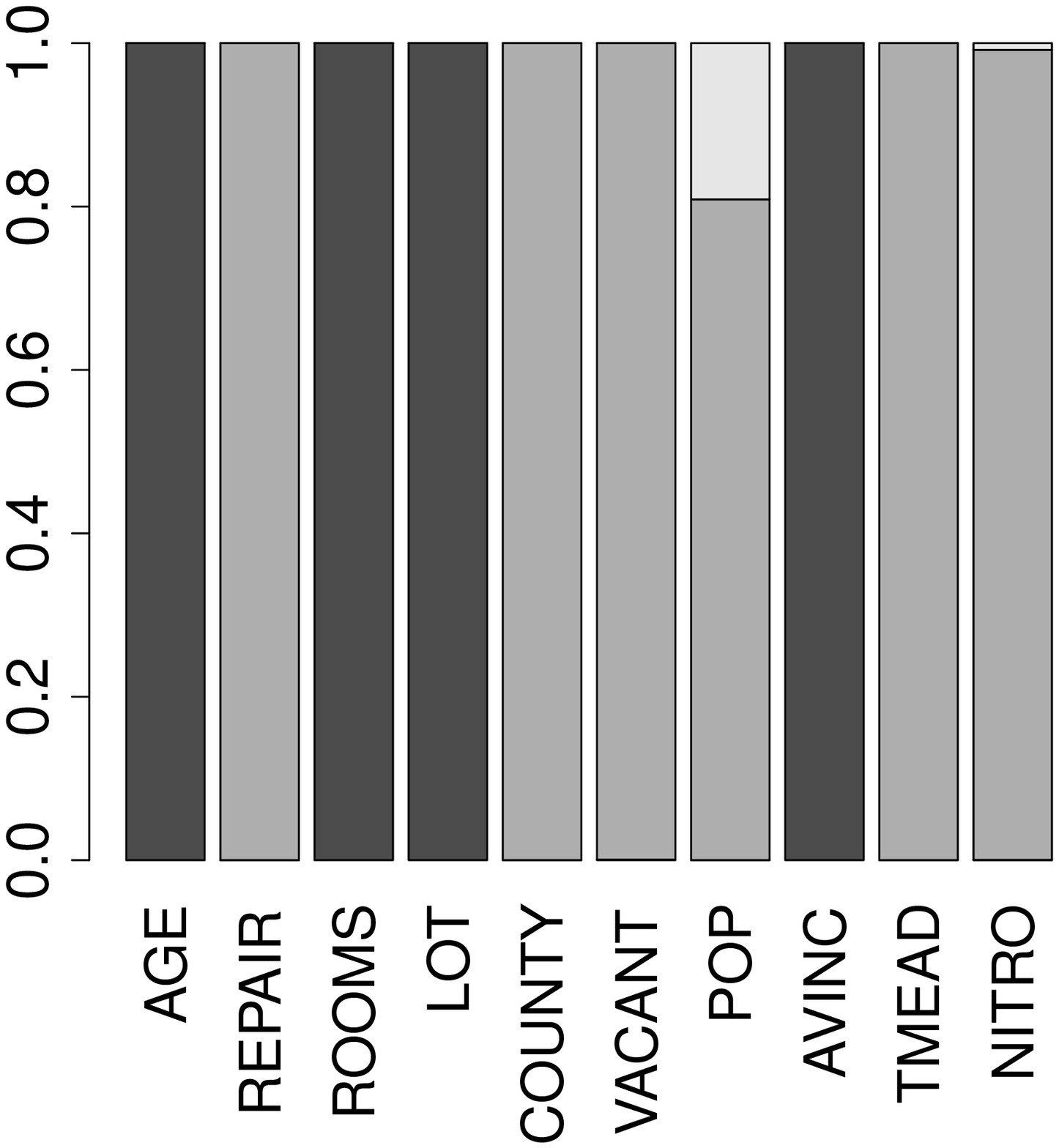}}
\subfigure[$\tau=0.1$]{\includegraphics[width=1.8in,trim=0 30 0 0]{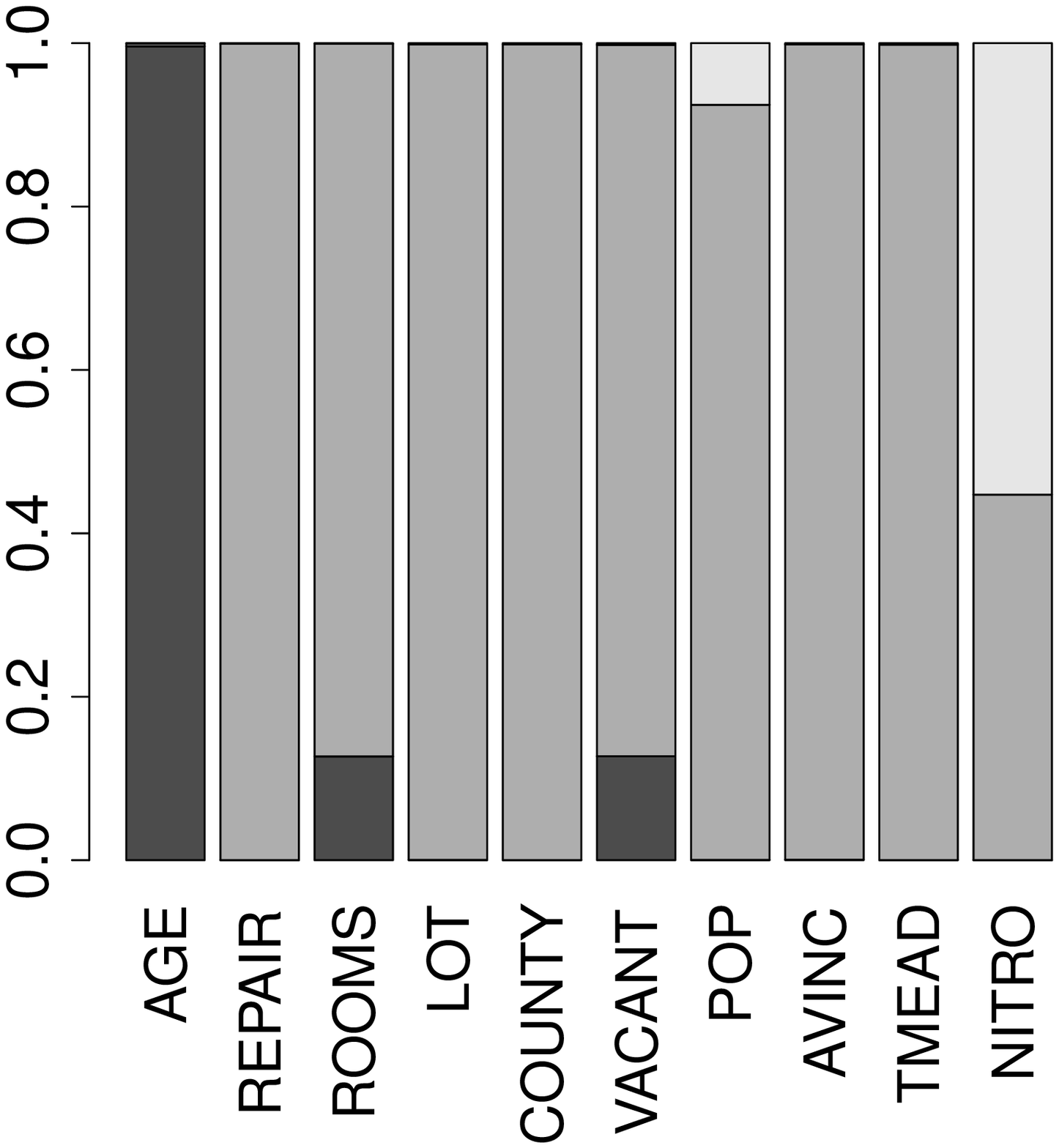}}
\subfigure[$\tau=0.3$]{\includegraphics[width=1.8in,trim=0 30 0 0]{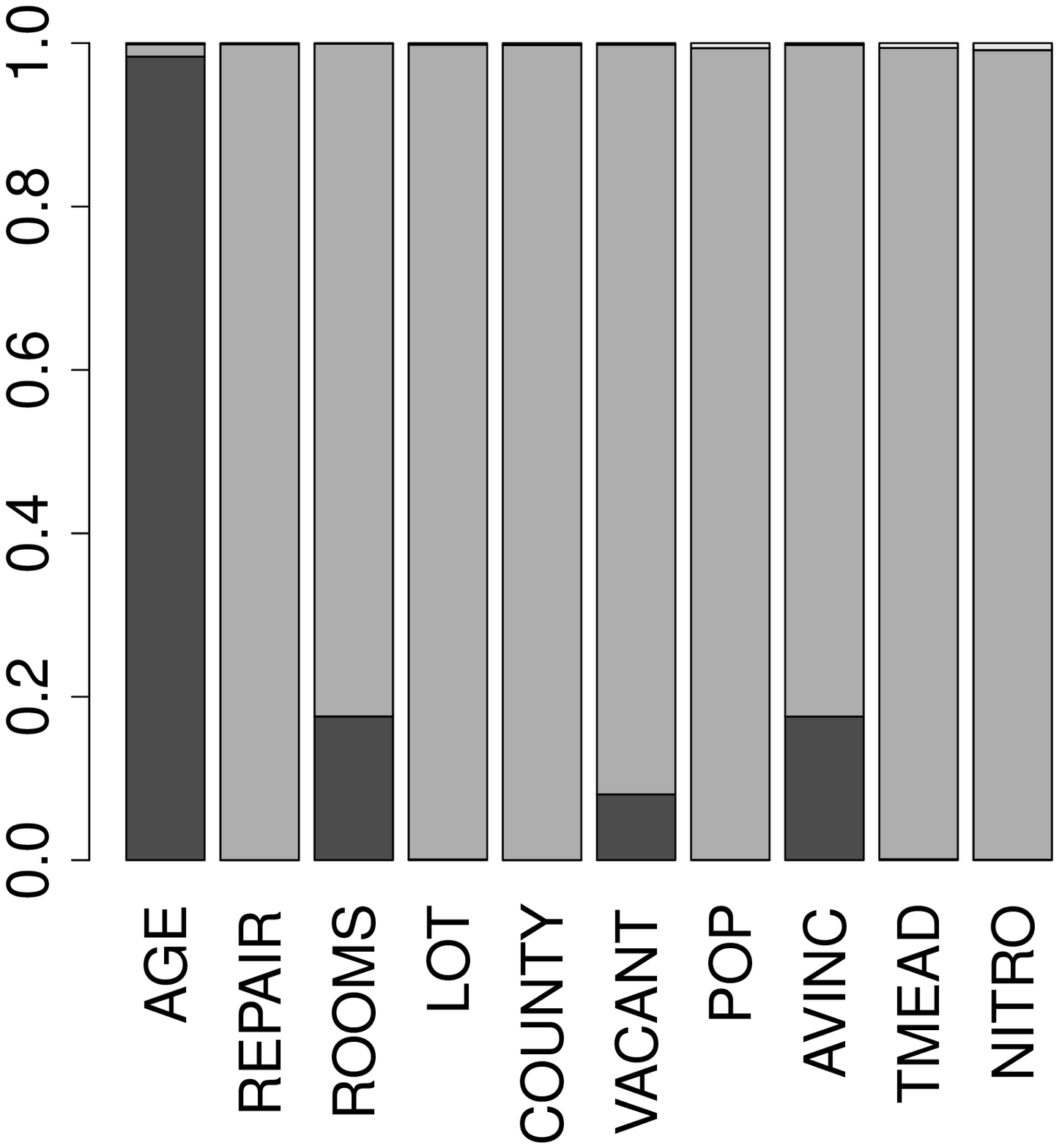}}
}
\centerline{
\subfigure[$\tau=0.5$]{\includegraphics[width=1.8in,trim=0 30 0 0]{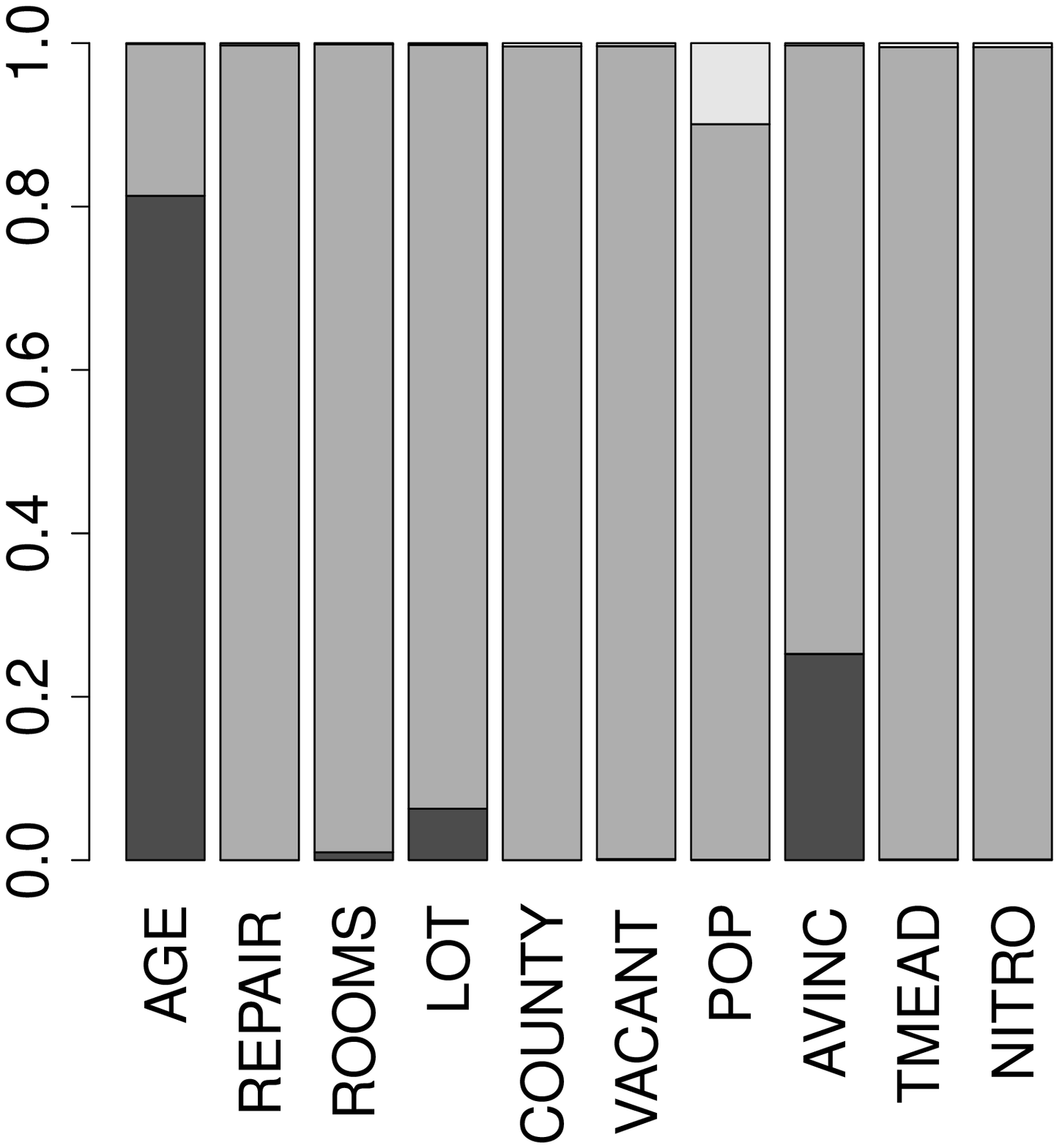}}
\subfigure[$\tau=0.7$]{\includegraphics[width=1.8in,trim=0 30 0 0]{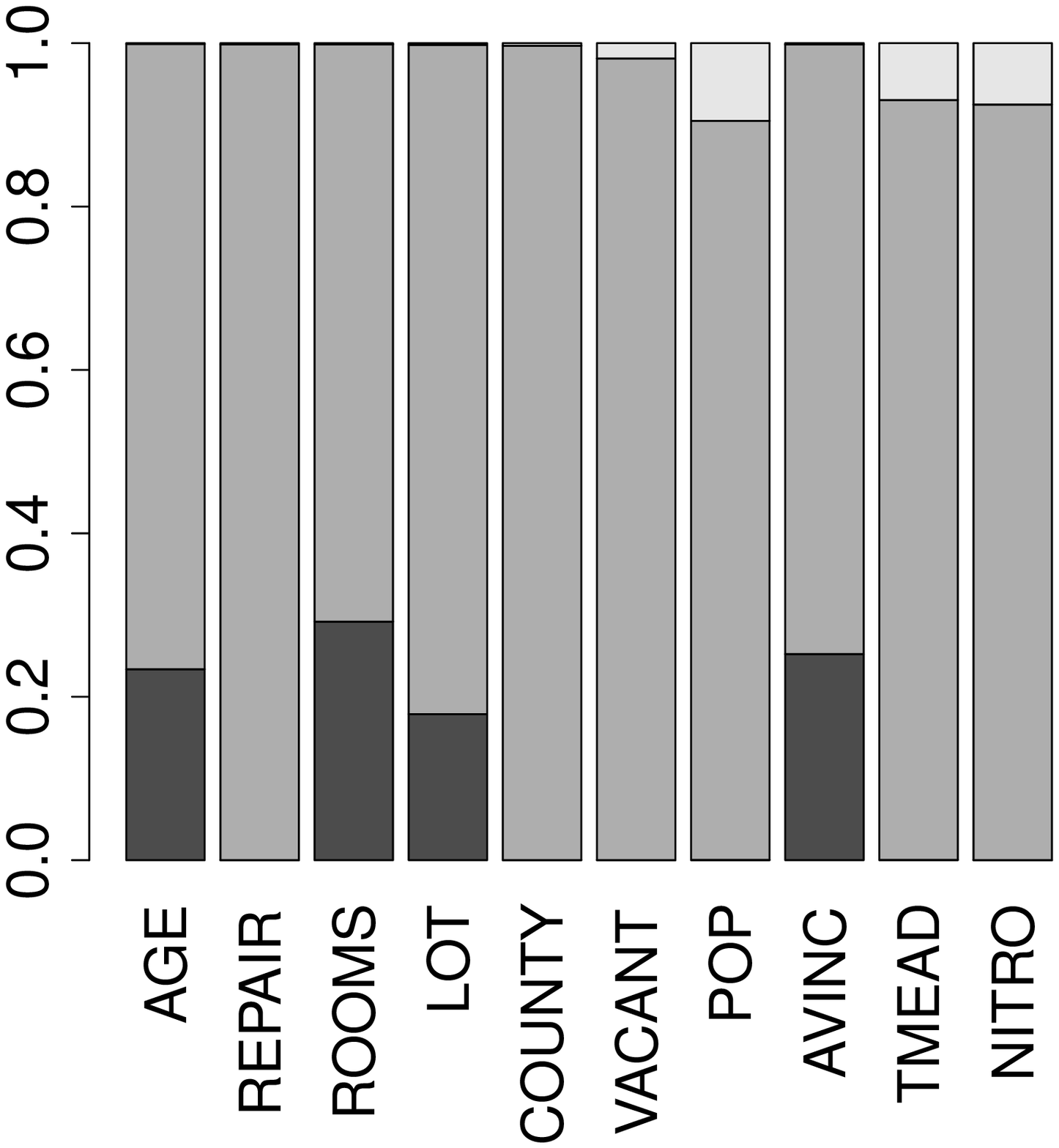}}
\subfigure[$\tau=0.9$]{\includegraphics[width=1.8in,trim=0 30 0 0]{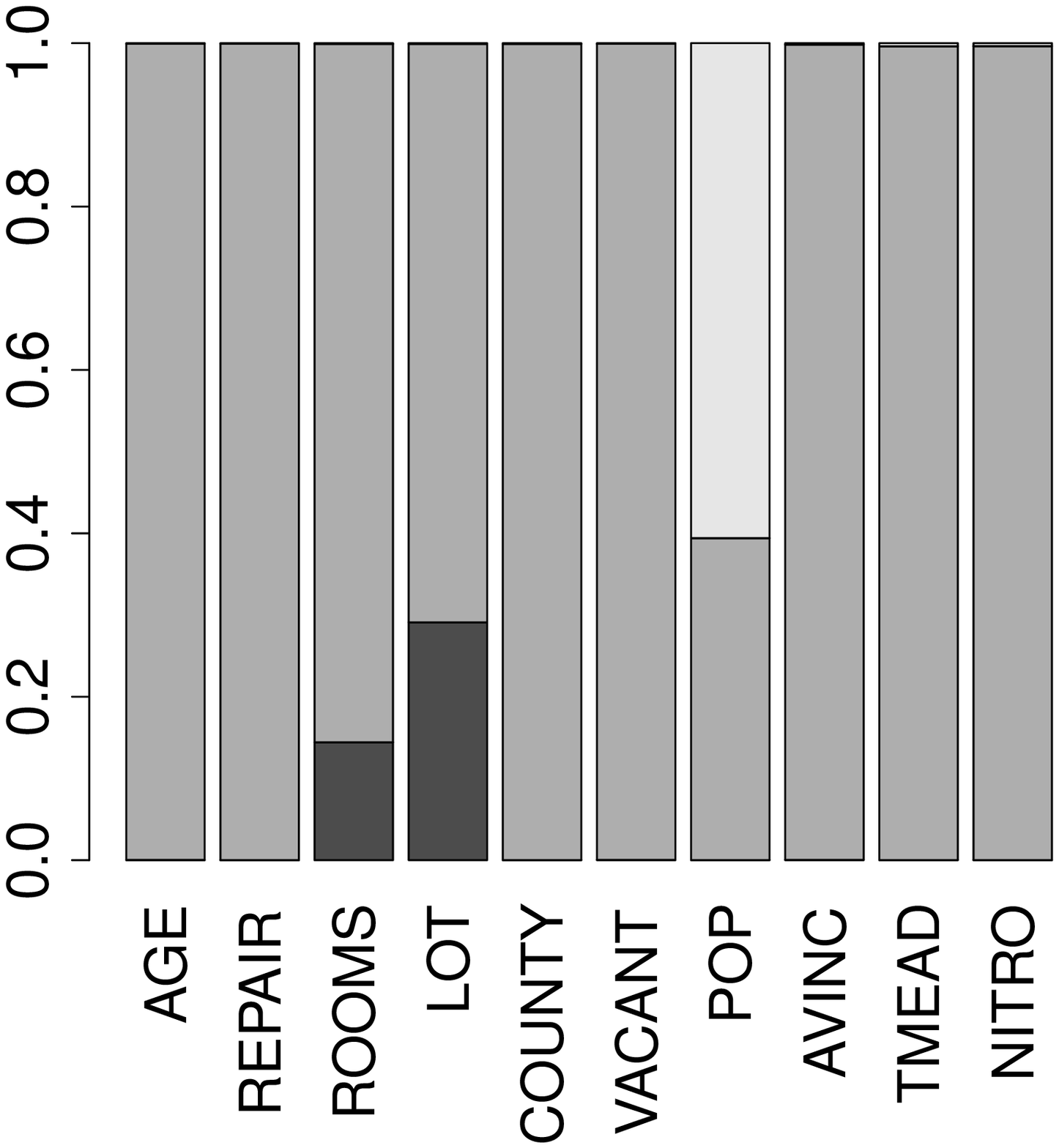}}
}
\caption{Component selection results for the housing price data. }
\label{indicator_posterior_house}
\end{figure}

\begin{figure}[htp]
\centerline{
\subfigure{\includegraphics[width=1.8in,trim=0 30 0 0]{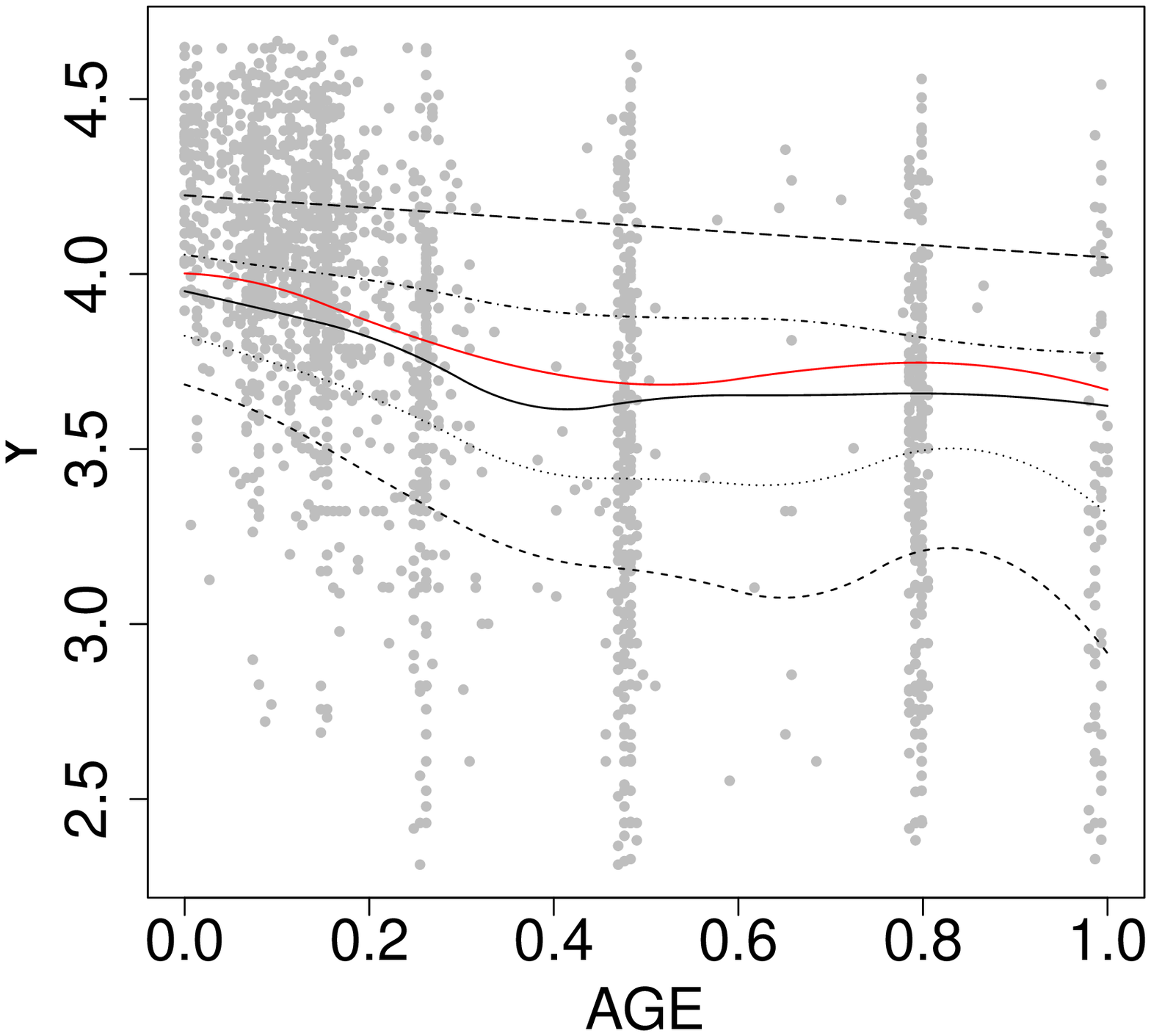}}
\hfil
\subfigure{\includegraphics[width=1.8in,trim=0 30 0 0]{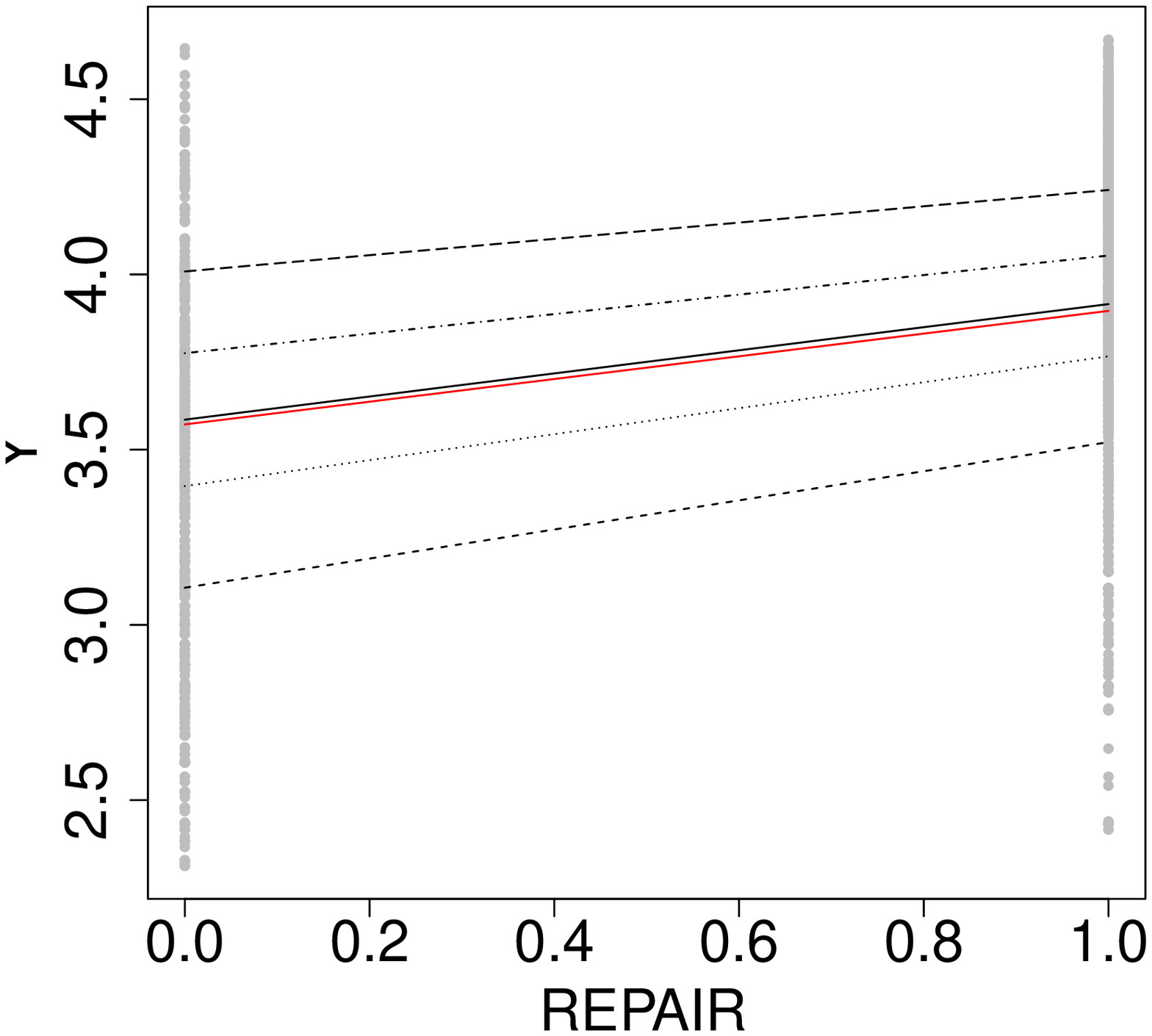}}
\hfil
\subfigure{\includegraphics[width=1.8in,trim=0 30 0 0]{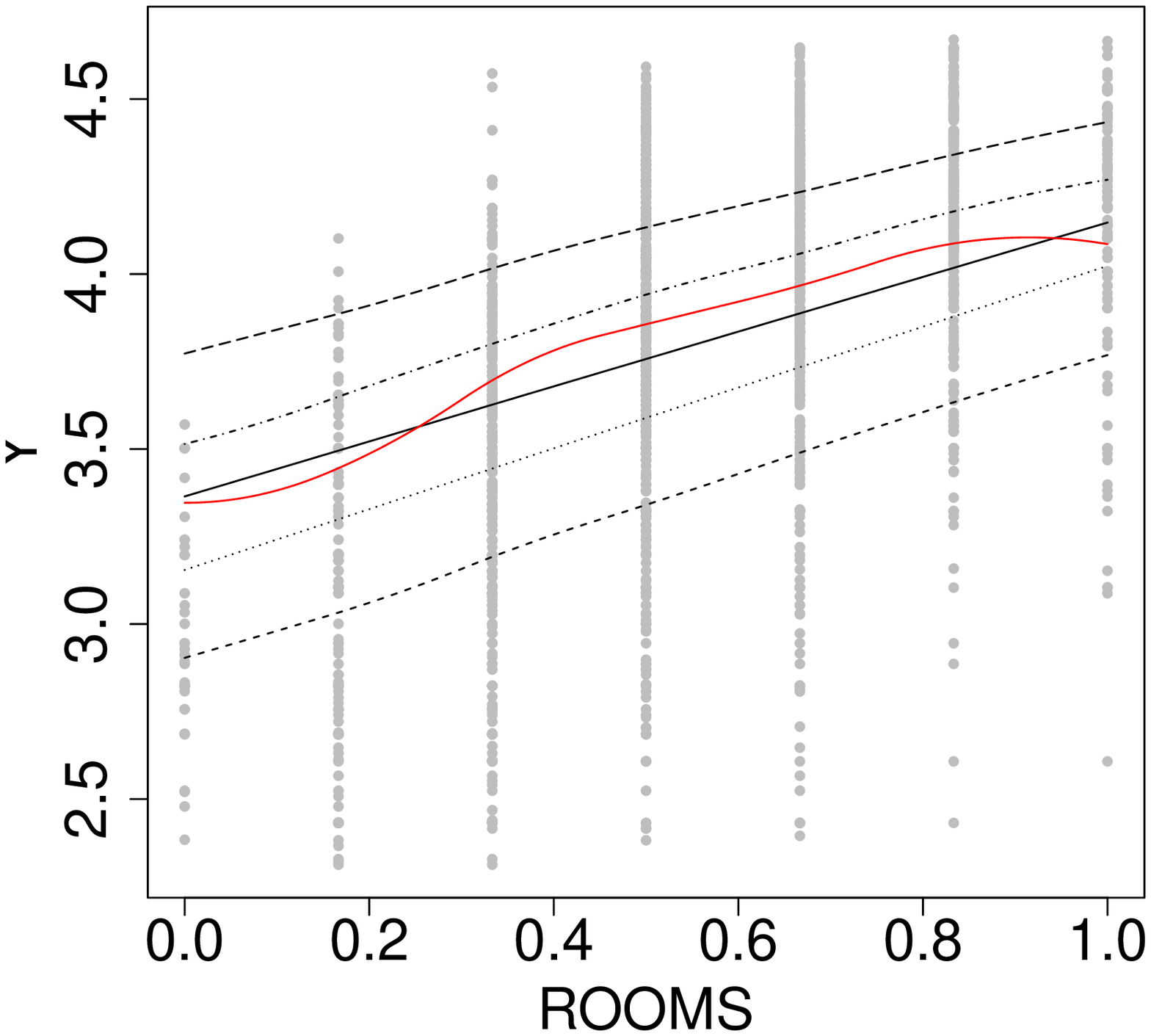}}
}
\vspace{-0.2in}
\centerline{
\subfigure{\includegraphics[width=1.8in,trim=0 30 0 0]{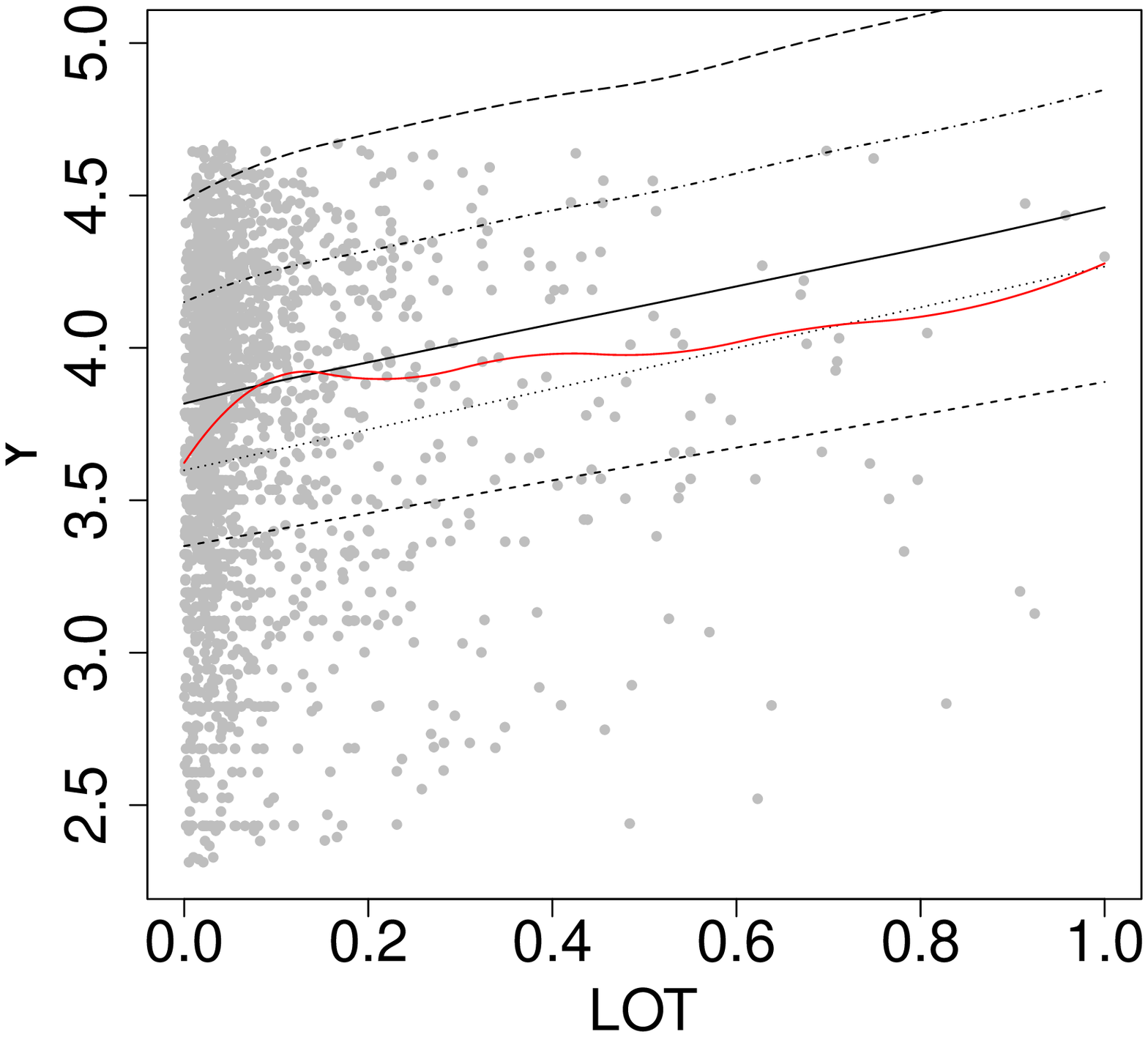}}
\hfil
\subfigure{\includegraphics[width=1.8in,trim=0 30 0 0]{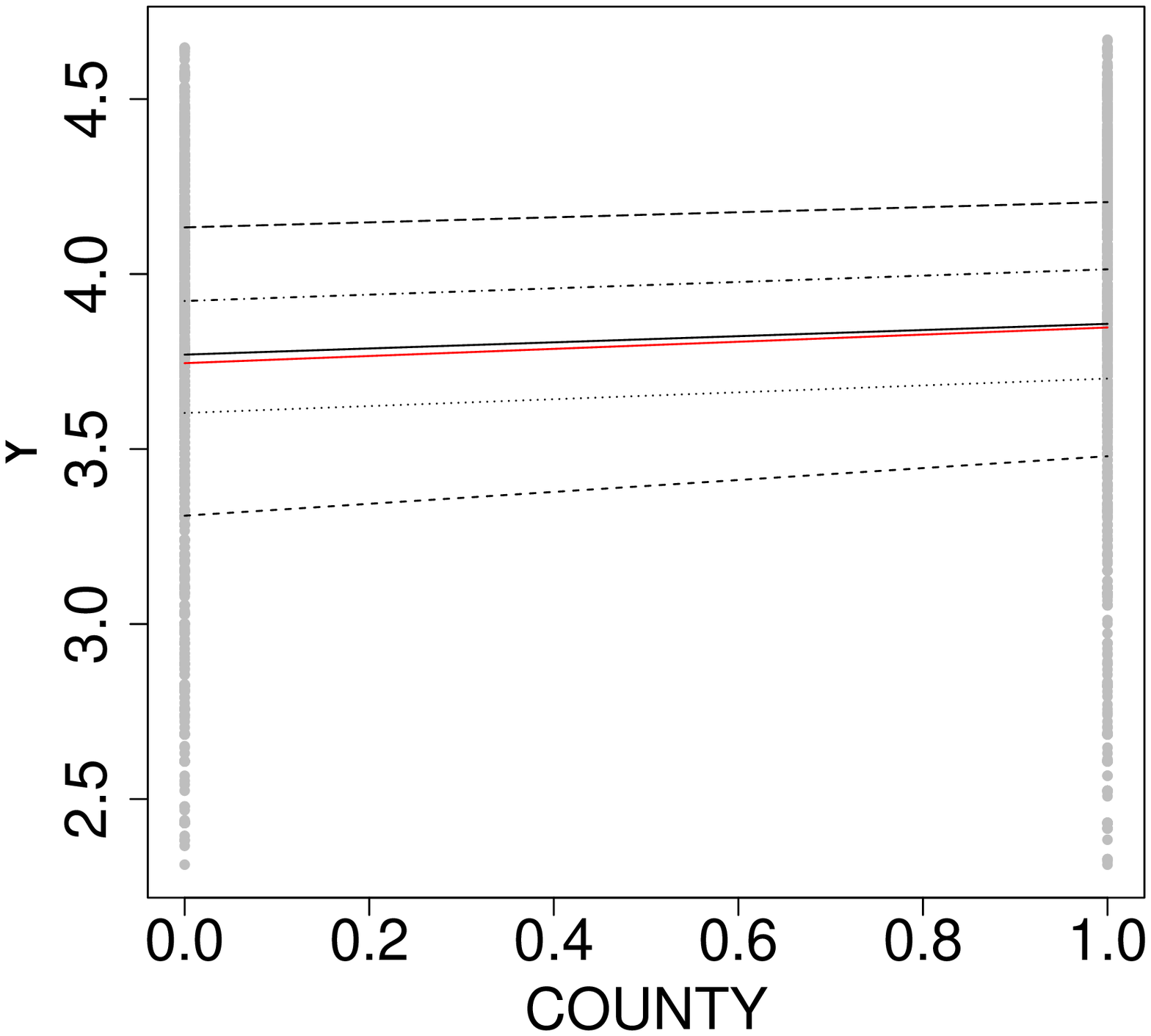}}
\hfil
\subfigure{\includegraphics[width=1.8in,trim=0 30 0 0]{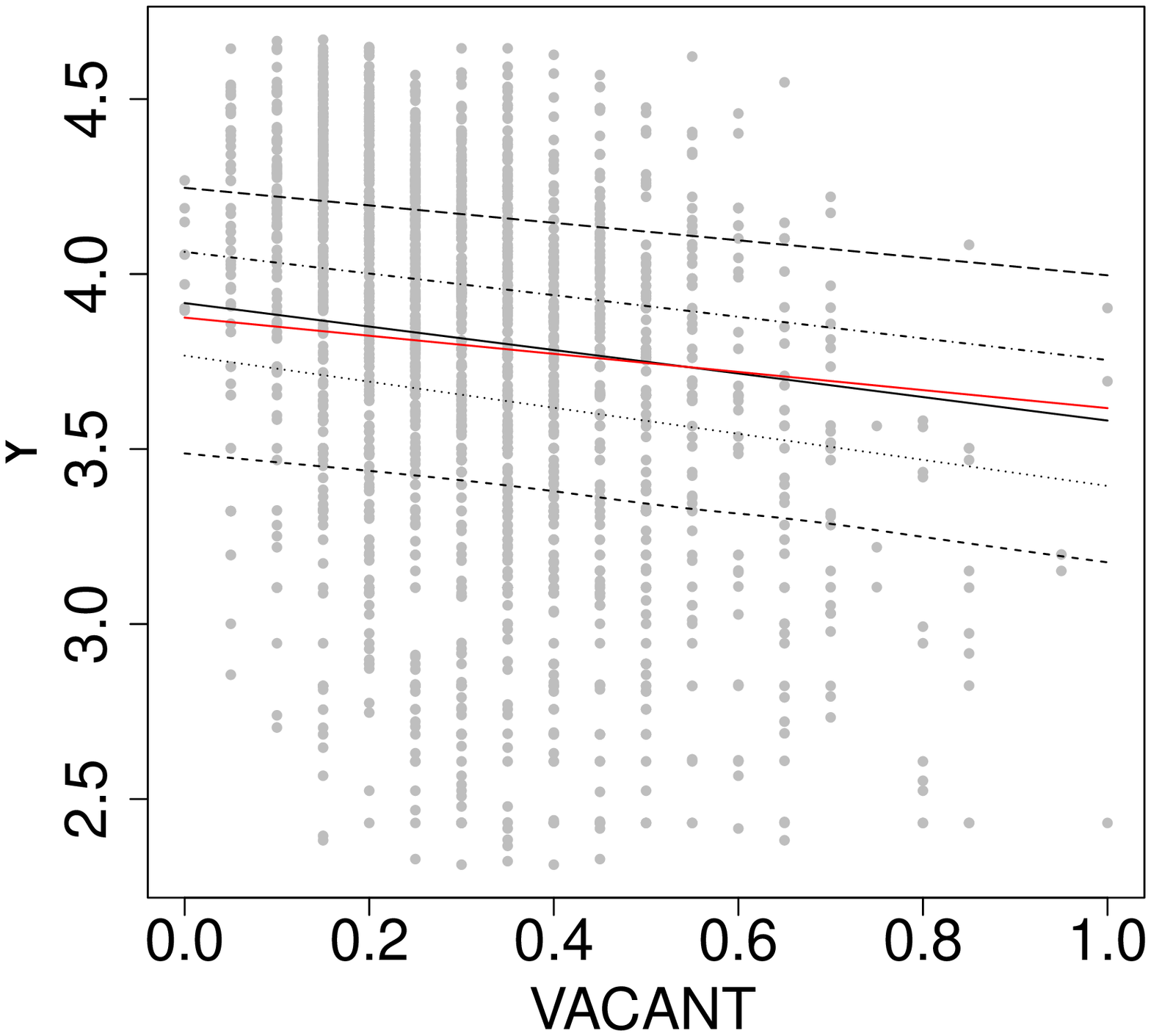}}
}
\vspace{-0.2in}
\centerline{
\subfigure{\includegraphics[width=1.8in,trim=0 30 0 0]{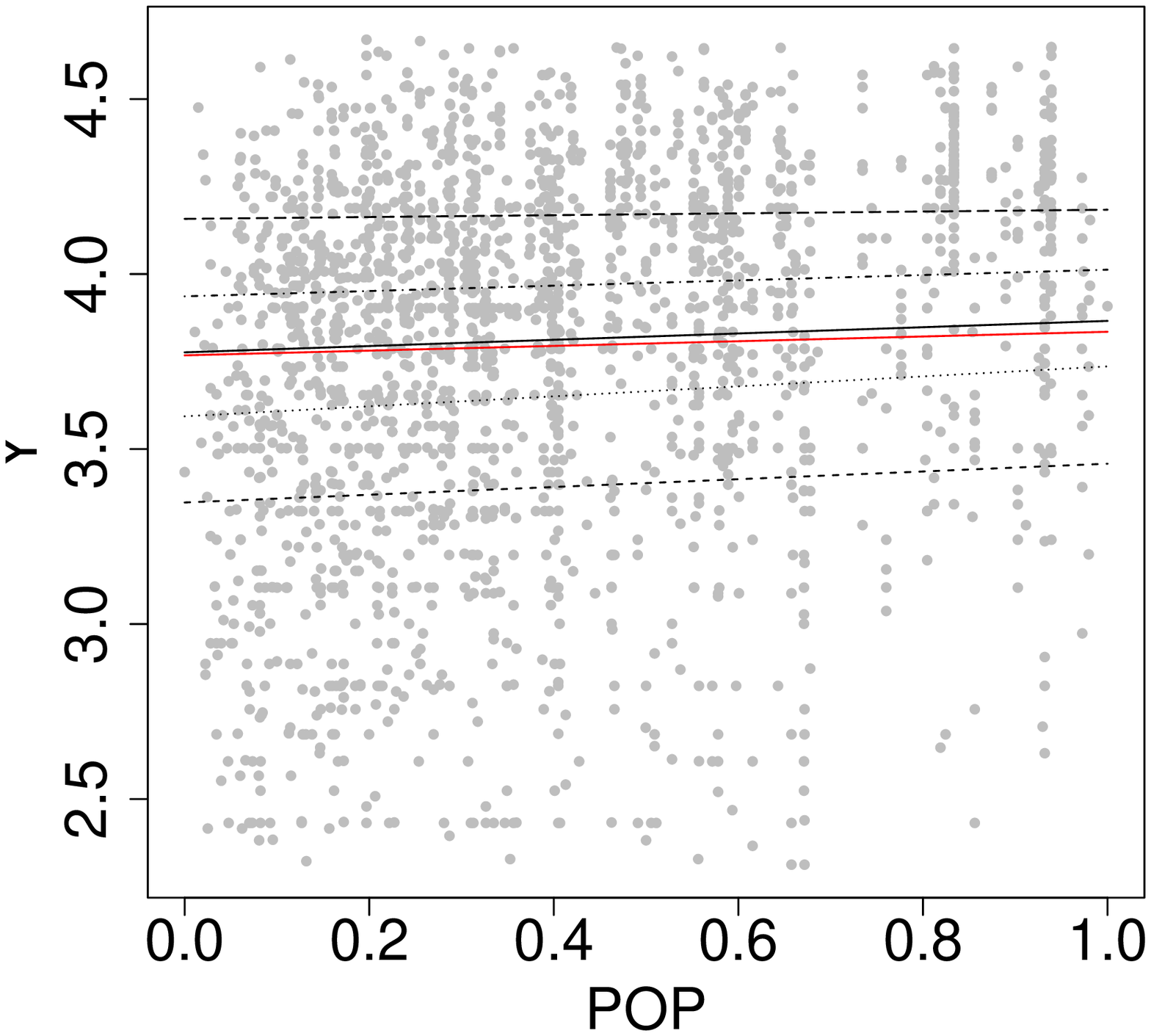}}
\hfil
\subfigure{\includegraphics[width=1.8in,trim=0 30 0 0]{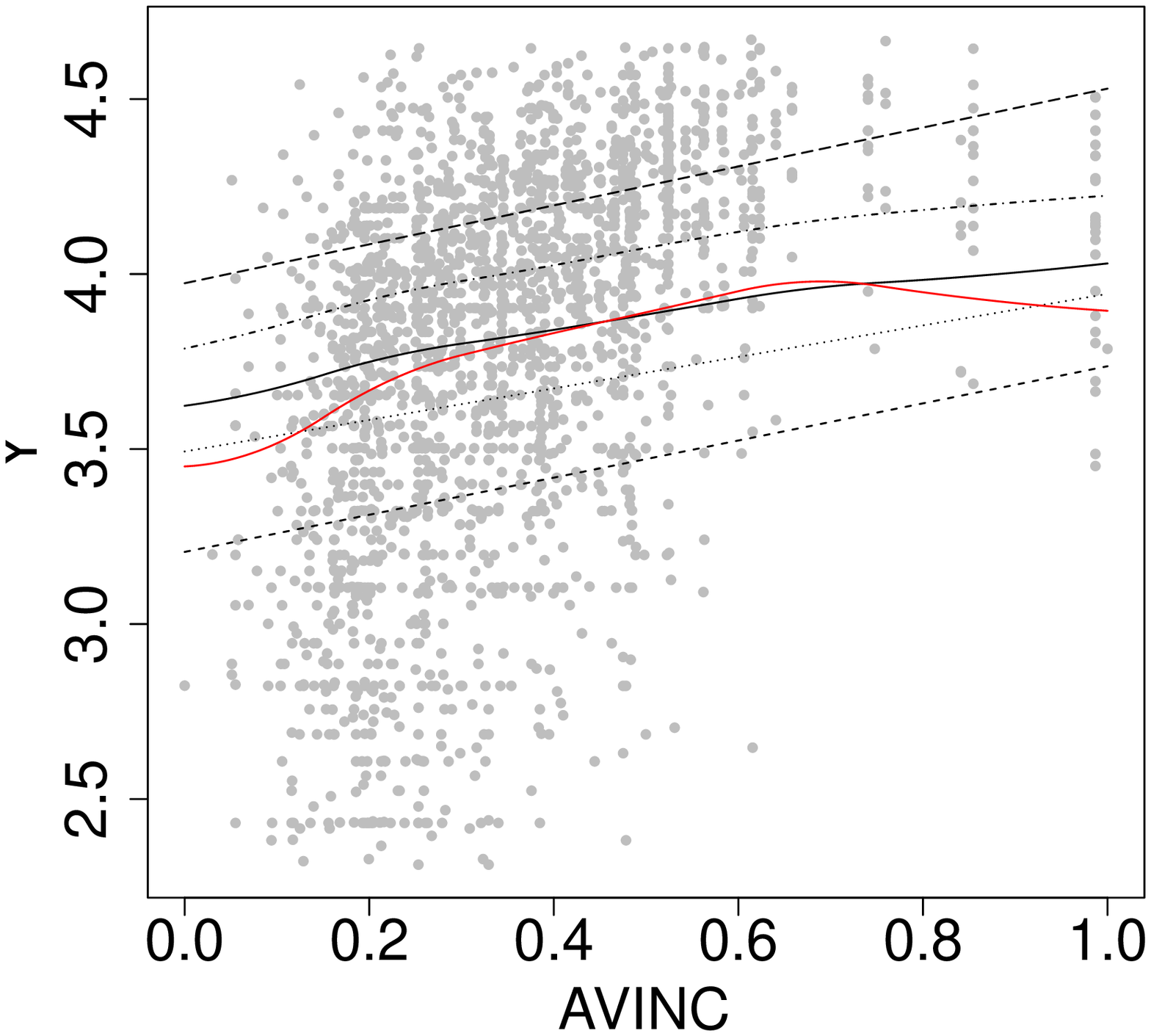}}
}
\vspace{-0.2in}
\centerline{
\subfigure{\includegraphics[width=1.8in,trim=0 30 0 0]{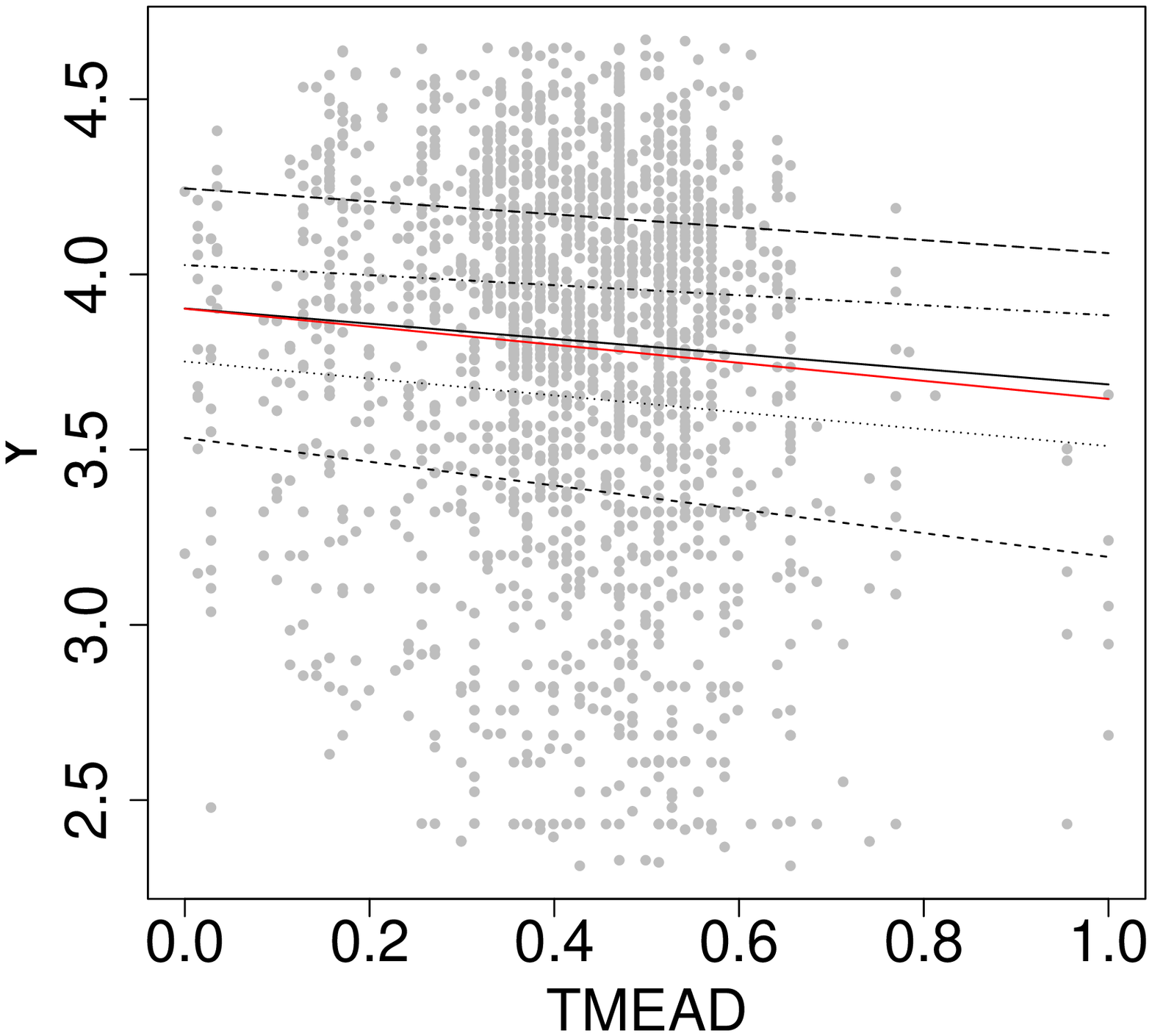}}
\hfil
\subfigure{\includegraphics[width=1.8in,trim=0 30 0 0]{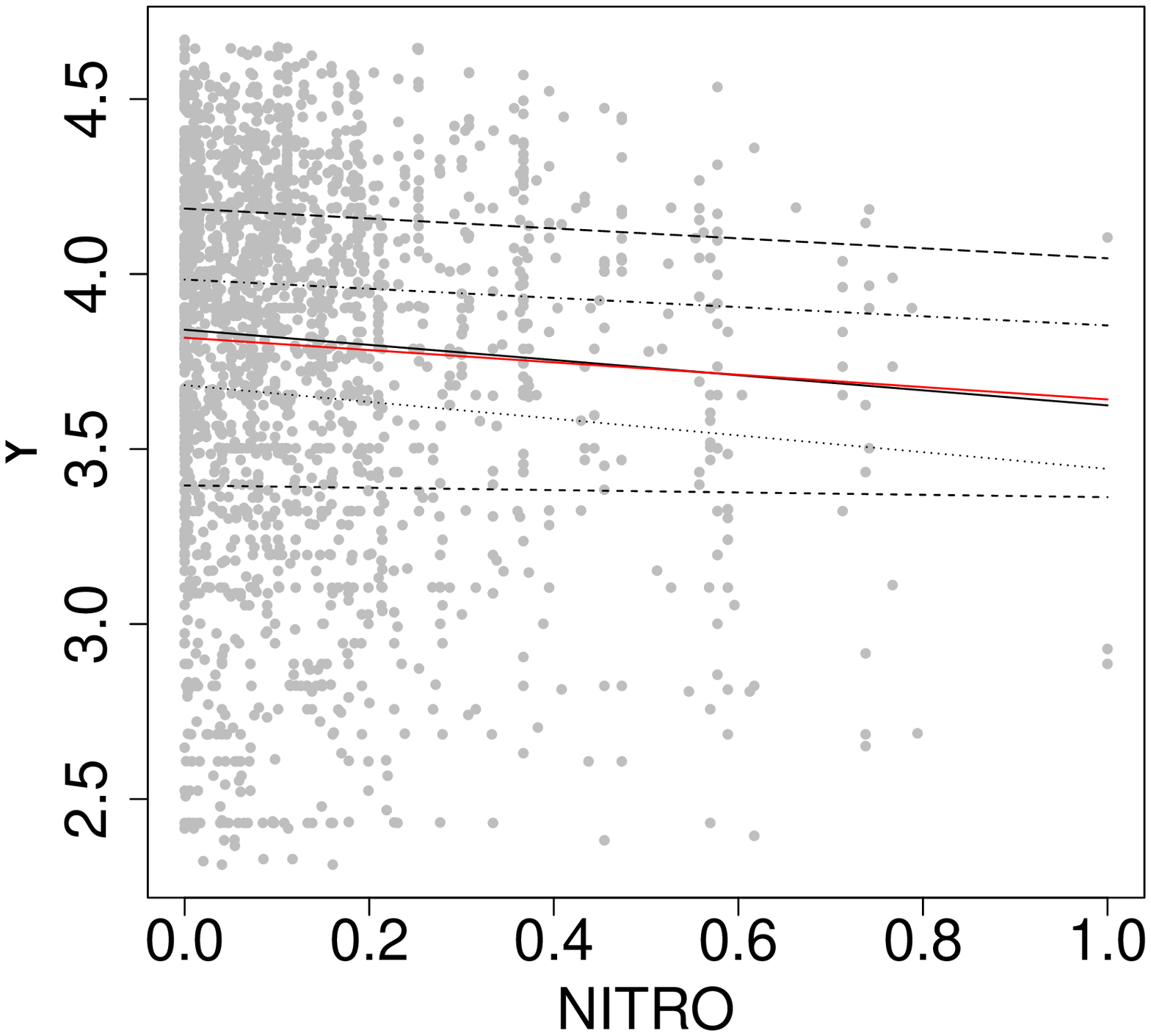}}
}
\caption{The fitted regression functions for housing price data when one covariate varies and others are fixed at 0.5, at quantile levels $\tau=0.1, 0.3, 0.5, 0.7, 0.9$. The red solid lines are the fitted curves of mean regression.}
\label{fact_plot_house}
\end{figure}

We display the barplots in Figure \ref{indicator_posterior_house} and the fitted regression functions in Figure \ref{fact_plot_house}. The results show that all the components are identified as  nonzero components and 4 of them are nonlinear for mean regression. For quantile regression at the 5 quantile levels, we identified 9, 10, 10, 10 and 9 nonzero components, and 1, 1, 1, 0, 0 nonlinear components respectively. Examining the influence of the physical characteristics of houses on prices at different levels, we find that a larger number of rooms, a bigger lot size and the fact that a house has been repaired are factors contributing to an increase in the price of a house, while an older age has a negative impact on the price. Similarly, the characteristics of the surrounding districts show their effects on house prices that conform to our expectations. For example, the price of  houses located in the districts of the most urbanized country of Brittany (Ille-et-Vilaine) is higher, and the price of residential houses located in districts with lower housing vacancy rates is higher. For the two environmental indicators, we reveal their negative effects on house prices. The effects of covariates on responses at different quantile levels show some degree of heterogeneity. For example, the effect of AGE is linear at quantile 0.9, but is nonlinear at lower quantiles; the negative effect of NITRO is not obvious at quantile 0.1, but more obvious at upper quantiles.

\section{Discussions}
In this article, we have proposed a Bayesian quantile regression method for partially linear additive models, which explicitly models components that have linear and nonlinear effects. As detailed in the Appendix, we designed an efficient MCMC algorithm for posterior inferences. With simulation studies, we illustrated the  empirical performances of our proposed quantile regression approach and demonstrated the efficacy of the two sets of indicator variables. The method can automatically determine the type of component effects. The performance of the proposed approach in our simulations is quite encouraging, even when $p$ is large.

Note that the error distributions we use in the simulation examples are very different from the asymmetric Laplace distribution.  Our simulation results show that our methods perform well even when the errors are not generated based on their assumed generating mechanism. Theoretically it has been recently demonstrated that Bayesian quantile regression based on the asymmetric Laplace distribution is consistent even when the error distribution is misspecified \cite{sriram13}. \cite{li2010bayesian} and \cite{hu2011bayesian} have also discussed this problem and showed that Bayesian methods are not sensitive to this assumption.  

Finally, using asymmetric Laplace distribution is only one among a few possible alternatives for Bayesian quantile regression. A problem of the current approach is that regression functions at different quantile levels are separately estimated, which can cause the quantile curves at different levels to cross each other. Other approaches may be able to address this issue, but we choose to use asymmetric Laplace distribution for its simplicity and computational convenience.


\section*{Appendix: MCMC algorithm details}
The joint distribution of all the variables is 
\begin{equation*}
\begin{split}
&p(\boldsymbol{\alpha}, \{\boldsymbol{\beta}_j\}, \boldsymbol{E},\mu,\delta_{0}| \boldsymbol{y}, \boldsymbol{x})\\
\propto&\exp\{-\frac{1}{2}(\boldsymbol{y}-\mu\boldsymbol{1}_n-\boldsymbol{B}_{0}\boldsymbol{\alpha}-\mathop{\sum}\limits_{j=1}\limits^{p}\boldsymbol{B}_{j} \boldsymbol{\beta}_{j}-k_{1}\boldsymbol{e})^{T}\boldsymbol{E}^{-1}\\
&(\boldsymbol{y}-\mu\boldsymbol{1}_n-\boldsymbol{B}_{0}\boldsymbol{\alpha}-\mathop{\sum}\limits_{j=1}\limits^{p}\boldsymbol{B}_{j} \boldsymbol{\beta}_{j}-k_{1}\boldsymbol{e})\}\times\det[\boldsymbol{E}]^{-1/2}\\
&\times p(\boldsymbol{\alpha})\times \mathop{\prod}\limits_{j=1}\limits^{p}p(\boldsymbol{\beta}_j)\times p(\delta_{0})\times\mathop{\prod}\limits_{i=1}\limits^{n}p(e_i)\times p(\mu),\\
\end{split}
\end{equation*}
where $p(\boldsymbol{\beta}_j)$, $p(\boldsymbol{\alpha})$, $p(e_i)$, $p(\delta_{0})$ and $p(\mu)$ are the prior distributions of $\boldsymbol{\beta}_j$, $\boldsymbol{\alpha}$, $e_i$, $\delta_{0}$, and $\mu$ respectively.

We use the Metropolis-within-Gibbs algorithm to sample from the posterior distribution.  We integrate out $\alpha_j$ in step 3 and $\boldsymbol{\beta}_j$ in step 5 to improve mixing of the Markov Chain. The posterior distribution of each variable is as follows ($\sim$ denotes all variables except the one to be sampled):

\begin{enumerate}
\item Sample $p(\alpha_j|\sim)=p (\alpha_j|\boldsymbol{y}^{*},\boldsymbol{e}, \delta_{0}, \sigma_j^{2},\gamma^{(\alpha)}_j ),~ j=1, \ldots, p$,  from the
  conditional distribution of $\alpha_j$,
\begin{equation*}
\label{alpha_posterior_qr}
\begin{split}
p(\alpha_j|\boldsymbol{y}^{*},\boldsymbol{e},\delta_{0}, \sigma_j^{2},\gamma^{(\alpha)}_j=1
)&\sim N (\mu_j, \xi_j^{2}),\\
p(\alpha_j|\boldsymbol{y}^{*},\boldsymbol{e},\delta_{0}, \sigma_j^{2},\gamma^{(\alpha)}_j=0
)&=0,
\end{split}
\end{equation*} 
where $\boldsymbol{y}^{*}=\boldsymbol{y}-\mu\boldsymbol{1}_n-\mathop{\sum}\limits_{i\neq
  j}\limits^{p} \alpha_i\boldsymbol{B}_{i0}-\mathop{\sum}\limits_{i=1
  }\limits^{p} \boldsymbol{B}_{i} \boldsymbol{\beta}_{i}-k_{1}\boldsymbol{e}$, $\xi^{2}_j
  =(\boldsymbol{B}^{T}_{j0}\boldsymbol{E}^{-1}\boldsymbol{B}_{j0}+\frac{1}{\sigma_j^{2}})^{-1}$
  and $\mu_j=\xi^{2}_j \boldsymbol{B}^{T}_{j0}
  \boldsymbol{E}^{-1}\boldsymbol{y}^{*}$.

  \item Sample $p(\mu|\sim)=p (\mu|\boldsymbol{y}^{*},\boldsymbol{e},\delta_{0})$, from the conditional distribution of $\mu$,
\begin{equation*}
\label{b0_posterior_qr}
\begin{split}
p(\mu|\boldsymbol{y}^{*},\boldsymbol{e}, \delta_{0})&\sim N (\mu_0, \xi_0^{2}),\\
\end{split}
\end{equation*} 
where $\xi^{2}_0=k_2\delta_{0}(\mathop{\sum}\limits_{i=1}\limits^{n}e_i^{-1})^{-1}$, $\mu_0=\xi^{2}_0\boldsymbol{1}^{T}_{n}
  \boldsymbol{E}^{-1}\boldsymbol{y}^{*}$, and $\boldsymbol{y}^{*}=\boldsymbol{y}-\mathop{\sum}\limits_{i=1}\limits^{p} \alpha_i\boldsymbol{B}_{i0}-\mathop{\sum}\limits_{i=1
  }\limits^{p} \boldsymbol{B}_{i} \boldsymbol{\beta}_{i}-k_1\boldsymbol{e}$.

  \item Sample $p(\gamma^{(\alpha)}_j|\sim)=p(\gamma^{(\alpha)}_j|\boldsymbol{y}^{*},
  \boldsymbol{e},\delta_{0}, \sigma_j^{2}), ~j=1, \ldots, p$, from its conditional posterior after
  integrating over $\alpha_j$, $p(\gamma^{(\alpha)}_j=1|\boldsymbol{y}^{*},
  \boldsymbol{e},\delta_{0}, \sigma_j^{2})=\frac{1}{1+h}$, with
  
\begin{equation*}
\begin{split}
h~=&~h_1\frac{p(\gamma_j^{(\alpha)}=0|\gamma^{(\alpha)}_{i\neq
    j})}{p(\gamma_j^{(\alpha)}=1|\gamma^{(\alpha)}_{i\neq
    j})},\\
h_1=&~\exp\{-\frac{1}{2}(\boldsymbol{B}^{T}_{j0}\boldsymbol{E}^{-1}
\boldsymbol{y}^{*})^{2}(\boldsymbol{B}^{T}_{j0}\boldsymbol{E}^{-1}
\boldsymbol{B}_{j0}+\frac{1}{\sigma^2_j})^{-1}\}\\
&\times (\sigma_j^{2}\boldsymbol{B}^{T}_{j0}\boldsymbol{E}^{-1}
\boldsymbol{B}_{j0}+1)^{\frac{1}{2}},\\
\end{split}
\end{equation*} 

where $p(\gamma^{(\alpha)}_j=0|\gamma^{(\alpha)}_{i\neq
  j})=(p-q_{\gamma^{(\alpha)}_{0}})/(p+1)$ and  $p(\gamma^{(\alpha)}_j=1|\gamma^{(\alpha)}_{i\neq
  j})=(1+q_{\gamma^{(\alpha)}_{0}})/(p+1)$ with
$\gamma^{(\alpha)}_{0}=(\gamma^{(\alpha)}_1,\ldots,\gamma^{(\alpha)}_{j-1},0,\gamma^{(\alpha)}_{j+1},\ldots,\gamma^{(\alpha)}_{p})^{T}$,
and $\boldsymbol{y}^{*}=\boldsymbol{y}-\mu\boldsymbol{1}_n-\mathop{\sum}\limits_{i\neq
  j}\limits^{p} \alpha_i\boldsymbol{B}_{i0}-\mathop{\sum}\limits_{i=1
  }\limits^{p} \boldsymbol{B}_{i} \boldsymbol{\beta}_{i}-k_{1}\boldsymbol{e} $.\\

\item Sample $p(\boldsymbol{\beta}_j|\sim)=p(\boldsymbol{\beta}_j|\boldsymbol{y}^{*},
 \boldsymbol{e}, \delta_{0}, \tau^{2}_j, \gamma^{(\boldsymbol{\beta})}_j), ~j=1, \ldots, p$, 
 \begin{equation*}
\begin{split}
p(\boldsymbol{\beta}_j|\boldsymbol{y}^{*},\boldsymbol{e},\delta_{0}, \tau_j^{2},\gamma^{(\boldsymbol{\beta})}_j=1
)&\sim N (\boldsymbol{\mu}_j, \boldsymbol{\Sigma}_j),\\
p(\boldsymbol{\beta}_j|\boldsymbol{y}^{*},\boldsymbol{e},\delta_{0}, \tau_j^{2},\gamma^{(\boldsymbol{\beta})}_j=0
)&=0,
\end{split}
\end{equation*} 
where  $\boldsymbol{y}^{*}=\boldsymbol{y}-\mu\boldsymbol{1}_n-\mathop{\sum}\limits_{i=1
  }\limits^{p} \alpha_i \boldsymbol{B}_{i0}-\mathop{\sum}\limits_{i\neq j
  }\limits^{p} \boldsymbol{B}_{i} \boldsymbol{\beta}_{i}-k_1\boldsymbol{e}$,
  $\boldsymbol{\mu}_j=\boldsymbol{\Sigma}_j
  \boldsymbol{B}^{T}_j \boldsymbol{E}^{-1}\boldsymbol{y}^{*}$, 
  $\boldsymbol{\Sigma}_j=(\boldsymbol{B}^{T}_j \boldsymbol{E}^{-1}
  \boldsymbol{B}_j+\frac{1}{\tau^{2}_j}\boldsymbol{\Omega}_j)^{-1}$. \\

  \item Sample $p(\gamma^{(\boldsymbol{\beta})}_j|\sim)=p(\gamma^{(\boldsymbol{\beta})}_j|\boldsymbol{y}^{*},\boldsymbol{e},\delta_{0}, \tau_j^{2}),~ j=1, \ldots, p$, from its conditional posterior after
  integrating ove $\boldsymbol{\beta}_j$, $p(\gamma^{(\boldsymbol{\beta})}_j=1|\boldsymbol{y}^{*},\boldsymbol{e},\delta_{0}, \tau_j^{2})=\frac{1}{1+h}$, with
\begin{equation*}
\begin{split}
h~=&~h_{1}\frac{p(\gamma_j^{(\boldsymbol{\beta})}=0|~\gamma^{(\boldsymbol{\beta})}_{i\neq
 j})}{p(\gamma_j^{(\boldsymbol{\beta})}=1|~\gamma^{(\boldsymbol{\beta})}_{i\neq j})},\\
h_1=&~\small{\exp\{-\frac{1}{2}[\boldsymbol{y}^{*T}\boldsymbol{E}^{-1}\boldsymbol{B}_j(\boldsymbol{B}_j^{T}\boldsymbol{E}^{-1}\boldsymbol{B}_j+\frac{1}{\tau_j^{2}}\boldsymbol{\Omega}_j)^{-1}\boldsymbol{B}_j^T \boldsymbol{E}^{-1}\boldsymbol{y}^{*}]\}}\\
&\times \det [\frac{1}{\tau^{2}_{j}}\boldsymbol{\Omega}_j]^{-\frac{1}{2}}\times \det [\boldsymbol{B}_j^T\boldsymbol{E}^{-1}
 \boldsymbol{B}_j+\frac{1}{\tau_j^{2}}\boldsymbol{\Omega}_j]^{\frac{1}{2}},\\
 \end{split}
\end{equation*}
where  $\boldsymbol{y}^{*}=\boldsymbol{y}-\mu\boldsymbol{1}_n-\mathop{\sum}\limits_{i=1
  }\limits^{p} \alpha_i \boldsymbol{B}_{i0}-\mathop{\sum}\limits_{i\neq j
  }\limits^{p} \boldsymbol{B}_{i} \boldsymbol{\beta}_{i}-k_1\boldsymbol{e}$, $p(\gamma^{(\boldsymbol{\beta})}_j=0|\gamma^{(\boldsymbol{\beta})}_{i\neq
  j})=(p-q_{\gamma^{(\boldsymbol{\beta})}_{0}})/(p+1)$ and  $p(\gamma^{(\boldsymbol{\beta})}_j=1|\gamma^{(\boldsymbol{\beta})}_{i\neq
  j})=(1+q_{\gamma^{(\boldsymbol{\beta})}_{0}})/(p+1)$ with
$\gamma^{(\boldsymbol{\beta})}_{0}=(\gamma^{(\boldsymbol{\beta})}_1,\ldots,\gamma^{(\boldsymbol{\beta})}_{j-1},0,\gamma^{(\boldsymbol{\beta})}_{j+1},\ldots,\gamma^{(\boldsymbol{\beta})}_{p})^{T}$.

\item Sample $\delta_0$,
\begin{equation*}
\label{delta_posterior_qr}
\begin{split}
\delta_{0}&\sim IG (a_1+3n/2,
\nu),\\
\nu&=a_2+((2k_2e_i)^{-1}\mathop{\sum}\limits_{i=1}\limits^{n}(y_{i}-\mathop{\sum}\limits_{j=1}\limits^{p}\alpha_{j} B_{j0}(x_{ij})\\
&-\mathop{\sum}\limits_{j=1}\limits^{p}\mathop{\sum}\limits_{k=1}\limits^{K}\beta_{jk}B_{jk}(x_{ij})-k_{1}e_{i})^{2}+e_{i}).\\
\end{split}
\end{equation*}

\item Sample $\sigma^{2}_j,~ j=1,\ldots,p$, and
  $\tau^{2}_j, ~j=1,\ldots,p$, from their conditional
  posterior distributions if $\gamma^{(\alpha)}_j$, $\gamma^{(\boldsymbol{\beta})}_{j}\neq 0$,
 \begin{equation*}
\label{sigma_posterior_qr}
\begin{split}
\sigma^{2}_j&\sim IG (a_1+1/2,
a_2+(\alpha_j^{2}/2)),\\
\tau^2_j&\sim IG(a_1+K
/2, a_2+(\boldsymbol{\beta}_j^{T} \boldsymbol{\Omega}_j\boldsymbol{\beta}_j/2)).\\
\end{split}
\end{equation*}
Otherwise they are generated from their priors.

\item The full conditional distribution of $e_{i}, ~i=1,\ldots, n$ is a generalized inverse Gaussian distribution $(GIG)$,
\begin{equation*}
\label{E_posterior_qr}
\begin{split}
p(e_{i}|\delta_0,\nu_i)\sim& GIG(\frac{1}{2},
\sqrt{\frac{(y_{i}-\nu_i)^{2}}{k_{2}\delta_0}},\sqrt{\frac{k_{1}^{2}}{k_{2}\delta_0}+\frac{2}{\delta_0}}),\\
\nu_i=&y_{i}-\mu-\mathop{\sum}\limits_{j=1}\limits^{p}\alpha_{j} B_{j0}(x_{ij})-\mathop{\sum}\limits_{j=1}\limits^{p}\mathop{\sum}\limits_{k=1}\limits^{K}\beta_{jk}B_{jk}(x_{ij}),
\end{split}
\end{equation*}
where the probability density function of $GIG(\rho,m,n)$ is 
\begin{equation*}
\label{GIG}
\begin{split}
f(x|\rho,m,n)&=\frac{(n/m)^{\rho}}{2K_{\rho}(mn)}x^{\rho-1}\exp\{-\frac{1}{2}(m^{2}x^{-1}+n^{2}x)\},\\
&x>0,-\infty<\rho<\infty,m\geq 0,n\geq 0,
\end{split}
\end{equation*}
and $K_{\rho}$ is the modified Bessel function of the third kind (\cite{barndorff2001non}).\\
\end{enumerate}

\newpage
\bibliographystyle{elsart-harv}
\bibliography{reference2}

\begin{thebibliography}{31}
\expandafter\ifx\csname natexlab\endcsname\relax\def\natexlab#1{#1}\fi
\expandafter\ifx\csname url\endcsname\relax
  \def\url#1{\texttt{#1}}\fi
\expandafter\ifx\csname urlprefix\endcsname\relax\def\urlprefix{URL }\fi

\bibitem[{Abrevaya(2001)}]{abrevaya2001effects}
Abrevaya, J., 2001. The effects of demographics and maternal behavior on the
  distribution of birth outcomes. Empirical Economics 26~(1), 247--257.

\bibitem[{Barndorff-Nielsen and Shephard(2001)}]{barndorff2001non}
Barndorff-Nielsen, O., Shephard, N., 2001. {Non-Gaussian
  Ornstein--Uhlenbeck-based models and some of their uses in financial
  economics}. Journal of the Royal Statistical Society: Series B (Statistical
  Methodology) 63~(2), 167--241.

\bibitem[{Bontemps et~al.(2008)Bontemps, Simioni, and
  Surry}]{bontemps2008semiparametric}
Bontemps, C., Simioni, M., Surry, Y., 2008. Semiparametric hedonic price
  models: assessing the effects of agricultural nonpoint source pollution.
  {Journal of Applied Econometrics} 23~(6), 825--842.

\bibitem[{Buchinsky(1994)}]{buchinsky1994changes}
Buchinsky, M., 1994. {Changes in the US wage structure 1963-1987: Application
  of quantile regression}. Econometrica: Journal of the Econometric Society
  62~(2), 405--458.

\bibitem[{Cade and Noon(2003)}]{cade2003gentle}
Cade, B.~S., Noon, B.~R., 2003. A gentle introduction to quantile regression
  for ecologists. Frontiers in Ecology and the Environment 1~(8), 412--420.

\bibitem[{Chib and Jeliazkov(2006)}]{chib2006inference}
Chib, S., Jeliazkov, I., 2006. {Inference in semiparametric dynamic models for
  binary longitudinal data}. Journal of the American Statistical Association
  101~(474), 685--700.

\bibitem[{Cripps et~al.(2005)Cripps, Carter, and Kohn}]{cripps2005variable}
Cripps, E., Carter, C., Kohn, R., 2005. Variable selection and covariance
  selection in multivariate regression models. Handbook of Statistics 25~(2),
  519--552.

\bibitem[{De~Gooijer and Zerom(2003)}]{de2003additive}
De~Gooijer, J., Zerom, D., 2003. On additive conditional quantiles with
  high-dimensional covariates. Journal of the American Statistical Association
  98~(461), 135--146.

\bibitem[{George and McCulloch(1993)}]{george1993variable}
George, E., McCulloch, R., 1993. {Variable selection via Gibbs sampling}.
  Journal of the American Statistical Association 88~(423), 881--889.

\bibitem[{Goldstein and Smith(1974)}]{goldstein1974ridge}
Goldstein, M., Smith, A., 1974. {Ridge-type estimators for regression
  analysis}. Journal of the Royal Statistical Society. Series B (Statistical
  Methodology) 36~(2), 284--291.

\bibitem[{Horowitz and Lee(2005)}]{horowitz2005nonparametric}
Horowitz, J., Lee, S., 2005. Nonparametric estimation of an additive quantile
  regression model. Journal of the American Statistical Association 100~(472),
  1238--1249.

\bibitem[{Hu et~al.(2013)Hu, Gramacy, and Lian}]{hu2011bayesian}
Hu, Y., Gramacy, R., Lian, H., 2013. Bayesian quantile regression for
  single-index models. Statistics and Computing to appear.

\bibitem[{Huang et~al.(2010)Huang, Horowitz, and Wei}]{huang2010variable}
Huang, J., Horowitz, J., Wei, F., 2010. Variable selection in nonparametric
  additive models. Annals of Statistics 38~(4), 2282--2312.

\bibitem[{Koenker and Bassett(1978)}]{koenker1978regression}
Koenker, R., Bassett, G., 1978. Regression quantiles. Econometrica: Journal of
  the Econometric Society 46~(1), 33--50.

\bibitem[{Kozumi and Kobayashi(2011)}]{kozumi2011gibbs}
Kozumi, H., Kobayashi, G., 2011. {Gibbs sampling methods for Bayesian quantile
  regression}. Journal of Statistical Computation and Simulation 81~(11),
  1565--1578.

\bibitem[{Li et~al.(2010)Li, Xi, and Lin}]{li2010bayesian}
Li, Q., Xi, R., Lin, N., 2010. Bayesian regularized quantile regression.
  Bayesian Analysis 5~(3), 533--556.

\bibitem[{Lian et~al.(2012)Lian, Chen, and Yang}]{lian2011identification}
Lian, H., Chen, X., Yang, J., 2012. Identification of partially linear
  structure in additive models with an application to gene expression
  prediction from sequences. Biometrics 68~(2), 437--445.

\bibitem[{Meier et~al.(2009)Meier, Van De~Geer, and
  B{\"u}hlmann}]{meier2009high}
Meier, L., Van De~Geer, S., B{\"u}hlmann, P., 2009. High-dimensional additive
  modeling. The Annals of Statistics 37~(6B), 3779--3821.

\bibitem[{Panagiotelis and Smith(2008)}]{anagiotelis2008bayesian}
Panagiotelis, A., Smith, M., 2008. Bayesian identification, selection and
  estimation of semiparametric functions in high-dimensional additive models.
  Journal of Econometrics 143~(2), 291--316.

\bibitem[{Ravikumar et~al.(2009)Ravikumar, Lafferty, Liu, and
  Wasserman}]{ravikumar2009sparse}
Ravikumar, P., Lafferty, J., Liu, H., Wasserman, L., 2009. Sparse additive
  models. Journal of the Royal Statistical Society: Series B (Statistical
  Methodology) 71~(5), 1009--1030.

\bibitem[{Scheipl et~al.(2012)Scheipl, Fahrmeira, and Kneib}]{scheipl12}
Scheipl, F., Fahrmeira, L., Kneib, T., 2012. {Posterior consistency of Bayesian
  quantile regression based on the misspecified asymmetric Laplace density}.
  Journal of the American Statistical Association 107~(500), 1518--1532.

\bibitem[{Shively et~al.(1999)Shively, Kohn, and Wood}]{shively1999variable}
Shively, T., Kohn, R., Wood, S., 1999. Variable selection and function
  estimation in additive nonparametric regression using a data-based prior.
  Journal of the American Statistical Association 94~(447), 777--794.

\bibitem[{Smith and Kohn(1996)}]{smith1996nonparametric}
Smith, M., Kohn, R., 1996. {Nonparametric regression using Bayesian variable
  selection}. Journal of Econometrics 75~(2), 317--343.

\bibitem[{Sriram et~al.(2013)Sriram, Ramamoorthi, and Ghosh}]{sriram13}
Sriram, K., Ramamoorthi, R., Ghosh, P., 2013. {Posterior consistency of
  Bayesian quantile regression based on the misspecified asymmetric Laplace
  density}. Bayesian analysis 8~(2), 1--26.

\bibitem[{Tan(2010)}]{tan2010no}
Tan, C., 2010. No one true path: uncovering the interplay between geography,
  institutions, and fractionalization in economic development. Journal of
  Applied Econometrics 25~(7), 1100--1127.

\bibitem[{van Dyk and Park(2008)}]{van2008partially}
van Dyk, D., Park, T., 2008. {Partially collapsed Gibbs samplers}. Journal of
  the American Statistical Association 103~(482), 790--796.

\bibitem[{Yau et~al.(2003)Yau, Kohn, and Wood}]{yau2003bayesian}
Yau, P., Kohn, R., Wood, S., 2003. Bayesian variable selection and model
  averaging in high-dimensional multinomial nonparametric regression. Journal
  of Computational and Graphical Statistics 12~(1), 23--54.

\bibitem[{Yu and Lu(2004)}]{yu2004local}
Yu, K., Lu, Z., 2004. Local linear additive quantile regression. Scandinavian
  Journal of Statistics 31~(3), 333--346.

\bibitem[{Yu and Moyeed(2011)}]{yu2001bayesian}
Yu, K., Moyeed, R.~A., 2011. Bayesian quantile regression. {Statistics and
  Probability Letters} 54~(4), 437--447.

\bibitem[{Yue and Rue(2011)}]{yue2011bayesian}
Yue, Y., Rue, H., 2011. Bayesian inference for additive mixed quantile
  regression models. Computational Statistics and Data Analysis 55~(1), 84--96.

\bibitem[{Zhang et~al.(2011)Zhang, Cheng, and Liu}]{zhang2011linear}
Zhang, H., Cheng, G., Liu, Y., 2011. {Linear or nonlinear? Automatic structure
  discovery for partially linear models}. Journal of the American Statistical
  Association 106~(495), 1099--1112.

\end{thebibliography}

\end{document}